\documentclass{ieeeaccess}
\usepackage{cite}
\usepackage{amsmath,amssymb,amsfonts}
\usepackage{graphicx}%
\usepackage{dcolumn}%
\usepackage{bm}%
\usepackage{qcircuit}
\usepackage{braket}
\usepackage{xspace}
\usepackage{algorithmic,algorithm}
\usepackage{multirow}
\usepackage{subcaption}
\usepackage{caption}
\usepackage{lscape}
\usepackage{textcomp}
\def\BibTeX{{\rm B\kern-.05em{\sc i\kern-.025em b}\kern-.08em
    T\kern-.1667em\lower.7ex\hbox{E}\kern-.125emX}}
\newcommand{\st}{ST\xspace}
\newcommand{\tcx}{RT3\xspace}
\newcommand{\fcx}{RT4\xspace}
\newcommand{\itcx}{IRT3\xspace}
\newcommand{\ifcx}{IRT4\xspace}
\newcommand{\grt}{GRT\xspace}
\newcommand{\igrt}{IGRT\xspace}
\newcommand{\pgrt}{PGRT\xspace}
\newcommand{\ctrlq}{CTRL\xspace}
\newcommand{\compq}{COMP\xspace}

\begin{document}
\doi{10.1109/TQE.2021.3136195}

\title{Efficient Construction of a Control Modular Adder on a Carry-Lookahead Adder Using Relative-phase Toffoli Gates}%

\author{
\uppercase{Kento~Oonishi}\authorrefmark{1,2},
\uppercase{Tomoki~Tanaka}\authorrefmark{3,4},
\uppercase{Shumpei~Uno}\authorrefmark{4,5},
\uppercase{Takahiko~Satoh}\authorrefmark{4,6},
\uppercase{Rodney~Van~Meter}\authorrefmark{4,7}\IEEEmembership{Senior Member, IEEE},
\uppercase{Noboru Kunihiro}\authorrefmark{8},
}

\address[1]{Graduate School of Information Science and Technology, The University of Tokyo, 7-3-1 Hongo, Bunkyo-ku, Tokyo, 113-8656, Japan (email: oonishi059@gmail.com)}
\address[2]{The Graduate School of Science and Technology, Keio University, 3-14-1 Hiyoshi, Kohoku, Yokohama, Kanagawa, 223-8522, Japan}
\address[3]{Mitsubishi UFJ Financial Group, Inc. and MUFG Bank, Ltd., 2-7-1 Marunouchi, Chiyoda-ku, Tokyo, 100-8388, Japan}
\address[4]{Quantum Computing Center, Keio University, 3-14-1 Hiyoshi, Kohoku-ku, Yokohama, Kanagawa, 223-8522, Japan}
\address[5]{Mizuho Research \& Technologies, Ltd., 2-3 Kanda-Nishikicho, Chiyoda-ku, Tokyo, 101-8443, Japan}
\address[6]{Graduate  School  of  Media  and  Governance, Keio University SFC, 5322, Endo, Fujisawa, Kanagawa 252-0882 Japan}
\address[7]{Faculty of Environment and Information Studies, Keio University SFC, 5322, Endo, Fujisawa, Kanagawa 252-0882 Japan}
\address[8]{University of Tsukuba, 1-1-1 Tennodai, Tsukuba, Ibaraki, 305-8573, Japan}

\tfootnote{The first author is supported by a JSPS Fellowship for Young Scientists.
This work was supported by JSPS KAKENHI Grant Numbers JP20J11754, JP21H03440, MEXT Quantum Leap Flagship Program Grant Number JPMXS0118067285, and JST CREST Grant Number JPMJCR14D6, Japan.
This paper is written based on Chapter~6 of the dissertation by Oonishi~\cite{Oon2020}.
}

\markboth
{Oonishi \headeretal: Efficient Construction of a Control Modular Adder on a Carry-Lookahead Adder Using Relative-phase Toffoli Gates}
{Oonishi \headeretal: Efficient Construction of a Control Modular Adder on a Carry-Lookahead Adder Using Relative-phase Toffoli Gates}

\corresp{Corresponding author: Kento Oonishi (email: oonishi059@gmail.com).}

\begin{abstract}
Control modular addition is a core arithmetic function, and we must consider the computational cost for actual quantum computers to realize efficient implementation.
To achieve a low computational cost in a control modular adder, we focus on minimizing KQ, defined by the product of the number of qubits and the depth of the circuit.
In this paper, we construct an efficient control modular adder with small KQ by using relative-phase Toffoli gates in two major types of quantum computers:
Fault-Tolerant Quantum Computers~(FTQ) on the Logical layer and Noisy Intermediate-Scale Quantum Computers~(NISQ).
We give a more efficient construction compared to Van Meter and Itoh's, based on a carry-lookahead adder.
In FTQ, $T$ gates incur heavy cost due to distillation, which fabricates ancilla for running $T$ gates with high accuracy but consumes a lot of specially prepared ancilla qubits and a lot of time.
Thus, we must reduce the number of $T$ gates.
We propose a new control modular adder that uses only 20\% of the number of $T$ gates of the original.
Moreover, when we take distillation into consideration, we find that we minimize $\text{KQ}_{T}$ (the product of the number of qubits and $T$-depth) by running $\Theta\left(n / \sqrt{\log n} \right)$ $T$ gates simultaneously.
In NISQ, CNOT gates are the major error source.
We propose a new control modular adder that uses only 35\% of the number of CNOT gates of the original.
Moreover, we show that the $\text{KQ}_{\text{CX}}$ (the product of the number of qubits and CNOT-depth) of our circuit is 38\% of the original.
Thus, we realize an efficient control modular adder, improving prospects for the efficient execution of arithmetic in quantum computers.
\end{abstract}


\titlepgskip=-15pt

\maketitle

\section{Introduction}

Recently, functional but imperfect quantum computers have emerged, called Noisy Intermediate-Scale Quantum Computers~(NISQ)~\cite{Pre2018}, with machines from IBM~\cite{corcoles2019challenges,IBMQE}, Google~\cite{Goo2019}, Rigetti~\cite{SCZ2016}, IonQ~\cite{MDB2020}, and Honeywell~\cite{MPD2020} all accessible via the web.

Many researchers have constructed simple quantum circuits for NISQ machines. Researchers at IBM implemented the first $15 = 3 \times 5$ factoring circuit on a liquid NMR machine in 2001~\cite{VSB2001}. Since then, researchers have implemented Shor's algorithm on a variety of machines~\cite{PMO2009,LBC2012,MLL2012,MNM2016,lu2007demonstration,lanyon2007experimental}, though care must be taken not to extrapolate too far from these demonstrations~\cite{smolin2013oversimplifying}. Researchers have also demonstrated small instances of Grover's algorithm~\cite{BGH2020,SOV2020}.

However, we cannot realize large-scale quantum computation on NISQ, due to the high error rate.
These errors propagate as the calculation proceeds, and we cannot extract the correct result.
Thus, we must reduce the error rate in quantum computers.
To realize computation with high accuracy, research on Fault-Tolerant Quantum Computers~(FTQ) is proceeding~\cite{DMN2013,Pre1998,Ste1998}.

Jones et al.~\cite{JMF2012} proposed a method for constructing FTQ as a layered architecture.
Specifically, we conduct the accurate computation on the Logical layer, which is achieved using large numbers of physical qubits with errors.

However, $T$~gates impose an additional cost when run on FTQ.
By the Gottesman-Knill theorem~\cite{NC2002}, we can conduct classical simulation of quantum circuits composed only of Clifford gates, but to realize universal quantum computation, we require non-Clifford gates such as a $T$~gate, taking us into a realm that cannot be simulated classically.
We achieve high-fidelity $T$ gates by incorporating distillation~\cite{FSG2009}, which requires a lot of logical qubits and a lot of time; research on optimization of distillation is being carried out~\cite{GF2019}.
In FTQ, we may realize large-scale quantum algorithms such as Shor's algorithm~\cite{Sho1999} and Grover's algorithm~\cite{Gro1996}.
Shor's algorithm is of particular interest \emph{if} it can implemented effectively, because it solves the factorization problem or the discrete logarithm problem in polynomial time, breaking the security of current cryptosystems, such as RSA~\cite{RSA1978} or elliptic curve~\cite{Kob1987,Mil1985}, whose security is based on the factorization problem or the discrete logarithm problem, respectively.

In Shor's algorithm, the control modular exponentiation step dominates the total cost, leading many researchers to study its construction~\cite{BCDP1996,DRGG2016,FDH2004,GE2019,HJN2020,MS2012,PG2014,RNSL2017,VI2005,VBE1996,Zal1998}.
The control modular exponentiation calculates 
\begin{equation}
\label{eq:exponentiation}
\ket{j} \ket{1} \to \ket{j} \ket{a^{j} \bmod N}.
\end{equation}
In eq.(\ref{eq:exponentiation}), $a$, $N$ and $j$ are non-negative integers satisfying $a < N$.

One strategy realizes a control modular exponentiation by the repeated calling of control modular additions.
More precisely, this construction is realized by following two steps:
\begin{enumerate}
\item Decomposition of a control modular exponentiation into control modular multipliers
\item Decomposition of a control modular multiplier into control modular adders
\end{enumerate}

The first decomposition is based on the following equation where $j$ can be expressed in $n$-bit, namely $j = \left(j_{n-1} \ldots j_{0}\right)_{2}$:
\begin{equation}
\label{eq:decompex}
a^{j} \bmod N = \displaystyle \prod_{l=0}^{n-1} \left(a^{2^{l}} \bmod N\right)^{j_{l}}  \bmod N
\end{equation}
In eq.(\ref{eq:decompex}), $a^{j}$ is decomposed into $n$-times multiplications, namely $a^{2^{l}j_{l}}$.
For example, a exponentiation $2^{5}$ is decomposed into $2^{5} = 2^{101_{2}} = 2^{4} \times 2^{1}$.

Next, we consider the second decomposition.
In quantum computation, a multiplication $ka$ is based on the following operation:
\begin{equation}
\label{eq:decompmul}
\ket{k}\ket{0} 
\to \ket{k}\ket{0 + ka}
\to \ket{k-a^{-1}(ka)}\ket{ka}
= \ket{0} \ket{ka}
\end{equation}
The operation of (\ref{eq:decompmul}) requires an addition by the result of multiplication.
This addition can be decomposed as follows, where $a$ and $b$ are non-negative integers less than $N$, and $k$ is a $n$-bit number expressed as $k = \left(k_{n-1} \ldots k_{0}\right)_{2}$:
\begin{equation}
\label{eq:decompmul2}
b + ak \bmod N = 
b + \displaystyle \sum_{l=0}^{n-1} \left(a2^{k_{l}} \bmod N\right) \bmod N
\end{equation}
For example, $6 \cdot 5$ is decomposed into 
$6 \cdot 5 
= 6 \cdot (4 + 1) 
= 6\cdot 4 + 6 \cdot 1$, because $5 = 101_{2} = 4 + 1$.

From the above discussion, a control modular exponentiation is decomposed into control modular adders.
Thus, if we reduce the cost of a control modular adder, the total cost of Shor's algorithm will shrink.
In this paper, we focus on the efficient construction of a control modular addition.

\subsection{Background}

\begin{figure}
\[
\Qcircuit @C=1em @R=.7em {
  \lstick{\ket{x}} & \qw & \ctrl{1}  & \rstick{\ket{x}} \qw \\
  \lstick{\ket{b}} & {/} \qw & \gate{\text{\rotatebox{90}{ModADD}}} & \rstick{\ket{b+ax \bmod N}} \qw
}
\]
\caption{\label{fig:Madd} 
Overview of a control modular adder.
The first register has a single qubit which is used as a control bit.
The second register has $n$ qubits which are used to store the result.
$a$ and $N$ are $n$-bit classical numbers.
}
\end{figure}
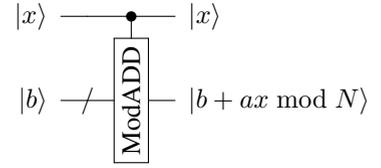

A control modular addition is defined by a control qubit $x$ and $n$-bit numbers $a, b$, and $N$.
$a$ and $b$ satisfy $0 \leq a,b \leq N-1$, and $a$ and $N$ are classical numbers.
A control modular addition calculates
\begin{equation}
\label{eq:Madd}
\ket{x}\ket{b} \to \ket{x}\ket{b+xa \bmod N}.
\end{equation}
An overview is shown in Figure~\ref{fig:Madd}.

However, the optimal construction of a control modular adder is not obvious.
A control modular adder is constructed from simple adders~\cite{DRGG2016,FDH2004,HJN2020,RNSL2017,VI2005,VBE1996,Zal1998}, and there are many kinds of adders~\cite{CDKM2004,DKRS2006,Dra2000,Gid2018,TTK2010,VI2005,VBE1996}.
Previous constructions follow similar overall structure, but differ in detail.
We need to determine which combination is the best.

This paper focuses on minimizing KQ~\cite{Ste2003} to construct a circuit with low execution cost.
KQ is defined by the product of the number of qubits and the depth of the circuit.
Minimizing KQ benefits both FTQ~\cite{JMF2012} and NISQ~\cite{SOV2020}.
Much previous research focuses only on depth or the number of qubits, but reducing only one metric improves only one performance. We believe KQ more accurately represents the total resource consumption, especially for deep circuits, capturing the total "qubit-time steps" of a circuit.

This paper proposes our new circuit based on Van Meter and Itoh's construction~\cite{VI2005}, which uses three of Draper et al.'s carry-lookahead adders~\cite{DKRS2006}.
Van Meter and Itoh's construction has small KQ values than the other constructions but has room for further minimization of KQ.
For example, Thapliyal et al.~\cite{TMK2020} proposed a means of minimizing the number of $T$~gates in a carry-lookahead adder by using Gidney's relative-phase Toffoli gates~\cite{Gid2018}.
Thapliyal et al.'s construction reduces KQ in FTQ, but similar optimization can be applied to NISQ by Maslov's relative-phase Toffoli gates~\cite{Mas2016}.
Thus, we reduce KQ by applying relative-phase Toffoli gates on the Van Meter-Itoh construction.

\subsection{Our Contribution}

\begin{figure}
\centering
\includegraphics[keepaspectratio, scale=0.25]{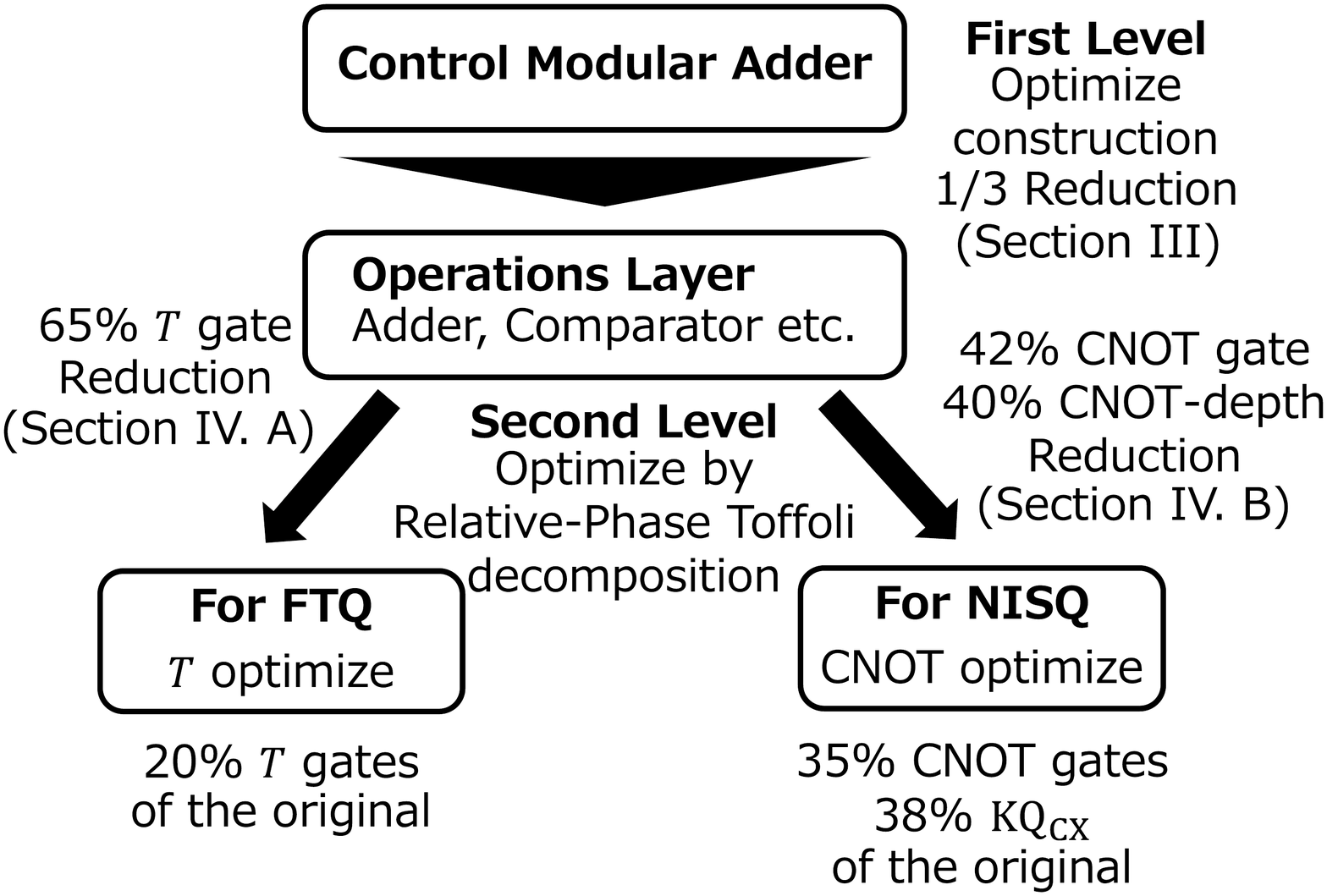}
\caption{\label{fig:Intro} 
Optimization of a control modular adder.
In first-level optimization, we optimize the construction of a control modular adder.
In second-level optimization, we minimize KQ for FTQ or NISQ by using relative-phase Toffoli gates.}
\end{figure}

In this study, we propose a method for optimizing a control modular adder based on a carry-lookahead adder for both FTQ and NISQ.
We apply two-level optimization on the original Van Meter-Itoh construction~\cite{VI2005} as in Figure~\ref{fig:Intro}.

In first-level optimization, we optimize the construction of a control modular adder (Section~III).
Specifically, we optimize by focusing on the efficiency of the comparator in a carry-lookahead adder and reduce some control operations by taking advantage of the classicality of $a$ and $N$.

In second-level optimization, we minimize KQ for FTQ or NISQ by using relative-phase Toffoli gates (Section~IV).
In this study, we assumed all qubits are connected, without considering the physical or logical topology~\cite{CV2012,FDH2004,HNY2011}.
We assume full connectivity because some current NISQ machines, such as IonQ~\cite{MDB2020} and Honeywell~\cite{MPD2020}, realize full connectivity. Future work must consider the problem of mapping circuits to other NISQ machines and to FTQ machines.

First, we clarify the definition of KQ in each device, because the cost of gates is different in FTQ or NISQ. For many implementations, the most expensive gates are $T$~gates~\cite{JMF2012} CNOT gates~\cite{corcoles2019challenges,IBMQE}, respectively. 
We define $\text{KQ}_{T}$ for use with FTQ and $\text{KQ}_{\text{CX}}$ on NISQ as the product of the number of qubits and $T$-depth or CNOT-depth, respectively.
Next, we use Gidney's relative-phase Toffoli gates~\cite{Gid2018} in FTQ circuits and Maslov's relative-phase Toffoli gates~\cite{Mas2016} in NISQ circuits, instead of the standard Toffoli gates.
However, the construction for FTQ does not consider the cost of distillation, and there is a trade-off between $T$-depth and the number of $T$~gates running simultaneously.
We show that we achieve smallest $\text{KQ}_{T}$ when we run $\Theta\left(n / \sqrt{\log n} \right)$ $T$ gates simultaneously.

\section{Preliminaries}

In this paper, we optimize a carry-lookahead adder by replacing Toffoli gates with relative-phase Toffoli gates.
To maintain an accurate calculation, we must consider the role of Toffoli gates well.
Moreover, we reduce computational costs by decomposing Toffoli gates into single-qubit gates and CNOT gates.

In subsection~A, we explain the quantum gate set used in this paper.
Next, to clarify the role of Toffoli gates, we review Draper et al.'s carry-lookahead adder~\cite{DKRS2006} briefly in subsection~B.
We explain $T$-minimization~\cite{TMK2020} by using Gidney's relative-phase Toffoli gates~\cite{Gid2018} in subsection~C.
We review the general construction of a control modular adder in subsection~D.

\subsection{Quantum Gate Set}

In this paper, we use the following:
\begin{itemize}
\item \textbf{Clifford gates}: $X$~gate, $Y$~gate, $Z$~gate, $H$~gate, $S$~gate, CNOT~gate
\item \textbf{non-Clifford gates}: $T$~gate
\end{itemize}
The CNOT gate is a two-qubit gate, and the others are one-qubit gates.
We express $X$ gates as $\oplus$ in the circuit.

In this paper, we focus on two gates: $T$ and CNOT,
\begin{align}
T =
\begin{bmatrix}
1 & 0 \\
0 & \text{exp}\left(\dfrac{\text{i}\pi}{4}\right)
\end{bmatrix}
, \text{CNOT} =
\begin{bmatrix}
1 & 0 & 0 & 0 \\
0 & 1 & 0 & 0 \\
0 & 0 & 0 & 1 \\
0 & 0 & 1 & 0
\end{bmatrix}.
\end{align}

\subsection{Draper et al.'s Carry-lookahead Adder}

In this subsection, we use the same notations in Draper et al.'s paper~\cite{DKRS2006}.
First, we explain the calculation of $a+b$ when $a$ and $b$ are $n$-bit numbers.
We express $a$ as $a_{n-1} a_{n-2} \ldots a_{0}$ and $b$ as $b_{n-1} b_{n-2} \ldots b_{0}$, where $a_{i}$ and $b_{i}$ are $0$ or $1$.
To calculate $a + b$, we employ a carry $c_{i}$.
Carry $c_{i}$ is defined as an overflow from the $(i-1)$-th bit to the $i$-th bit.
In more detail, we define $c_{i}$ as
\begin{align}
\label{eq:carry}
c_{i} =
\begin{cases}
0 & \text{if } i = 0 \\
\left\lfloor\dfrac{a_{i-1} + b_{i-1} + c_{i-1}}{2}\right\rfloor & \text{otherwise}
\end{cases}
\end{align}
Then, $\left(a+b\right)_{i}$, the $i$-th bit of $a+b$, is calculated as
\begin{align}
\left(a+b\right)_{i}
= a_{i} \oplus b_{i} \oplus c_{i}.
\end{align}
Thus, we need carries to calculate an addition.

Now, we give a brief explanation of a carry-lookahead adder.
Before calculating an addition, we determine the propagation of a carry from the $i$-th bit to the $j$-th bit as a function of the following three conditions:
\begin{itemize}
\item \textbf{propagate:} A carry is propagated from the $i$-th bit to the $j$-th bit. Namely, $c_{j} = c_{i}$.
\item \textbf{generate:} A carry is generated in the $j$-th bit, namely $c_{j} = 1$, regardless of the value of $c_{i}$.
\item \textbf{kill:} A carry is killed in the $j$-th bit, namely $c_{j} = 0$, regardless of the value of $c_{i}$.
\end{itemize}

To calculate the propagation, we define two functions $p\left[i, j\right], g\left[i, j\right] \in \left\{0, 1\right\}$.
$p\left[i, j\right]$ is true when the carry from the $i$-th bit to the $j$-th bit should be propagated.
Similarly, $g\left[i, j\right]$ is true when the carry out at the $j$th bit is true independent of the value of the carry in at the $i$ bit.
We do not need a separate function for \textbf{kill}, as its value can be inferred from $p$ and $g$.
By using these functions, we can calculate the propagation state over a wider span.
Specifically, when $i < k < j$,
\begin{align}
\label{eq:renewalP}
p\left[i ,j\right] &= p\left[i, k\right] \wedge p\left[k, j\right], \\
\label{eq:renewalG}
g\left[i, j\right]
&= g\left[k, j\right]
\oplus \left(g\left[i, k\right] \wedge p\left[k, j\right] \right),
\end{align}
where $\wedge$ is Boolean AND, and $\oplus$ is Boolean XOR.
By using these properties, we calculate $c_{j} = g\left[0, j\right]$.

Now, we explain Draper et al.'s carry-lookahead adder for $\ket{a}\ket{b} \to \ket{a}\ket{b+a}$.
This requires an additional $n$ qubits for the carry register $\ket{c}$ and $n$ qubits for register $\ket{p}$, containing $p\left[i, j\right]$.
Thus, a carry-lookahead adder requires $4n$ qubits.

\begin{figure}
\[
\Qcircuit @C=.2em @R=.2em {
  \lstick{\ket{a_{0}}} & \qw & \qw
  & \multigate{12}{\text{\rotatebox{90}{Initialization}}}
  & \multigate{12}{\text{\rotatebox{90}{P}}}
  & \multigate{12}{\text{\rotatebox{90}{G}}}
  & \multigate{12}{\text{\rotatebox{90}{C}}}
  & \multigate{12}{\text{\rotatebox{90}{Inverse P}}}
  & \qw 
  & \multigate{9}{\text{\rotatebox{90}{Erasing Carry}}}
  & \qw 
  & \qw
  & \rstick{\ket{a_{0}}}
  \\
  \lstick{\ket{b_{0}}} & \qw & \qw
  & \ghost{\text{\rotatebox{90}{Initialization}}}
  & \ghost{\text{\rotatebox{90}{P}}}
  & \ghost{\text{\rotatebox{90}{G}}}
  & \ghost{\text{\rotatebox{90}{C}}}
  & \ghost{\text{\rotatebox{90}{Inverse P}}}
  & \qw
  & \ghost{\text{\rotatebox{90}{Erasing Carry}}}
  & \qw 
  & \qw
  & \rstick{\ket{\left(a + b\right)_{0}}}
  \\
  \lstick{\ket{c_{1}}} & \qw & \qw
  & \ghost{\text{\rotatebox{90}{Initialization}}}
  & \ghost{\text{\rotatebox{90}{P}}}
  & \ghost{\text{\rotatebox{90}{G}}}
  & \ghost{\text{\rotatebox{90}{C}}}
  & \ghost{\text{\rotatebox{90}{Inverse P}}}
  & \ctrl{2}
  & \ghost{\text{\rotatebox{90}{Erasing Carry}}}
  & \qw 
  & \qw
  & \rstick{\ket{c_{1}}}
  \\ 
  \lstick{\ket{a_{1}}} & \qw & \qw
  & \ghost{\text{\rotatebox{90}{Initialization}}}
  & \ghost{\text{\rotatebox{90}{P}}}
  & \ghost{\text{\rotatebox{90}{G}}}
  & \ghost{\text{\rotatebox{90}{C}}}
  & \ghost{\text{\rotatebox{90}{Inverse P}}}
  & \qw
  & \ghost{\text{\rotatebox{90}{Erasing Carry}}}
  & \qw 
  & \qw
  & \rstick{\ket{a_{1}}}
  \\
  \lstick{\ket{b_{1}}} & \qw & \qw
  & \ghost{\text{\rotatebox{90}{Initialization}}}
  & \ghost{\text{\rotatebox{90}{P}}}
  & \ghost{\text{\rotatebox{90}{G}}}
  & \ghost{\text{\rotatebox{90}{C}}}
  & \ghost{\text{\rotatebox{90}{Inverse P}}}
  & \targ
  & \ghost{\text{\rotatebox{90}{Erasing Carry}}}
  & \qw 
  & \qw
  & \rstick{\ket{\left(a + b\right)_{1}}}
  \\
  \lstick{\ldots} & {/} \qw & \qw
  & \ghost{\text{\rotatebox{90}{Initialization}}}
  & \ghost{\text{\rotatebox{90}{P}}}
  & \ghost{\text{\rotatebox{90}{G}}}
  & \ghost{\text{\rotatebox{90}{C}}}
  & \ghost{\text{\rotatebox{90}{Inverse P}}}
  & \qw
  & \ghost{\text{\rotatebox{90}{Erasing Carry}}}
  & \qw
  & \qw & \rstick{\ldots}
  \\ 
  \lstick{\ket{c_{n-2}}} & \qw & \qw
  & \ghost{\text{\rotatebox{90}{Initialization}}}
  & \ghost{\text{\rotatebox{90}{P}}}
  & \ghost{\text{\rotatebox{90}{G}}}
  & \ghost{\text{\rotatebox{90}{C}}}
  & \ghost{\text{\rotatebox{90}{Inverse P}}}
  & \ctrl{2}
  & \ghost{\text{\rotatebox{90}{Erasing Carry}}}
  & \qw 
  & \qw
  & \rstick{\ket{c_{n-2}}} \\ 
  \lstick{\ket{a_{n-2}}} & \qw & \qw
  & \ghost{\text{\rotatebox{90}{Initialization}}}
  & \ghost{\text{\rotatebox{90}{P}}}
  & \ghost{\text{\rotatebox{90}{G}}}
  & \ghost{\text{\rotatebox{90}{C}}}
  & \ghost{\text{\rotatebox{90}{Inverse P}}}
  & \qw
  & \ghost{\text{\rotatebox{90}{Erasing Carry}}}
  & \qw 
  & \qw
  & \rstick{\ket{a_{n-2}}} \\ 
  \lstick{\ket{b_{n-2}}} & \qw & \qw
  & \ghost{\text{\rotatebox{90}{Initialization}}}
  & \ghost{\text{\rotatebox{90}{P}}}
  & \ghost{\text{\rotatebox{90}{G}}}
  & \ghost{\text{\rotatebox{90}{C}}}
  & \ghost{\text{\rotatebox{90}{Inverse P}}}
  & \targ
  & \ghost{\text{\rotatebox{90}{Erasing Carry}}}
  & \qw 
  & \rstick{\ket{\left(a + b\right)_{n-2}}} \\ 
  \lstick{\ket{c_{n-1}}} & \qw & \qw
  & \ghost{\text{\rotatebox{90}{Initialization}}}
  & \ghost{\text{\rotatebox{90}{P}}}
  & \ghost{\text{\rotatebox{90}{G}}}
  & \ghost{\text{\rotatebox{90}{C}}}
  & \ghost{\text{\rotatebox{90}{Inverse P}}}
  & \ctrl{2}
  & \ghost{\text{\rotatebox{90}{Erasing Carry}}}
  & \qw 
  & \qw
  & \rstick{\ket{c_{n-1}}}
  \\ 
  \lstick{\ket{a_{n-1}}} & \qw & \qw
  & \ghost{\text{\rotatebox{90}{Initialization}}}
  & \ghost{\text{\rotatebox{90}{P}}}
  & \ghost{\text{\rotatebox{90}{G}}}
  & \ghost{\text{\rotatebox{90}{C}}}
  & \ghost{\text{\rotatebox{90}{Inverse P}}}
  & \qw
  & \qw
  & \qw
  & \qw
  & \rstick{\ket{a_{n-1}}}
  \\ 
  \lstick{\ket{b_{n-1}}} & \qw & \qw
  & \ghost{\text{\rotatebox{90}{Initialization}}}
  & \ghost{\text{\rotatebox{90}{P}}}
  & \ghost{\text{\rotatebox{90}{G}}}
  & \ghost{\text{\rotatebox{90}{C}}}
  & \ghost{\text{\rotatebox{90}{Inverse P}}}
  & \targ
  & \qw
  & \qw
  & \qw
  & \rstick{\ket{\left(a + b\right)_{n-1}}}
  \\  
  \lstick{\ket{c_{n}}} & \qw & \qw
  & \ghost{\text{\rotatebox{90}{Initialization}}}
  & \ghost{\text{\rotatebox{90}{P}}}
  & \ghost{\text{\rotatebox{90}{G}}}
  & \ghost{\text{\rotatebox{90}{C}}}
  & \ghost{\text{\rotatebox{90}{Inverse P}}}
  & \qw
  & \qw
  & \qw
  & \qw
  & \rstick{\ket{\left(a + b\right)_{n}}}
}
\]
\caption{\label{fig:Draper_add}
An block level figure of Draper et al.'s carry-lookahead adder.
In this figure, we sort qubits from the lowest qubits to the highest qubits, which is different from Figure~\ref{fig:Madd}.
$\ket{c_{i}}$ is given as $\ket{0}$ at the beginning of this circuit and these are cleared to $\ket{0}$ after Erasing Carry.
The detailed circuit is shown in Appendix~A.
}
\end{figure}
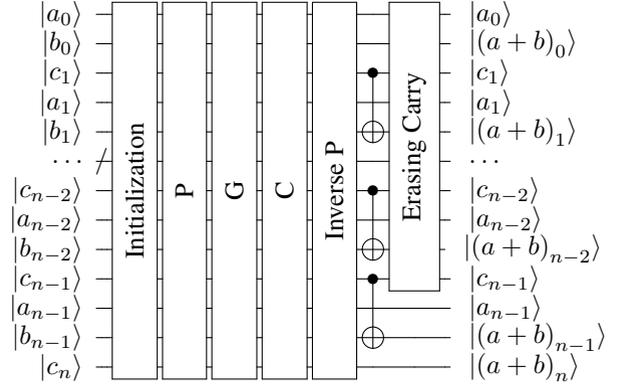

Now, we explain the implementation briefly.
This implementation consists of five phases, Initialization, P-rounds, G-rounds, C-rounds, and inverse P-rounds.
In each round,
\begin{itemize}
\item \textbf{Initialization}: we calculate $g\left[i, i+1\right]$ in $\ket{c_{i+1}}$ and $p\left[i, i+1\right]$ in $\ket{b_{i}}$,
\item \textbf{P-rounds}: we calculate the $p$-function and write result in $\ket{p}$,
\item \textbf{G-rounds}: we calculate $\ket{c_{2^{k}}}~\left(k \in \mathbb{N}\right)$ by calculating some $g$-function,
\item \textbf{C-rounds}: we calculate all carry $\ket{c}$ by calculating some $g$-function,
\end{itemize}
and we clean $\ket{p}$ in inverse P-rounds.
After inverse P-rounds, we calculate each bit of $a + b$ by using these carries $\ket{c}$.
In this calculation, we run P-rounds and G-rounds simultaneously, and we run C-rounds and inverse P-rounds simultaneously.
However, the value of carries remain on $\ket{c}$.
Thus, we must clean $\ket{c}$ to $\ket{0}$ except for $c_{n}$.
Draper et al. found that the value of carries $c_{i}$ except $c_{n}$ in $a + b$ is the same in $a + \left(2^{n} - 1 - a - b\right)$.
Therefore, we erase carries by performing the addition $a + \left(2^{n} - 1 - a - b\right)$ on the lower $n - 1$ qubits.
The block level circuit is shown in Figure~\ref{fig:Draper_add}.

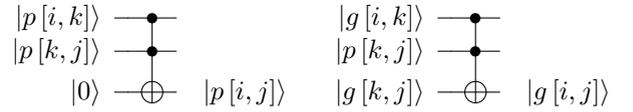
\begin{figure}
\begin{minipage}{0.49\linewidth}
\[
\Qcircuit @C=.5em @R=1em {
  \lstick{\ket{p\left[i, k\right]}} & \qw & \ctrl{1} & \qw \\
  \lstick{\ket{p\left[k, j\right]}} & \qw & \ctrl{1} & \qw \\ 
  \lstick{\ket{0}} & \qw & \targ & \qw & \rstick{\ket{p\left[i, j\right]}} 
}
\]
    \subcaption{\label{fig:CalcP}
    Calculation circuit of $p[i, j]$ as eq.~(\ref{eq:renewalP}).
    }
  \end{minipage}
  \begin{minipage}{0.49\linewidth}
\[
\Qcircuit @C=.5em @R=1em {
  \lstick{\ket{g\left[i, k\right]}} & \qw & \ctrl{1} & \qw \\
  \lstick{\ket{p\left[k, j\right]}} & \qw & \ctrl{1} & \qw \\ 
  \lstick{\ket{g\left[k, j\right]}} & \qw & \targ & \qw & \rstick{\ket{g\left[i, j\right]}} 
}
\]
    \subcaption{\label{fig:CalcG}
    Calculation circuit of $g[i, j]$ as eq.~(\ref{eq:renewalG}).
    }
  \end{minipage}
\caption{\label{fig:Toffoli_Calc}
Calculation circuit of $p\left[i, j\right]$ and $g\left[i, j\right]$.
}
\end{figure}

As noted above, a carry-lookahead adder is mainly constructed by a calculation on $p$ and $g$.
We calculate $p$ and $g$ with eq.~(\ref{eq:renewalP}) or~(\ref{eq:renewalG}) respectively, and those are implemented by Toffoli gates as shown in Figure~\ref{fig:Toffoli_Calc}.
The detailed explanation of Draper et al.'s adder, including which $p$-function or $g$-function we calculate, is given in Appendix~A. 
In total, a carry-lookahead adder requires $10n$ Toffoli gates and $4n$ CNOT gates.
Moreover, the Toffoli depth is $4 \log n$.

Up to this point, we have explained the construction of an adder.
Draper et al. also proposed other operations, such as a subtractor and a comparator, based on their adder.
The number of gates and the depth in a subtractor is almost the same as those in an adder.
In a comparator, the number of gates is 60\% of an adder and the depth is 50\% of an adder.
Draper et al. implement a comparator using only Initialization, P-rounds, G-rounds, and their inverses.
More precisely, Draper et al. regard $a$ and $b$ as $2^{\left\lceil \log n \right\rceil}$-bit numbers by padding $0$ in higher bits, but we do not use these qubits.
If we calculate $p\left[i,j\right]$ or $g\left[i,j\right]$ when $i \leq n-1$ and $j \geq n$, we calculate $p\left[i,n\right]$ or $g\left[i,n\right]$ respectively.
Then, we calculate $g\left[0, n \right]$ after G-rounds.

\subsection{$T$-count Minimization of a Carry-lookahead Adder}

\begin{figure}
\[
\Qcircuit @C=.5em @R=.3em {
 & \qw & \qw & \qw & \ctrl{2} & \qw & \qw & \qw & \ctrl{2}
   & \qw & \ctrl{1} & \qw & \ctrl{1} & \gate{T} & \qw  \\
 & \qw & \ctrl{1} & \qw & \qw & \qw & \ctrl{1} & \qw & \qw
  & \gate{T^{\dag}} & \targ & \gate{T^{\dag}} & \targ & \gate{S} & \qw   \\
 & \gate{H} & \targ & \gate{T^{\dag}} & \targ & \gate{T} & \targ
  & \gate{T^{\dag}} & \targ
   & \gate{T} & \gate{H} & \qw & \qw & \qw & \qw 
}
\]
\caption{\label{fig:toffoli_NC} 
The standard decomposition of a Toffoli gate~\cite{NC2002}.
We call this decomposition \st. 
The control bits are the first and second qubits, and the target bit is the third qubit.
This calculation preserves the phase.
}
\end{figure}
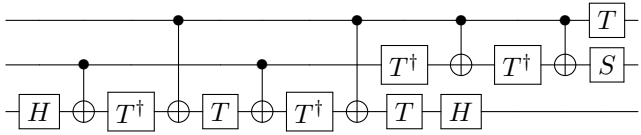

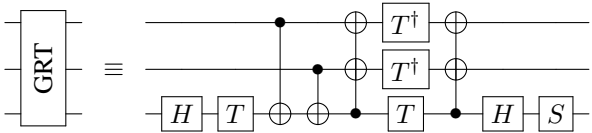
\begin{figure}
\[
\Qcircuit @C=.6em @R=.3em {
  & \multigate{2}{\text{\rotatebox{90}{GRT}}} & \qw & & & & & \qw & \qw & \ctrl{2} & \qw & \targ & \gate{T^{\dag}} & \targ & \qw & \qw & \qw \\
  & \ghost{\text{\rotatebox{90}{GRT}}} & \qw & & \mbox{$\equiv$} & & & \qw & \qw & \qw & \ctrl{1} & \targ & \gate{T^{\dag}} & \targ & \qw & \qw & \qw \\
  & \ghost{\text{\rotatebox{90}{GRT}}} & \qw & & & & & \gate{H} & \gate{T} & \targ & \targ & \ctrl{-2} & \gate{T} & \ctrl{-2} & \gate{H} & \gate{S} & \qw
}
\]
\caption{\label{fig:toffoli_Gid} 
Gidney's relative-phase Toffoli gate~\cite{Gid2018} given by the unitary matrix~(\ref{eq:Gidmat}). 
We call this decomposition \grt.
The control bits are the first and second qubits, and the target bit is the third qubit.
This calculation preserves the phase only when the target qubit is $\ket{0}$ on input.
}
\end{figure}

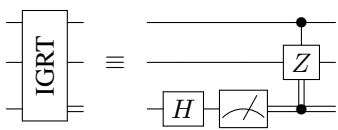
\begin{figure}
\[
\Qcircuit @C=.6em @R=.4em {
  & \multigate{2}{\text{\rotatebox{90}{IGRT}}} & \qw & & & & & \qw & \qw & \ctrl{1} & \qw \\
  & \ghost{\text{\rotatebox{90}{IGRT}}} & \qw & & \mbox{$\equiv$} & & & \qw & \qw & \gate{Z} & \qw \\
  & \ghost{\text{\rotatebox{90}{IGRT}}} & \cw & & & & & \gate{H} & \meter & \control \cw \cwx & \cw
}
\]
\caption{\label{fig:toffoli_invGid} 
Inverse of Gidney's relative-phase Toffoli gate~\cite{Gid2018}.
We call this decomposition \igrt.
This calculation preserves the phase when we input $\ket{000}, \ket{010}, \ket{100}$, or $\ket{111}$, which are outputs of \grt having valid phase.
Control-Z is a Clifford gate, and we use no $T$ gate.
}
\end{figure}

Thapliyal et al.~\cite{TMK2020} proposed $T$-count minimization by using relative-phase Toffoli gates.
The standard Toffoli gate~(\st)~\cite{NC2002} decomposition is given in Figure~\ref{fig:toffoli_NC}.
However, we can calculate correctly even if we replace some Toffoli gates with Gidney's relative-phase Toffoli
gate~(\grt) or its inverse~(\igrt)~\cite{Gid2018}.
\grt is shown in Figure~\ref{fig:toffoli_Gid} and the corresponding unitary matrix of \grt in the computational basis is
\begin{align}
\label{eq:Gidmat}
\begin{bmatrix}
1 & 0 & 0 & 0 & 0 & 0 & 0 & 0 \\
0 & i & 0 & 0 & 0 & 0 & 0 & 0 \\
0 & 0 & 1 & 0 & 0 & 0 & 0 & 0 \\
0 & 0 & 0 & -i & 0 & 0 & 0 & 0 \\
0 & 0 & 0 & 0 & 1 & 0 & 0 & 0 \\
0 & 0 & 0 & 0 & 0 & -i & 0 & 0 \\
0 & 0 & 0 & 0 & 0 & 0 & 0 & -i \\
0 & 0 & 0 & 0 & 0 & 0 & 1 & 0 \\
\end{bmatrix},
\end{align}
and we calculate correctly when the target bit is $\ket{0}$.
\igrt is shown in Figure~\ref{fig:toffoli_invGid}.
In the carry-lookahead adder, as in many circuits, we must clean our ancilla qubits, returning them to a known, disentangled state, typically $\ket{0}$.
In this case, we can reduce our cost by measuring the ancilla on \igrt.
Gidney's paper~\cite{Gid2018} shows that using measurement reduces 2~$T$ gates. Using measurement is better because one accurate $T$ gate requires many measurements.

Thapliyal et al. proposed two constructions.
The first construction replaced Toffoli gates in Initialization and P-rounds with \grt, and Toffoli gates in the inverse rounds with \igrt.
Other Toffoli gates are replaced with \st.
Thapliyal et al. call this construction qubit-optimize.
The number of qubits is $4n$ and the number of $T$~gates is $40n$.

\begin{figure}
\[
\Qcircuit @C=1em @R=1em {
  & \ctrl{1} & \qw & & & & & &
  & \multigate{3}{\text{\rotatebox{90}{GRT}}} 
  & \qw 
  & \multigate{3}{\text{\rotatebox{90}{IGRT}}} 
  & \qw \\
  & \ctrl{3} & \qw & & & & & &
  & \ghost{\text{\rotatebox{90}{GRT}}} 
  & \qw 
  & \ghost{\text{\rotatebox{90}{IGRT}}} 
  & \qw  \\
  & & & & \mbox{$\equiv$} & & & &
  & 
  & 
  & 
  & \\
  \lstick{\ket{0}} & \qw & \qw & & & & & &
  \lstick{\ket{0}} 
  & \ghost{\text{\rotatebox{90}{GRT}}} 
  & \ctrl{1} 
  & \ghost{\text{\rotatebox{90}{IGRT}}} 
  & \cw \\
  \lstick{\ket{c_{i}}} & \targ & \qw & & & & & &
  \lstick{\ket{c_{i}}} 
  & \qw 
  & \targ 
  & \qw 
  & \qw
}
\]
\caption{\label{fig:PGRT} 
Replacing Toffoli gates in G-rounds and C-rounds in a $T$-optimized carry-lookahead adder.
We call this decomposition \pgrt.
We replace the first Toffoli gate with \grt and the second Toffoli gate with \igrt.
The third qubit is an ancilla qubit.
This qubit is measured as part of executing \igrt and will be $\ket{0}$ after running \pgrt.
}
\end{figure}
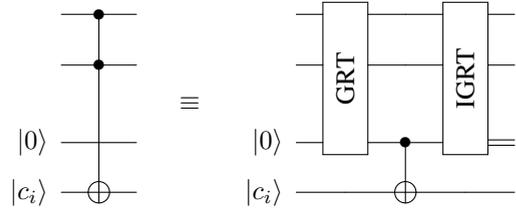

The second construction replaced all Toffoli gates with \grt or \igrt, increasing the required number of ancilla qubits.
Thapliyal et al. call this construction $T$-optimize.
Specifically, we replace Toffoli gates in Initialization, P-rounds, and the inverse of these similarly as the first construction.
We replace Toffoli gates in G-rounds and C-rounds by the pair of \grt and \igrt as in Figure~\ref{fig:PGRT}.
We call these gates \pgrt, where P is the abbreviation of \lq\lq pair''.
In this construction, Thapliyal et al. claim that the number of qubits is $6n$ and the number of $T$ gates is $20n$.
However, we recalculated these results and our results differ from results in~\cite{TMK2020}.
In our result, the number of qubits is $4.5n$ and the number of $T$ gates is $28n$.
The difference in the number of qubits occurs from our method for preparing ancilla qubits.
Thapliyal et al. prepare new ancilla qubits for G-rounds and C-rounds respectively, while they recycle ancilla qubits for P-rounds.
We apply this to G-rounds and C-rounds similarly.

\subsection{The General Construction of a Control Modular Adder}

\begin{figure}
\[
\Qcircuit @C=1em @R=.7em {
  \text{\ctrlq} &&&& \lstick{\ket{x}} & \qw & \ctrl{1}  & \ctrl{1} & \ctrl{1} & \rstick{\ket{x}} \qw \\
  &&&& \lstick{\ket{b}} & {/} \qw & \multigate{1}{\text{\rotatebox{90}{Comp}}} & \gate{\text{\rotatebox{90}{Add}}} & \multigate{1}{\text{\rotatebox{90}{Comp}}} & \rstick{\ket{b+ax \bmod N}} \qw  \\
  \text{\compq} &&&& \lstick{\ket{0}} & \qw & \ghost{\text{\rotatebox{90}{Comp}}} & \ctrl{-1} & \ghost{\text{\rotatebox{90}{Comp}}} & \rstick{\ket{0}} \qw
}
\]
\caption{\label{fig:Zal_Madd} 
The general construction of a control modular adder.
Add means an adder, and Comp means a comparator.
\ctrlq has a single qubit which is used to hold the value of the control.
$\ket{b}$ has $n$ qubits which are used to hold the result of a control modular addition.
\compq has a single qubit which is used to hold the result of a comparison. $a$ and $N$ are classical numbers.
}
\end{figure}
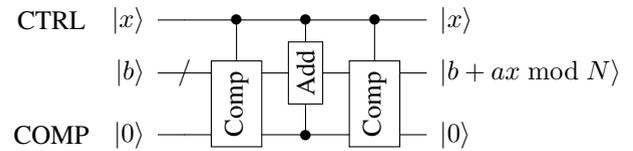

In this subsection, we explain the calculation of
\begin{align}
\ket{x}\ket{b}\ket{0} \to \ket{x}\ket{b+ax \bmod N}\ket{0}.
\end{align}
The general construction of a control modular adder is shown in Figure~\ref{fig:Zal_Madd}.
The first register has a single qubit which is used to hold the value of the control.
We call this the \ctrlq qubit.
The second register has $n$ qubits which are used to hold the result of a control modular addition.
The third register has a single qubit which is used to hold the result of a comparison temporarily.
We call this the \compq qubit.
Specifically, we determine whether we subtract $N$ or not based on \compq.
We conduct a comparator with one control qubit and an adder with two control qubits, and we write these as a C-comparator and a CC-adder, respectively.

To execute a control modular adder, we conduct operations in this order:
\begin{enumerate}
\item 
We compare the second register $\ket{b}$ and the classical value $N-a$.
If $b \geq N-a$, namely $a+b \geq N$, we flip \compq.
\item 
If both \ctrlq and \compq are $1$, we subtract $N - a$ from the second register.
If \ctrlq is $1$ and \compq is $0$, we add $a$.
Otherwise, we add no value.
\item 
If the second register is strictly less than $a$, we flip \compq.
\end{enumerate}

\section{First-Level Optimization: Our Construction of a Control Modular Adder}

In this section, we explain first-level optimization on the original construction~\cite{VI2005}.
In the general construction, a comparator has about $1/2$ the depth of a carry-lookahead adder.
Thus, by constructing a carry-lookahead adder using the same general construction, the depth is about the same as $2$ adders, because a carry-lookahead adder is composed of two comparators and one adder.
In the original construction, we use $3$ adders.
Thus, we use only $2/3$ of KQ of the original construction when comparing two constructions of a control modular adder. The original construction applies two optimizations on repeating control modular adders. Our construction will be more efficient by adopting the same optimizations with some overhead, but the detail, such as the amount of overhead, should be evaluated in future work.

Based on the above discussion, we need to give the construction of
\begin{itemize}
\item C-comparator~(subsection~A)
\item CC-adder~(subsection~B)
\end{itemize}
on a carry-lookahead adder.
In this construction, we do not decompose Toffoli gates, because the decomposition of Toffoli gates is different in FTQ or NISQ respectively.
Thus, we leave Toffoli gates as they are, and we consider the decomposition of Toffoli gates in Section IV.

In our construction, we consider the classicality of $a$ and $N$ as described by Markov and Saeedi~\cite{MS2012} to realize higher efficiency.
Moreover, we give a construction of C-comparator that is not given in the original paper.
By doing these, we propose a circuit construction of a control modular adder.

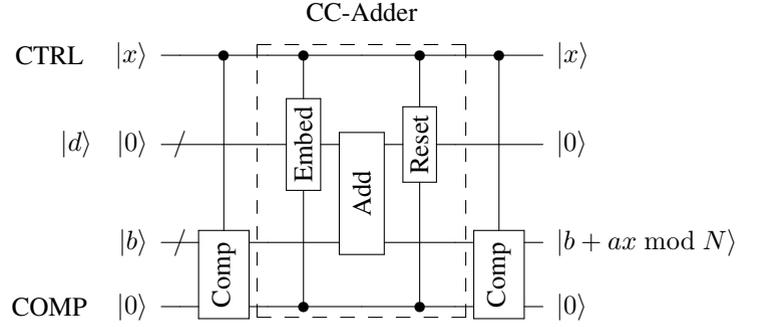
\begin{figure}
\leftline{
$
\Qcircuit @C=.7em @R=1.5em {
  &&&&&&&&&&& \mbox{CC-Adder} \\
  \mbox{\ctrlq} &&&&&& \lstick{\ket{x}} & \qw & \ctrl{2} & \qw & \ctrl{1} &\qw & \ctrl{1} & \qw & \ctrl{2} & \rstick{\ket{x}} \qw \\
   &&& \lstick{\ket{d}} &&& \lstick{\ket{0}} & {/} \qw & \qw & \qw & \gate{\text{\rotatebox{90}{Embed}}} & \multigate{1}{\text{\rotatebox{90}{Add}}} & \gate{\text{\rotatebox{90}{Reset}}} & \qw & \qw & \rstick{\ket{0}} \qw \\
  &&&&&& \lstick{\ket{b}} & {/} \qw & \multigate{1}{\text{\rotatebox{90}{Comp}}} & \qw & \qw & \ghost{\text{\rotatebox{90}{Add}}} & \qw & \qw & \multigate{1}{\text{\rotatebox{90}{Comp}}} & \rstick{\ket{b+ax \bmod N}} \qw  \\
  \mbox{\compq} &&&&&& \lstick{\ket{0}} & \qw & \ghost{\text{\rotatebox{90}{Comp}}} & \qw & \ctrl{-2} & \qw & \ctrl{-2} & \qw \gategroup{2}{10}{5}{14}{.8em}{--} &  \ghost{\text{\rotatebox{90}{Comp}}} & \rstick{\ket{0}} \qw
}
$
}
\caption{\label{fig:Our_Madd} 
Our construction of a control modular adder based on Figure~\ref{fig:Zal_Madd}.
A CC-adder is constructed by embedding, an adder, and resetting.
Then, we add the second register $\ket{d}$ as an $n$-qubit ancilla for embedding the value based on \ctrlq.
The carry register $\ket{c}$ with $n$ qubits and the $p$-function register $\ket{p}$ with $n$ qubits are not represented in this figure for visibility.
In a C-comparator, we do not use the second register.
In total, our control modular adder requires $4n+2$ qubits.
}
\end{figure}

Based on Figure~\ref{fig:Zal_Madd}, we construct our circuit as shown in Figure~\ref{fig:Our_Madd}.
We add the second $n$-qubit ancilla register for embedding value with \ctrlq.
In addition to these registers, we use the carry register $\ket{c}$ with $n$ qubits and the $p$-function register $\ket{p}$ with $n$ qubits to realize the carry-lookahead adder, not represented in Figure~\ref{fig:Our_Madd}.
Thus, our control modular adder requires $4n+2$ qubits.
The number of gates and the depth is given in Table~\ref{tab:gate_count_of_ModAdder}, and the breakdown of this is given in Table~\ref{tab:detail_gate_count_of_ModAdder} in Appendix~B.
Now, we explain the C-comparator and the CC-adder briefly.

\begin{table*}
\caption{\label{tab:gate_count_of_ModAdder} 
Gate count and depth of our proposed control modular adder.
The breakdown of this is shown in Table~\ref{tab:detail_gate_count_of_ModAdder} in Appendix~B.
}
\centering
\begin{tabular}{c|cc|cc}
&\multicolumn{2}{c|}{Count} & \multicolumn{2}{c}{Depth} \\ \hline
Operation & Toffoli & CNOT & Toffoli & CNOT \\\hline
C-comparator (twice) & $4n$ & $n$ & $2 \log n$ & $O(1)$ \\
CC-adder & $9.5n$ & $4.75n$ & $4 \log n$ & $2 \log n$ \\\hline
Total & $17.5n$ & $6.75n$ & $8\log n$ & $2 \log n$
\end{tabular}
\end{table*}

\subsection{Construction of a C-comparator}

In a C-comparator, only \compq is changed and other qubits do not change.
Thus, to implement a C-comparator, it is sufficient that we add control operations only on the gates including \compq and remain other gates.

In our construction of a control modular adder, we use two types of C-comparators.
In the first C-comparator, we flip \compq if \ctrlq is $1$ and $b \geq N-a$.
In the final C-comparator, we flip \compq if \ctrlq is $1$ and $b < a$.
In both cases, we judge whether $b \geq d$ or $b < d$ with a classical value of $d$.

\begin{figure}
\[
\Qcircuit @C=.5em @R=.5em {
& &
  & 
  & 
  & 
  & \mbox{} &  & \mbox{} \\
\lstick{\text{\ctrlq}} & \qw & \qw
  & \qw
  & \qw
  & \qw
  & \qw & \ctrl{7} & \qw 
  & \qw
  & \qw
  & \qw
  & \qw 
  & \qw
  \\
  \lstick{\ket{d_{0}}} & \qw & \qw
  & \multigate{6}{\text{\rotatebox{90}{Initialization}}}
  & \multigate{6}{\text{\rotatebox{90}{P}}}
  & \multigate{6}{\text{\rotatebox{90}{G}}}
  & \qw & \qw & \qw
  & \multigate{6}{\text{\rotatebox{90}{Inverse G}}}
  & \multigate{6}{\text{\rotatebox{90}{Inverse P}}}
  & \multigate{6}{\text{\rotatebox{90}{Inverse Initialization}}}
  & \qw 
  & \qw
  \\
  \lstick{\ket{b_{0}}} & \qw & \qw
  & \ghost{\text{\rotatebox{90}{Initialization}}}
  & \ghost{\text{\rotatebox{90}{P}}}
  & \ghost{\text{\rotatebox{90}{G}}}
  & \qw & \qw & \qw
  & \ghost{\text{\rotatebox{90}{Inverse G}}}
  & \ghost{\text{\rotatebox{90}{Inverse P}}}
  & \ghost{\text{\rotatebox{90}{Inverse Initialization}}}
  & \qw 
  & \qw
  \\
  \lstick{\ket{c_{1}}} & \qw & \qw
  & \ghost{\text{\rotatebox{90}{Initialization}}}
  & \ghost{\text{\rotatebox{90}{P}}}
  & \ghost{\text{\rotatebox{90}{G}}}
  & \qw & \qw & \qw
  & \ghost{\text{\rotatebox{90}{Inverse G}}}
  & \ghost{\text{\rotatebox{90}{Inverse P}}}
  & \ghost{\text{\rotatebox{90}{Inverse Initialization}}}
  & \qw 
  & \qw
  \\
  \lstick{\ldots} & {/} \qw & \qw
  & \ghost{\text{\rotatebox{90}{Initialization}}}
  & \ghost{\text{\rotatebox{90}{P}}}
  & \ghost{\text{\rotatebox{90}{G}}}
  & \qw & \qw & \qw
  & \ghost{\text{\rotatebox{90}{Inverse G}}}
  & \ghost{\text{\rotatebox{90}{Inverse P}}}
  & \ghost{\text{\rotatebox{90}{Inverse Initialization}}}
  & \qw 
  & \qw
  \\ 
  \lstick{\ket{d_{n-1}}} & \qw & \qw
  & \ghost{\text{\rotatebox{90}{Initialization}}}
  & \ghost{\text{\rotatebox{90}{P}}}
  & \ghost{\text{\rotatebox{90}{G}}}
  & \qw & \qw & \qw
  & \ghost{\text{\rotatebox{90}{Inverse G}}}
  & \ghost{\text{\rotatebox{90}{Inverse P}}}
  & \ghost{\text{\rotatebox{90}{Inverse Initialization}}}
  & \qw 
  & \qw
  \\ 
  \lstick{\ket{b_{n-1}}} & \qw & \qw
  & \ghost{\text{\rotatebox{90}{Initialization}}}
  & \ghost{\text{\rotatebox{90}{P}}}
  & \ghost{\text{\rotatebox{90}{G}}}
  & \qw & \qw & \qw
  & \ghost{\text{\rotatebox{90}{Inverse G}}}
  & \ghost{\text{\rotatebox{90}{Inverse P}}}
  & \ghost{\text{\rotatebox{90}{Inverse Initialization}}}
  & \qw 
  & \qw
  \\  
  \lstick{\ket{c_{n}}} & \qw & \qw
  & \ghost{\text{\rotatebox{90}{Initialization}}}
  & \ghost{\text{\rotatebox{90}{P}}}
  & \ghost{\text{\rotatebox{90}{G}}}
  & \qw & \ctrl{1} & \qw
  & \ghost{\text{\rotatebox{90}{Inverse G}}}
  & \ghost{\text{\rotatebox{90}{Inverse P}}}
  & \ghost{\text{\rotatebox{90}{Inverse Initialization}}}
  & \qw 
  & \qw \\
  \lstick{\text{\compq}} & \qw & \qw
  & \qw
  & \qw
  & \qw
  & \qw & \targ & \qw
  & \qw
  & \qw
  & \qw
  & \qw 
  & \qw \\
  & &
  & 
  & 
  & 
  & \mbox{} &  & \mbox{}
  \gategroup{1}{7}{11}{9}{.5em}{.}
}
\]
\caption{\label{fig:OComp}
Block-level view of our construction of a C-comparator.
In this figure, we sort qubits from the low-order qubits to the high-order qubits, top to bottom.
This circuit is symmetric about the Toffoli gate surrounded by a dotted box.
$\ket{c_{i}}$ is given as $\ket{0}$ at the beginning of this circuit and these are cleared back to $\ket{0}$ after the computation.
The example circuits are shown in Figure~\ref{fig:CFcomp} and~\ref{fig:CLcomp} in Appendix~B.
}
\end{figure}
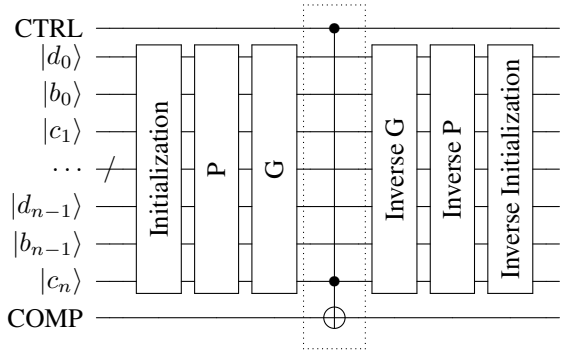

We construct these operations taking advantage of the classicality of $d$.
The intuitive explanation of this operation is that we calculate $b + \left(2^{n} - d \right)$ and check whether there is an overflow in the $n$-th bit.
Specifically,
\begin{align}
b + \left(2^{n} - d \right) = 2^{n} + (b - d)
\end{align}
and there is an overflow when $b \geq d$.
This construction is similar to previous constructions by Markov and Saeedi~\cite{MS2012}, but slightly different from them because our construction does not require $X$ gates on $\ket{b}$.
The number of gates and the depth is given in Table~\ref{tab:gate_count_of_ModAdder}.
The detailed construction is given in Appendix~B.
The block level construction of our C-comparator is given in Figure~\ref{fig:OComp}, and the example circuits are shown in Figure~\ref{fig:CFcomp} and~\ref{fig:CLcomp}.

\subsection{Construction of a CC-adder}

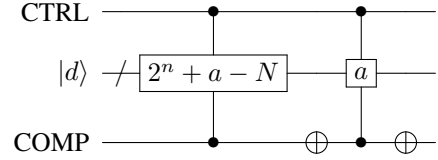
\begin{figure}
\[
\Qcircuit @C=.7em @R=1.5em {
  \lstick{\text{\ctrlq}} & \qw & \ctrl{1} & \qw & \ctrl{1} & \qw & \qw \\ 
  \lstick{\ket{d}} & \qw {/} & \gate{2^{n}+a-N} & \qw & \gate{a} & \qw & \qw \\
  \lstick{\text{\compq}} & \qw & \ctrl{-1} & \targ & \ctrl{-1} & \targ & \qw
}
\]
\caption{\label{fig:embed} 
Block-level diagram of the embedding circuit.
We omit $\ket{b}$ in Figure~\ref{fig:Our_Madd}.
We embed $2^{n}+a-N$ or $a$ on $\ket{d}$ based on \ctrlq and \compq.
The example circuit of the embedding is shown in Figure~\ref{fig:Padd} in Appendix~B.
}
\end{figure}

In a CC-adder, we embed values before and after an adder, similar to a C-adder~\cite{VBE1996}.
Based on this construction, we apply optimization by considering the classicality of $a$ and $N$.
From this point forward, we mainly focus on embedding on $\ket{d}$.
In a CC-adder, we conduct the following:
\begin{itemize}
\item If \ctrlq is $1$ and \compq is $1$, we add $a$ and subtract $N$.
This operation can be realized by adding $2^{n}+a-N$ and disregarding the calculation of a carry $c_{n}$.
\item If \ctrlq is $1$ and \compq is $0$, we add $a$.
\item Otherwise, we add no value.
\end{itemize}
Thus, the embedding is conducted as in Figure~\ref{fig:embed}.
The resetting is conducted by inverting the embedding circuit.

After embedding, we apply a standard adder.
Then, we conduct two optimizations as follows:
\begin{itemize}
\item disregarding gates including $\ket{g\left[0, n\right]}$.
\item eliminating gates in Initialization where we know the control bit is $0$.
\end{itemize}
The number of gates and the depth is given in Table~\ref{tab:gate_count_of_ModAdder}.
The detailed construction and example circuit of a CC-adder are given in Appendix~B.

\section{Second-Level Optimization: Constructing a Control Modular Adder for FTQ and NISQ devices}

In this section, we explain our second-level optimization.
We evaluate the computational cost for both FTQ on the logical layer, and NISQ, focusing on the decomposition of Toffoli gates.
We define KQ more specifically for FTQ and NISQ and minimize this value.
For FTQ, we minimize the number of $T$~gates by using Gidney's relative-phase Toffoli gates.
However, this construction does not take into consideration the cost of distillation.
We take into account the cost of distillation by finding the maximal number of $T$ gates which should be run simultaneously, optimizing $\text{KQ}_{T}$.
For NISQ, we apply Maslov's relative-phase Toffoli gates with a small number of CNOT gates~\cite{Mas2016} and minimize $\text{KQ}_{\text{CX}}$.
By doing these, we propose a control modular adder that is more efficient than Van Meter and Itoh~\cite{VI2005}, called the original construction in this section.
In the following discussion, we disregard the rounds with $O(1)$ gates.
In this section, we explain the optimization for FTQ in subsection~A and the optimization for NISQ in subsection~B.

\subsection{Computational Cost on the FTQ Logical Layer}

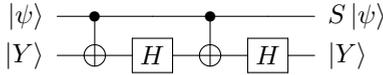
\begin{figure}
\[
\Qcircuit @C=1em @R=.7em {
  \lstick{\ket{\psi}} & \ctrl{1} & \qw & \ctrl{1} & \qw & \rstick{S\ket{\psi}} \qw  \\
  \lstick{\ket{Y}} & \targ & \gate{H} & \targ & \gate{H} & \rstick{\ket{Y}} \qw
}
\]
\caption{\label{fig:Sgate} 
Running an $S$ gate~\cite{JMF2012}. The second qubit is $\ket{Y} = \left(\ket{0}+\text{i}\ket{1}\right)/\sqrt{2}$.
Assuming correct operation on top of error correction, this ancilla passes through the gate execution unmodified, allowing it to be reused.}
\end{figure}

Next, we consider the optimal circuit for FTQ on the Logical layer, using Jones et al.'s architecture as a model~\cite{JMF2012}.
This architecture, in common with other error corrected-architectures,provides a fundamental gate set consisting of $X$, $Y$, $Z$, CNOT, and $H$~gates, and measurement; here, we ignore qubit movement in the surface code.
To run an $S$~gate, we prepare an ancilla qubit $\ket{Y} = \left(\ket{0}+\text{i}\ket{1}\right)/\sqrt{2}$ and run the circuit shown in Figure~\ref{fig:Sgate}.
An $S^{\dagger}$~gate can be realized by the reverse circuit of Figure~\ref{fig:Sgate}.

\begin{figure}
\[
\Qcircuit @C=1em @R=.7em {
  \lstick{\ket{\psi}} & \ctrl{1} & \gate{M_{Z}} \cwx[1] \\
  \lstick{\ket{A}} & \targ & \gate{S} &  \rstick{T\ket{\psi}} \qw
}
\]
\caption{\label{fig:Tgate} 
Running a $T$ gate~\cite{JMF2012}. The second qubit $\ket{A} = \left(\ket{0}+\text{e}^{\text{i}\pi/4}\ket{1}\right)/\sqrt{2}$; the $\ket{A}$ state is consumed in the process, with the consequence that creation of high-fidelity $\ket{A}$ states is one factor limiting performance.}
\end{figure}
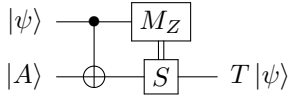

To achieve universal computation, we also need a non-Clifford gate; the choice of $T$ is typical.
To run a $T$~gate, we prepare an ancilla qubit $\ket{A} = \left(\ket{0}+\text{e}^{\text{i}\pi/4}\ket{1}\right)/\sqrt{2}$ and run the circuit shown in Figure~\ref{fig:Tgate}.
To run a $T^{\dagger}$~gate, we apply an $S^{\dagger}$~gate instead of a $S$ gate.
To realize accurate $T$~gates, we must prepare accurate $\ket{A}$ state defined by $\left(\ket{0} + \text{e}^{\text{i}\pi/4}\ket{1} \right)/\sqrt{2}$.
Preparing $\ket{A}$ is done by \emph{distillation}, as shown in Figure~\ref{fig:Agate} in Appendix~E.
This distillation circuit requires $15$~qubits and $6$~time steps, even assuming all of the CNOT gates can be implemented concurrently, but this is difficult to realize.
Distillation is an expensive operation, and its optimization is an ongoing topic of research~\cite{GF2019}.
Thus, a $T$~gate is the greatest factor in the cost of an FTQ circuit, leading us to focus on reducing the number of $T$~gates.

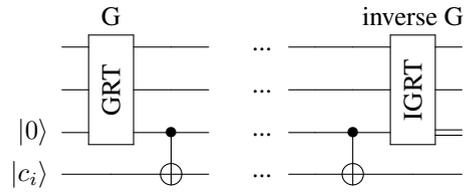
\begin{figure}
\[
\Qcircuit @C=1em @R=.7em {
  & \mbox{\text{G}} &&&&&&&& \mbox{\text{inverse G}} \\
  & \multigate{2}{\text{\rotatebox{90}{GRT}}} & \qw & \qw & \mbox{} & \mbox{...} & \mbox{} & \qw & \qw & \multigate{2}{\text{\rotatebox{90}{IGRT}}} & \qw \\
  & \ghost{\text{\rotatebox{90}{GRT}}} & \qw & \qw & \mbox{} & \mbox{...} & \mbox{} & \qw & \qw & \ghost{\text{\rotatebox{90}{IGRT}}} & \qw \\
  \lstick{\ket{0}}  & \ghost{\text{\rotatebox{90}{GRT}}} & \ctrl{1} & \qw & \mbox{} & \mbox{...} & \mbox{} & \qw & \ctrl{1} & \ghost{\text{\rotatebox{90}{IGRT}}} & \cw \\
  \lstick{\ket{c_{i}}} & \qw & \targ & \qw & \mbox{} & \mbox{...} & \mbox{} & \qw & \targ & \qw & \qw
}
\]
\caption{\label{fig:OurG} 
Our construction of G-rounds and inverse G-rounds in a C-comparator.
In Figure~\ref{fig:PGRT}, we apply \igrt after the first CNOT gate immediately in G-rounds and inverse G-rounds.
In our construction, we calculate the result of \grt in the third ancilla qubit and preserve this qubit until the corresponding Toffoli gate in inverse G-rounds.
Then, we clear this ancilla qubit by \igrt.
}
\end{figure}

We now minimize the number of $T$~gates on our control modular adder.
We adopt the Thapliyal construction with minor modifications, namely the replacement into relative-phase Toffoli gates, except G-rounds in a C-comparator. We employ \grt in G-rounds and \igrt in inverse G-rounds as Figure~\ref{fig:OurG}.
Our construction calculates correctly because Toffoli gates in G-rounds and inverse G-rounds are symmetric about the Toffoli gate surrounded by a dotted box as the block level circuit of a C-comparator shown in Figure~\ref{fig:OComp}.

Our construction requires an additional $n$ qubits to preserve in a C-comparator.
Fortunately, we do not use $n$ qubits for $\ket{d}$ in Figure~\ref{fig:Our_Madd}.
Thus, we realize this construction without an overhead of qubits.
We give example circuits as Figure~\ref{fig:CFcomp} or~\ref{fig:CLcomp} in Appendix~B.

The computational cost of our control modular adder is shown in Table~\ref{tab:FTQC_gate_count}, and the breakdown of constructions based on our construction is given in Table~\ref{tab:detail_FTQC_gate_count} in Appendix~D.
From Table~\ref{tab:FTQC_gate_count}, our construction requires $43n$ $T$ gates.
We call this construction a $T$-optimal control modular adder.
The original construction requires $30n$ Toffoli gates, which when implemented using \st (each requiring $7$ $T$~gates) gives $210n$~$T$ gates in total.
Thus, our construction requires only $43/210 \approx 20$\% $T$ of the number of $T$ gates of the original construction.

\begin{table*}
\caption{\label{tab:FTQC_gate_count} 
$T$-count of our control modular adder and prior work.
The latter four constructions are based on our construction proposed in Section III.
The breakdown of the latter four constructions is shown in Table~\ref{tab:detail_FTQC_gate_count} in Appendix~D.
}
\centering
\begin{tabular}{c|cc|c}
Construction & \#comparators & \#adders & Total $T$-count \\\hline
Van Meter and Itoh~\cite{VI2005} & $0$ & $3$ & $210n$ \\\hline 
Draper et al.~\cite{DKRS2006} & $2$ & $1$ & $122.5n$ \\
Thapliyal et al.~(qubit-optimize)~\cite{TMK2020} & $2$ & $1$ & $75n$ \\
Thapliyal et al.~($T$-optimize)~\cite{TMK2020} & $2$ & $1$ & $51n$ \\
\textbf{Ours} & $2$ & $1$ & $43n$
\end{tabular}
\end{table*}

Now, we focus on KQ of a $T$-optimal control modular adder.
In this circuit, we use $O(n)$ qubits and $O(\log n)$ depth, giving a KQ of $O(n \log n)$.
However, we do not consider the computational costs for distillation in this calculation.
We can trade space for time, with substantial flexibility, by allocating more qubits to ancilla \lq\lq factories'', corresponding to increasing the number of $T$ gates that are in concurrent execution~\cite{Ste1998,VLF2010}.

To realize an efficient circuit, we should consider the trade-off between the depth and the number of qubits allocated for distillation.
For example, Kim et al.~\cite{KLL2021} showed that it is possible to run Shor's algorithm with as little as 2\% of the qubits dedicated to distillation, but this construction runs only a single $T$ gate at at time. Since the circuit still requires $O(n)$ $T$ gates, this construct is unable to run in depth $O(\log n)$ and is instead still constrained to $O(n)$ depth. 
To realize smaller KQ, we must run many $T$ gates parallel.
However, there is an upper bound on the number of $T$~gates that can be usefully run in parallel, with the depth limited by the cascading reuse of the qubits. 
Paler and Basmadjian also consider this problem~\cite{PB2019}, and they have concluded that we must determine optimal scheduling methods for $T$ gates.
To realize an accurate estimate of the cost and to enable fair comparison with prior research, we must take into account the $T$ gate costs, including the space for distillation~\cite{IWP2008,JMF2012,VLF2010}, allowing a circuit to run at "the speed of data"~\cite{IWP2008}.

However, it is difficult to calculate computational costs for distillation precisely, because the cost depends on many architecture-specific parameters.
Instead of KQ, we define a new index $\text{KQ}_{T}$, defined as the product of the number of logical qubits and $T$-depth.
We define $n_{T}$ as the $T$-width, the upper-bound of the number of $T$~gates running simultaneously.
We assume that we require a constant $c_{g}$ logical qubits for the distillation step.
By calculating $n_{T}$ minimizing $\text{KQ}_{T}$, we reduce the computational cost of our control modular adder.

In the above discussion, our control modular adder uses $4n+2$ qubits for calculation, as explained in Section III.
In addition, we require ancilla qubits for running $n_{T}$ $T$~gates.
Specifically, to run one $T$ gate, we require one qubit $\ket{Y}$ for running $S$ gates and $c_{g}$ qubits for generating $\ket{A}$.
Thus, when we run $n_{T}$ $T$~gates simultaneously, we use the following qubits:
\begin{itemize}
\item $\ket{y}$~(Contains $\ket{Y}$ states) $n_{T}$ qubits
\item $\ket{g}$~(Generates $\ket{A}$ states) $c_{g}n_{T}$ qubits.
\end{itemize}
The number of qubits in $\ket{y}$ is given as $n_{T}$, because we consume one $S$ gate in each $T$ gate.
Then, the number of qubits is
\begin{align}
4n + (c_{g} + 1) n_{T} + 2.
\end{align}

\begin{figure}[h]
\centering
\includegraphics[keepaspectratio, scale=0.3]{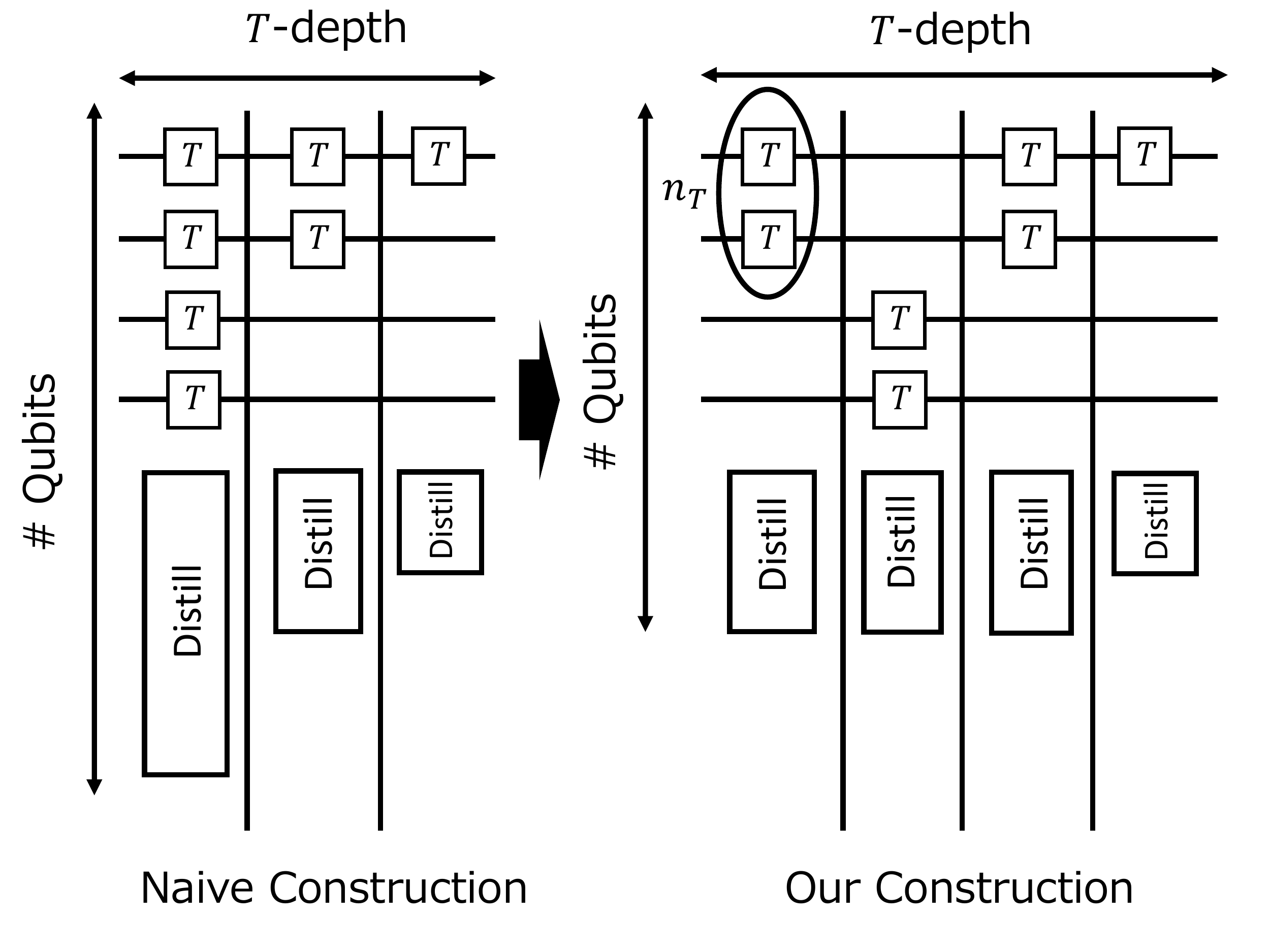}
\caption{\label{fig:KQT} 
Calculating $T$-depth.
Distill means distillation circuits.
In the naive construction, we run as many $T$ gates as possible.
In our construction, we restrict the upper-bound of the number of simultaneous $T$~gates to $n_{T}$.
When we reduce $n_{T}$, the total number of qubits is smaller and $T$-depth is larger.
}
\end{figure}

Now, we calculate $T$-depth.
To calculate $T$-depth, we assume that we run \grt with the same timing, and each \grt has $2$ $T$-depth from Figure~\ref{fig:toffoli_Gid}.
$T$-depth depends on $n_{T}$ as Figure~\ref{fig:KQT}.
Then, $T$-depth is
\begin{align}
\dfrac{86n}{n_{T}} + 12 \log n_{T} - 12.
\end{align}
The detailed calculation is given in Appendix~C.

Now, we minimize $\text{KQ}_{T}$ on $n_{T}$.
$\text{KQ}_{T}$ is
\begin{align}
\label{eq:KQ_Madd}
\left(4n + (c_{g} + 1)n_{T} + 2\right) \left( \dfrac{86n}{n_{T}} + 12 \log n_{T} - 12 \right).
\end{align}
We minimize this on $n_{T} > 0$.

Letting the expression in Eq.~\ref{eq:KQ_Madd} be $f\left(n_{T}\right)$, we see that
\begin{align}
\dfrac{\text{d}^{2} f\left(n_{T}\right)}{\text{d} n_{T}^{2}} > 0
\end{align}
in $n_{T} > 0$.
Thus, $f\left(n_{T}\right)$ is a convex function and it is sufficient to search for only one optimal value of $n_{T}$.
Then, the optimal value
\begin{align}
\label{eq:opt}
n_{T} = \sqrt{\dfrac{86}{3(c_{g} + 1)}} \dfrac{n}{\sqrt{\log n}}
\end{align}
Thus, $O\left(\dfrac{n}{\sqrt{\log n}}\right)$ $T$-width minimizes $\text{KQ}_{T}$.
Plugging this value into Eq.~\ref{eq:KQ_Madd}, 
\begin{align}
4n + (c_{g} + 1)n_{T} + 2 &\sim 4n \\
\dfrac{86n}{n_{T}} + 12 \log n_{T} - 12 &\sim 12 \log n
\end{align}
Therefore, the dominant term of $\text{KQ}_{T}$ is $48n\log n$.

\subsection{Optimization for NISQ}

Now, we propose a form of the control modular adder reducing CNOT gates.
To reduce this number, we review the decomposition of Toffoli gates into CNOT gates.
We use relative-phase Toffoli gates with differences in phase as in Figures~\ref{fig:toffoli_CNOT3} and~\ref{fig:toffoli_CNOT4}, proposed by Maslov~\cite{Mas2016}.
The corresponding unitary matrix of Figure~\ref{fig:toffoli_CNOT3} in the computational basis is
\begin{align}
\label{eq:3CXmat}
\begin{bmatrix}
1 & 0 & 0 & 0 & 0 & 0 & 0 & 0 \\
0 & 1 & 0 & 0 & 0 & 0 & 0 & 0 \\
0 & 0 & 1 & 0 & 0 & 0 & 0 & 0 \\
0 & 0 & 0 & 1 & 0 & 0 & 0 & 0 \\
0 & 0 & 0 & 0 & 1 & 0 & 0 & 0 \\
0 & 0 & 0 & 0 & 0 & -1 & 0 & 0 \\
0 & 0 & 0 & 0 & 0 & 0 & 0 & -i \\
0 & 0 & 0 & 0 & 0 & 0 & i & 0 \\
\end{bmatrix}.
\end{align}
This calculation changes the phase when we input $\ket{1}\ket{0}\ket{1}$, $\ket{1}\ket{1}\ket{0}$, or $\ket{1}\ket{1}\ket{1}$.
We call this relative-phase Toffoli gate \tcx, and we call its inverse \itcx.
The corresponding unitary matrix of Figure~\ref{fig:toffoli_CNOT4} in the computational basis is
\begin{align}
\label{eq:4CXmat}
\begin{bmatrix}
1 & 0 & 0 & 0 & 0 & 0 & 0 & 0 \\
0 & 1 & 0 & 0 & 0 & 0 & 0 & 0 \\
0 & 0 & 1 & 0 & 0 & 0 & 0 & 0 \\
0 & 0 & 0 & 1 & 0 & 0 & 0 & 0 \\
0 & 0 & 0 & 0 & 1 & 0 & 0 & 0 \\
0 & 0 & 0 & 0 & 0 & 1 & 0 & 0 \\
0 & 0 & 0 & 0 & 0 & 0 & 0 & -i \\
0 & 0 & 0 & 0 & 0 & 0 & -i & 0 \\
\end{bmatrix}.
\end{align}
This calculation changes the phase when both control bits are $1$.
We call this relative-phase Toffoli gate \fcx, and its inverse \ifcx.
By using these relative-phase Toffoli gates, we reduce the number of CNOT gates.
Next, we address which Toffoli gates can be replaced with relative-phase Toffoli gates.
\begin{figure}
\[
\Qcircuit @C=1em @R=.7em {
  & \qw & \qw & \qw & \qw & \ctrl{2} & \qw & \qw & \qw & \qw & \qw \\
  & \qw & \qw & \ctrl{1} & \qw & \qw & \qw & \ctrl{1} & \qw & \qw &\qw \\
  & \gate{H} & \gate{T} & \targ & \gate{T^{\dag}} & \targ & \gate{T} & \targ & \gate{T^{\dag}} & \gate{H} & \qw
}
\]
\caption{\label{fig:toffoli_CNOT3} 
A relative-phase Toffoli gate with 3 CNOT~(\tcx). 
This calculation changes the phase when we input $\ket{1}\ket{0}\ket{1}$, $\ket{1}\ket{1}\ket{0}$, and $\ket{1}\ket{1}\ket{1}$.
We call the inverse circuit of \tcx, \itcx.}
\end{figure}
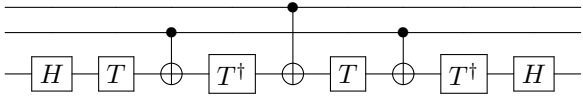
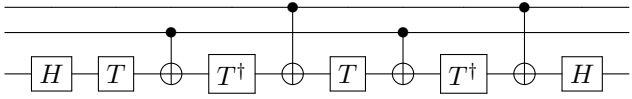
\begin{figure}
\[
\Qcircuit @C=1em @R=.7em {
  & \qw & \qw & \qw & \qw & \ctrl{2} & \qw & \qw & \qw & \ctrl{2} & \qw & \qw \\
  & \qw & \qw & \ctrl{1} & \qw & \qw & \qw & \ctrl{1} & \qw & \qw & \qw & \qw \\
  & \gate{H} & \gate{T} & \targ & \gate{T^{\dag}} & \targ & \gate{T} & \targ & \gate{T^{\dag}} & \targ & \gate{H} & \qw
}
\]
\caption{\label{fig:toffoli_CNOT4} 
A relative-phase Toffoli gate with 4 CNOT~(\fcx). 
This calculation change the phase when both control bits are $1$.
We call the inverse circuit of \fcx, \ifcx.}
\end{figure}

First, we consider which Toffoli gates can be replaced in a C-comparator.
The structure of a C-comparator is shown in Figure~\ref{fig:OComp}, and we give an example circuit in Figure~\ref{fig:CFcomp} or~\ref{fig:CLcomp} in Appendix~B.
In these figures, all Toffoli gates are symmetric about the Toffoli gate surrounded by a dotted box, in the middle of the circuit.
Thus, we can replace the Toffoli gates to the left of the dotted box by \tcx and those to the right of the box by \itcx.
Therefore, we can replace all of the Toffoli gates in a C-comparator except this middle one with \tcx or \itcx.

Next, we address which Toffoli gates can be replaced in a CC-adder, and find that those in P-rounds can be replaced by \tcx and those in inverse P-rounds by \itcx.
The other Toffoli gates used calculate the value of carries, and these carries are cleared after the calculation.
In the calculation of carries, the values of the control bits change between calculating a carry and erasing it, which would seem to rule out using anything but pure Toffoli gates.
However, looking more closely, we see that the value of a carry changes at most once, namely when both control bits are $\ket{1}$.
Thus, if we calculate correctly in the other situations, we can calculate and clear carries correctly.
\fcx satisfies this.
Therefore, we can replace Toffoli gates by \fcx in the Initialization, G-rounds, and C-rounds,
and we can replace Toffoli gates by \ifcx in the inverse rounds.

As a result, the cost of our control modular adder is shown in Table~\ref{tab:NISQ_gate_count} and~\ref{tab:NISQ_depth_count}.
The breakdown of those based on our construction are shown in Table~\ref{tab:detail_NISQ_gate_count} and~\ref{tab:detail_NISQ_depth_count} in Appendix~D.
From Table~\ref{tab:NISQ_gate_count} and~\ref{tab:NISQ_depth_count}, our construction is better in terms of both the number of CNOT gates and CNOT-depth.
Now, we compare our circuit to the original construction.

First, we compare the CNOT count.
Our construction requires $64.75n$ CNOT gates.
The original construction requires $30n$ Toffoli gates implemented by \st using $6$ CNOT gates, and we use an additional $4.5n$ CNOT gates in embedding or resetting.
Thus, the original construction requires $184.5n$ CNOT gates in total.
Therefore, our construction reduces the number of CNOT gates to only 35\% of the number in the original.

Next, we compare $\text{KQ}_{\text{CX}}$, defined as the product of the number of qubits and CNOT-depth.
Our construction requires $120n \log n$ $\text{KQ}_{\text{CX}}$.
The original construction requires $12 \log n$ Toffoli depth implemented by \st requiring $6$CNOT-depth, and we require $6 \log n$ CNOT-depth for the embedding step.
Thus, the original construction requires $78\log n$ CNOT-depth and $312n \log n$ $\text{KQ}_{\text{CX}}$.
Therefore, our construction requires only 38\% of the $\text{KQ}_{\text{CX}}$ of the original construction.

\begin{table*}
\caption{\label{tab:NISQ_gate_count} 
CNOT count of our control modular adder and prior work.
The latter four constructions are based on our construction proposed in Section III.
The breakdown of the latter four constructions is shown in Table~\ref{tab:detail_NISQ_gate_count} in Appendix~D.
}
\centering
\begin{tabular}{c|cc|c}
Construction & \#comparators & \#adders & Total CNOT count \\\hline
Van Meter and Itoh~\cite{VI2005} & $0$ & $3$ & $184.5n$ \\\hline 
Draper et al.~\cite{DKRS2006} & $2$ & $1$ & $111.75n$ \\
Thapliyal et al.~(qubit-optimize)~\cite{TMK2020} & $2$ & $1$ & $88n$ \\
Thapliyal et al.~($T$-optimize)~\cite{TMK2020} & $2$ & $1$ & $104n$ \\
\textbf{Ours} & $2$ & $1$ & $64.75n$
\end{tabular}
\end{table*}

\begin{table*}
\caption{\label{tab:NISQ_depth_count} 
$\text{KQ}_{\text{CX}}$ of our control modular adder and prior work.
The latter four constructions are based on our construction proposed in Section III.
The breakdown of the latter four constructions are shown in Table~\ref{tab:detail_NISQ_depth_count} in Appendix~D.
}
\centering
\begin{tabular}{c|cc|c}
Construction & \#qubits & The depth of the circuit & $\text{KQ}_{\text{CX}}$ \\\hline
Van Meter and Itoh~\cite{VI2005} & $4n$ & $78 \log n$ & $312n\log n$ \\\hline 
Draper et al.~\cite{DKRS2006} & $4n$ & $50 \log n$ & $200n \log n$ \\
Thapliyal et al.~(qubit-optimize)~\cite{TMK2020} & $4n$ & $50\log n$ & $200n \log n$ \\
Thapliyal et al.~($T$-optimize)~\cite{TMK2020} & $4.5n$ & $66 \log n$ & $297n \log n$ \\
\textbf{Ours} & $4n$ & $30 \log n$ & $120n\log n$
\end{tabular}
\end{table*}

\section{Conclusion and Future Work}

In this study, we proposed a method of optimizing a control modular adder based on a carry-lookahead adder~\cite{DKRS2006} and  Van Meter and Itoh's construction~\cite{VI2005}.
First, we show that the general construction given as Figure~\ref{fig:Zal_Madd} is about $2/3$ of the KQ of the original construction.
Then, we construct a more efficient circuit.
We evaluate the computational cost in FTQ and we show that our circuit requires only 20\% of the $T$ gates of the original.
Moreover, we show that our circuit achieves its minimum $\text{KQ}_{T}$ when we run $\Theta\left(\dfrac{n}{\sqrt{\log n}} \right)$ $T$ gates simultaneously.
Finally, we propose an efficient circuit for use in the NISQ era, and we show that our circuit requires only 35\% of the CNOT gates and 38\% $\text{KQ}_{\text{CX}}$ of the original.

In this work, we have focused on optimizing Toffoli gates by using relative-phase Toffoli gates.
However, in previous research~\cite{Mog2019,TJNA2013}, other researchers have used gates such as Fredkin and Peres gates.
These gates also may be simplified by replacing them with relative-phase gates.
Thus, we expect that those circuits would also show an improvement with these techniques applied.

In this paper, we have considered only the single control modular addition. In additional future work, the circuits that postpone and summarize multiple modular arithmetic operations, as proposed by Van Meter and Itoh~\cite{VI2005}, should be addressed using similar optimization techniques.
In addition, it is important to minimize KQ by reordering gates~\cite{MDM2008,PG2014}.

Our construction does not consider the architecture of quantum computers as linear nearest neighbor architecture~\cite{CV2012,FDH2004,HNY2011}.
Thus, in the next step, we will consider the appropriate architecture and additional cost for our construction.

Lastly, we focused only on the Logical layer of FTQ in this study. Future work, we must consider the mapping to physical qubits, as well as distillation protocols.

\appendix

\section{}

\subsection{Detailed Explanation of Draper et al.'s Carry-lookahead Adder}

Draper et al.'s carry-lookahead adder is given as follows:

\textbf{Initialization} ($n$ Toffoli gates and $n$ CNOT gates)

We calculate $g\left[i, i+1\right]$ and $p\left[i, i+1\right]$ $\left( 0 \leq i \leq n-1 \right)$, as follows:
\begin{align}
\label{eq:G1}
g\left[i, i+1\right] =
\begin{cases}
1 & \text{if } a_{i} = b_{i} = 1 \\
0 &  \text{otherwise}
\end{cases}
\end{align}
\begin{align}
\label{eq:P1}
p\left[i, i+1\right] =
\begin{cases}
1 & \text{if } a_{i} + b_{i} = 1 \\
0 &  \text{otherwise}
\end{cases}
\end{align}
The circuit calculating these is shown in Figure~\ref{fig:Init}.

\begin{figure}
\[
\Qcircuit @C=1em @R=.7em {
  \lstick{\ket{a_{i}}} &&& \qw & \ctrl{1} & \qw & \ctrl{1} & \qw & \rstick{\ket{a_{i}}} \\
  \lstick{\ket{b_{i}}} &&& \qw & \ctrl{1} & \qw & \targ & \qw & \rstick{\ket{p\left[i, i+1\right]}} \\ 
  \lstick{\ket{c_{i+1}}} & \ket{0} && \qw & \targ & \qw &\qw & \qw & \rstick{\ket{g\left[i, i+1\right]}} 
}
\]
\caption{\label{fig:Init}
A calculation circuit of $g\left[i, i+1\right]$ and $p\left[i, i+1\right]$ $\left( 0 \leq i \leq n-1 \right)$.
We use $\ket{c_{i+1}}$ as the third qubit.
We can run these gates simultaneously for $i = 0$ to $n-1$.
}
\end{figure}
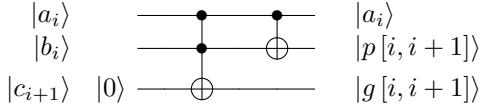

\textbf{P-rounds} ($n$ Toffoli gates and $\log n$ Toffoli depth)

We calculate the $p$-function by using eq.~(\ref{eq:renewalP}).
We use a parameter $t_{p}$ representing the range of the propagation of carry.
We increase $t_{p}$ from $1$ to $\left\lfloor \log n \right\rfloor - 1$.
In each $t_{p}$, we calculate 
$p\left[2^{t_{p}}i, 2^{t_{p}}\left(i+1\right)\right]$ $\left(1 \leq i \leq \left\lfloor n/2^{t_{p}} \right\rfloor - 1\right)$ 
by setting 
$\ket{p\left[2^{t_{p}}i, 2^{t_{p}}\left(i+1/2\right)\right]}$ 
and 
$\ket{p\left[2^{t_{p}}\left(i+1/2\right), 2^{t_{p}}\left(i+1\right)\right]}$ as the control qubits in in Toffoli gate in Figure~\ref{fig:CalcP}.
These Toffoli gates are applied simultaneously in each $t_{p}$.

\textbf{G-rounds} ($n$ Toffoli gates and $\log n$ Toffoli depth)

We calculate $\ket{c_{2^{k}}}~\left(k \in \mathbb{N} \cup \{0\}\right)$ by using eq.~(\ref{eq:renewalG}).
We use a parameter $t_{g}$ similar to the way we used it in P-rounds.
We increase $t_{g}$ from $1$ to $\left\lfloor \log n \right\rfloor$.
In each $t_{g}$, we calculate 
$g\left[2^{t_{g}}i, 2^{t_{g}}\left(i+1\right)\right]$ $\left(0 \leq i \leq \left\lfloor n/2^{t_{g}} \right\rfloor - 1\right)$
by setting 
$\ket{c_{2^{t_{g}}i+2^{t_{g}-1}}}$ and 
$\ket{p\left[2^{t_{g}}(i+1/2), 2^{t_{g}}\left(i+1 \right)\right]}$ 
as the control qubits and 
$\ket{c_{2^{t_{g}}\left(i+1\right)}}$ 
as the target qubit in Toffoli gate in Figure~\ref{fig:CalcG}.
These Toffoli gates are applied simultaneously in each $t_{g}$.
Moreover, G-rounds with $t_{g}$ can be run in parallel with former P-rounds with $t_{g}+1$.

\textbf{C-rounds} ($n$ Toffoli gates and $\log n$ Toffoli depth)

We calculate all carries $\ket{c}$ by using eq.~(\ref{eq:renewalG}).
We use a parameter $t_{c}$ similar to the way we used it in P-rounds.
We decrease $t_{c}$ from $\left\lfloor \log \left(2n/3\right) \right\rfloor$ to $1$.
In each $t_{c}$, we calculate 
$\ket{c_{2^{t_{c}}i+2^{t_{c}-1}}}$ $\left(1 \leq i \leq \left\lfloor \left(n-2^{t_{c}-1}\right)/2^{t_{c}} \right\rfloor - 1 \right)$
by setting
$\ket{c_{2^{t_{c}}i}}$ 
and $\ket{p\left[2^{t_{c}}i, 2^{t_{c}}i+2^{t_{c}-1}\right]}$
as the control qubits and 
$\ket{c_{2^{t_{c}}i+2^{t_{c}-1}}}$
as the target qubit in Toffoli gate in Figure~\ref{fig:CalcG}.
These Toffoli gates are applied simultaneously in each $t_{c}$.

\textbf{Inverse P-rounds} ($n$ Toffoli gates and $\log n$ Toffoli depth)

We apply the same gates as P-rounds in reverse order.
Rounds with $t_{p}$ can be run in parallel with former C-round with $t_{p}+1$.

\textbf{Calculating $\ket{a + b}$} ($n$ CNOT gates)

We calculate $\left(a + b \right)_{i}~\left(0 \leq i \leq n-2 \right)$ on $\ket{b_{i}}$.
We apply CNOT gates with the control qubit of $\ket{c_{i+1}}$ and the target qubit of $\ket{b_{i+1}}$.
These CNOT gates are applied simultaneously.

\textbf{Erasing Carry} ($5n$ Toffoli gates, $2n$ CNOT gates, and $2\log n$ Toffoli depth)

We erase all carries by applying the inverse circuit of $a + \left(2^{n} - 1 - a - b\right)$ on the lower $n - 1$ bits, as shown in Figure~\ref{fig:Erase_Carry}.
We apply gates before P-rounds and after inverse Initialization to erase carries.
We call these gates PE-rounds and inverse PE-rounds respectively.

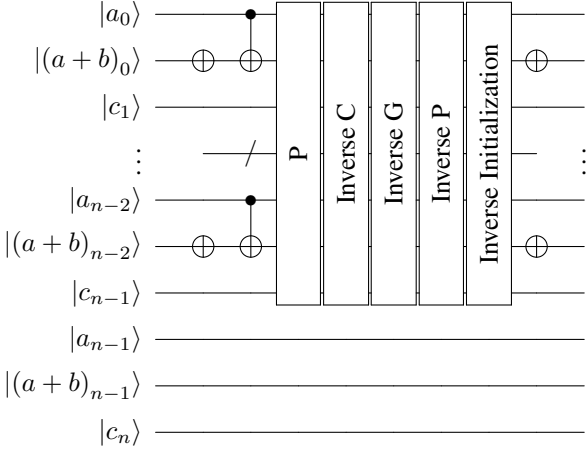
\begin{figure}
\[
\Qcircuit @C=.1em @R=.05em @! {
  \lstick{\ket{a_{0}}} & \qw & \ctrl{1} 
  & \multigate{6}{\text{\rotatebox{90}{P}}}
  & \multigate{6}{\text{\rotatebox{90}{Inverse C}}}
  & \multigate{6}{\text{\rotatebox{90}{Inverse G}}}
  & \multigate{6}{\text{\rotatebox{90}{Inverse P}}}
  & \multigate{6}{\text{\rotatebox{90}{Inverse Initialization}}}
  & \qw & \qw \\
  \lstick{\ket{\left(a + b\right)_{0}}} & \targ & \targ 
  & \ghost{\text{\rotatebox{90}{P}}}
  & \ghost{\text{\rotatebox{90}{Inverse C}}}
  & \ghost{\text{\rotatebox{90}{Inverse G}}}
  & \ghost{\text{\rotatebox{90}{Inverse P}}}
  & \ghost{\text{\rotatebox{90}{Inverse Initialization}}}
  & \targ & \qw \\
  \lstick{\ket{c_{1}}} & \qw & \qw 
  & \ghost{\text{\rotatebox{90}{P}}}
  & \ghost{\text{\rotatebox{90}{Inverse C}}}
  & \ghost{\text{\rotatebox{90}{Inverse G}}}
  & \ghost{\text{\rotatebox{90}{Inverse P}}}
  & \ghost{\text{\rotatebox{90}{Inverse Initialization}}}
  & \qw & \qw \\ 
  \lstick{\vdots} &  & {/} \qw 
  & \ghost{\text{\rotatebox{90}{P}}}
  & \ghost{\text{\rotatebox{90}{Inverse C}}}
  & \ghost{\text{\rotatebox{90}{Inverse G}}}
  & \ghost{\text{\rotatebox{90}{Inverse P}}}
  & \ghost{\text{\rotatebox{90}{Inverse Initialization}}}
  & \qw & \vdots \\ 
  \lstick{\ket{a_{n-2}}} & \qw & \ctrl{1}  
  & \ghost{\text{\rotatebox{90}{P}}}
  & \ghost{\text{\rotatebox{90}{Inverse C}}}
  & \ghost{\text{\rotatebox{90}{Inverse G}}}
  & \ghost{\text{\rotatebox{90}{Inverse P}}}
  & \ghost{\text{\rotatebox{90}{Inverse Initialization}}}
  & \qw & \qw \\ 
  \lstick{\ket{\left(a + b\right)_{n-2}}} & \targ & \targ 
  & \ghost{\text{\rotatebox{90}{P}}}
  & \ghost{\text{\rotatebox{90}{Inverse C}}}
  & \ghost{\text{\rotatebox{90}{Inverse G}}}
  & \ghost{\text{\rotatebox{90}{Inverse P}}}
  & \ghost{\text{\rotatebox{90}{Inverse Initialization}}}
  & \targ & \qw \\  
  \lstick{\ket{c_{n-1}}} & \qw & \qw
  & \ghost{\text{\rotatebox{90}{P}}}
  & \ghost{\text{\rotatebox{90}{Inverse C}}}
  & \ghost{\text{\rotatebox{90}{Inverse G}}}
  & \ghost{\text{\rotatebox{90}{Inverse P}}}
  & \ghost{\text{\rotatebox{90}{Inverse Initialization}}}
  & \qw & \qw \\ 
  \lstick{\ket{a_{n-1}}} & \qw & \qw 
  & \qw
  & \qw
  & \qw
  & \qw
  & \qw
  & \qw & \qw \\ 
  \lstick{\ket{\left(a + b\right)_{n-1}}} & \qw & \qw 
  & \qw
  & \qw
  & \qw
  & \qw
  & \qw
  & \qw & \qw \\  
  \lstick{\ket{c_{n}}} & \qw & \qw 
  & \qw
  & \qw
  & \qw
  & \qw
  & \qw
  & \qw & \qw
}
\]
\caption{\label{fig:Erase_Carry}
Erasing $\ket{c}$.
We apply gates only on the lower $n-1$ qubits of $\ket{a}$, $\ket{b}$, and $\ket{c}$.
We apply the same gates in omitted qubits $\ket{a_{i}}$, $\ket{\left(a+b\right)_{i}}$, and $\ket{c_{i+1}}$.
The P-rounds and inverse C-rounds can be run in parallel, as can the inverse G-rounds and inverse P-rounds.
We define PE-rounds as the gates before P-rounds, and inverse PE-rounds as the gates after inverse Initialization.
}
\end{figure}

Now, we show the example circuit of Draper et al.'s carry-lookahead adder as given in Figure~\ref{fig:Original_Adder}.
In this example, we define $a$ and $b$ as $6$-bit values, and we calculate $\ket{a}\ket{b} \to \ket{a}\ket{a + b}$.
In Figure~\ref{fig:Original_Adder}, in constrast to Figure~\ref{fig:Zal_Madd}, qubits are sorted from low order to high order.

\begin{figure}
\[
\Qcircuit @C=.08em @R=.7em {
& & & 
& &
& & &
& & &
& \mbox{\text{IP}} &
& &
& & &
& &
& \mbox{\text{IC}}& &
& & &
& \mbox{\text{IP}}&
& & &
& &
& \\
& \mbox{\text{Init}} & & 
& \mbox{\text{P}} &
& \mbox{\text{G}} & &
& \mbox{\text{C}} & &
& &
& &
& & &
& \mbox{\text{P}} &
& & &
& \mbox{\text{IG}} &  &
& &
& & \mbox{\text{IInit}} & 
& &
& \\
\lstick{\ket{a_{0}}} & \ctrl{1} & \ctrl{1} & \qw
& \qw & \qw
& \qw & \qw & \qw
& \qw & \qw & \qw
& \qw & \qw
& \qw & \qw
& \qw & \ctrl{1} & \qw
& \qw & \qw
& \qw & \qw & \qw
& \qw & \qw & \qw
& \qw & \qw
& \ctrl{1} & \ctrl{1} & \qw
& \qw & \qw
& \qw \\
\lstick{\ket{b_{0}}} & \ctrl{1} & \targ & \qw
& \qw & \qw
& \qw & \qw & \qw
& \qw & \qw & \qw
& \qw & \qw
& \qw & \qw
& \targ & \targ & \qw
& \qw & \qw
& \qw & \qw & \qw
& \qw & \qw & \qw
& \qw & \qw
& \targ & \ctrl{1} & \qw
& \targ & \qw
& \qw 
& \rstick{\ket{(a+b)_{0}}} \\
\lstick{\ket{c_{1}}} & \targ & \qw & \qw
& \qw & \qw
& \ctrl{2} & \qw & \qw
& \qw & \qw & \qw
& \qw & \qw
& \ctrl{2} & \qw
& \qw & \qw & \qw
& \qw & \qw
& \qw & \qw & \qw
& \qw & \ctrl{2} & \qw
& \qw & \qw
& \qw & \targ & \qw
& \qw & \qw
& \qw \\
\lstick{\ket{a_{1}}} & \ctrl{1} & \ctrl{1} & \qw
& \qw & \qw
& \qw & \qw & \qw
& \qw & \qw & \qw
& \qw & \qw
& \qw & \qw
& \qw & \ctrl{1} & \qw
& \qw & \qw
& \qw & \qw & \qw
& \qw & \qw & \qw
& \qw & \qw
& \ctrl{1} & \ctrl{1} & \qw
& \qw & \qw
& \qw \\
\lstick{\ket{b_{1}}} & \ctrl{1} & \targ & \qw
& \qw & \qw
& \ctrl{1} & \qw & \qw
& \qw & \qw & \qw
& \qw & \qw
& \targ & \qw
& \targ & \targ & \qw
& \qw & \qw
& \qw & \qw & \qw
& \qw & \ctrl{1} & \qw
& \qw & \qw
& \targ & \ctrl{1} & \qw
& \targ & \qw
& \qw 
& \rstick{\ket{(a+b)_{1}}} \\
\lstick{\ket{c_{2}}} & \targ & \qw & \qw
& \qw & \qw
& \targ & \ctrl{3} & \qw
& \qw & \ctrl{2} & \qw
& \qw & \qw
& \ctrl{2} & \qw
& \qw & \qw & \qw
& \qw & \qw
& \ctrl{2} & \qw & \qw
& \ctrl{3} & \targ & \qw
& \qw & \qw
& \qw & \targ & \qw
& \qw & \qw
& \qw \\
\lstick{\ket{a_{2}}} & \ctrl{1} & \ctrl{1} & \qw
& \qw & \qw
& \qw & \qw & \qw
& \qw & \qw & \qw
& \qw & \qw
& \qw & \qw
& \qw & \ctrl{1} & \qw
& \qw & \qw
& \qw & \qw & \qw
& \qw & \qw & \qw
& \qw & \qw
& \ctrl{1} & \ctrl{1} & \qw
& \qw & \qw
& \qw \\
\lstick{\ket{b_{2}}} & \ctrl{2} & \targ & \qw 
& \ctrl{1} & \qw
& \qw & \qw & \qw
& \qw & \ctrl{2} & \qw
& \ctrl{1} & \qw
& \targ & \qw
& \targ & \targ & \qw
& \ctrl{1} & \qw
& \ctrl{2} & \qw & \qw
& \qw & \qw & \qw
& \ctrl{1} & \qw
& \targ & \ctrl{2} & \qw
& \targ & \qw
& \qw
& \rstick{\ket{(a+b)_{2}}} \\
\lstick{\ket{p\left[2, 4\right]}} & \qw & \qw & \qw
& \targ & \qw
& \qw & \ctrl{4} & \qw
& \qw & \qw & \qw
& \targ & \qw
& \qw & \qw
& \qw & \qw & \qw
& \targ & \qw
& \qw & \qw & \qw
& \ctrl{4} & \qw & \qw
& \targ & \qw
& \qw & \qw & \qw
& \qw & \qw
& \qw \\
\lstick{\ket{c_{3}}} & \targ & \qw  & \qw
& \qw & \qw
& \ctrl{2} & \qw & \qw
& \qw & \targ & \qw
& \qw & \qw
& \ctrl{2} & \qw
& \qw & \qw & \qw
& \qw & \qw
& \targ & \qw & \qw
& \qw & \ctrl{2} & \qw
& \qw & \qw
& \qw & \targ & \qw
& \qw & \qw
& \qw \\
\lstick{\ket{a_{3}}} & \ctrl{1} & \ctrl{1} & \qw
& \qw & \qw
& \qw & \qw & \qw
& \qw & \qw & \qw
& \qw & \qw
& \qw & \qw
& \qw & \ctrl{1} & \qw
& \qw & \qw
& \qw & \qw & \qw
& \qw & \qw & \qw
& \qw & \qw
& \ctrl{1} & \ctrl{1} & \qw
& \qw & \qw
& \qw \\
\lstick{\ket{b_{3}}} & \ctrl{1} & \targ & \qw 
& \ctrl{-3} & \qw
& \ctrl{1} & \qw & \qw
& \qw & \qw & \qw
& \ctrl{-3} & \qw
& \targ & \qw
& \targ & \targ & \qw
& \ctrl{-3} & \qw
& \qw & \qw & \qw
& \qw & \ctrl{1} & \qw
& \ctrl{-3} & \qw
& \targ & \ctrl{1} & \qw
& \targ & \qw
& \qw 
& \rstick{\ket{(a+b)_{3}}} \\
\lstick{\ket{c_{4}}} & \targ & \qw & \qw
& \qw & \qw
& \targ & \targ & \qw
& \ctrl{3} & \ctrl{2} & \qw
& \qw & \qw
& \ctrl{2} & \qw
& \qw & \qw & \qw
& \qw & \qw
& \ctrl{2} & \qw & \qw
& \targ & \targ & \qw
& \qw & \qw
& \qw & \targ & \qw
& \qw & \qw
& \qw \\
\lstick{\ket{a_{4}}} & \ctrl{1} & \ctrl{1} & \qw
& \qw & \qw
& \qw & \qw & \qw
& \qw & \qw & \qw
& \qw & \qw
& \qw & \qw
& \qw & \ctrl{1} & \qw
& \qw & \qw
& \qw & \qw & \qw
& \qw & \qw & \qw
& \qw & \qw
& \ctrl{1} & \ctrl{1} & \qw
& \qw & \qw
& \qw \\
\lstick{\ket{b_{4}}} & \ctrl{2} & \targ & \qw
& \ctrl{1} & \qw
& \qw & \qw & \qw
& \qw & \ctrl{2} & \qw
& \ctrl{1} & \qw
& \targ & \qw
& \targ & \targ & \qw
& \qw & \qw
& \ctrl{2} & \qw & \qw
& \qw & \qw & \qw
& \qw & \qw
& \targ & \ctrl{2} & \qw
& \targ & \qw
& \qw 
& \rstick{\ket{(a+b)_{4}}} \\
\lstick{\ket{p\left[4, 6\right]}} & \qw & \qw & \qw
& \targ & \qw
& \qw & \qw & \qw
& \ctrl{4} & \qw & \qw
& \targ & \qw
& \qw & \qw
& \qw & \qw & \qw
& \qw & \qw
& \qw & \qw & \qw
& \qw & \qw & \qw
& \qw & \qw
& \qw & \qw & \qw
& \qw & \qw
& \qw \\
\lstick{\ket{c_{5}}} & \targ & \qw & \qw 
& \qw & \qw
& \ctrl{2} & \qw & \qw
& \qw & \targ & \qw
& \qw & \qw
& \ctrl{2} & \qw
& \qw & \qw & \qw
& \qw & \qw
& \targ & \qw & \qw
& \qw & \qw & \qw
& \qw & \qw
& \qw & \targ & \qw
& \qw & \qw
& \qw \\
\lstick{\ket{a_{5}}} & \ctrl{1} & \ctrl{1} & \qw
& \qw & \qw
& \qw & \qw & \qw
& \qw & \qw & \qw
& \qw & \qw
& \qw & \qw
& \qw & \qw & \qw
& \qw & \qw
& \qw & \qw & \qw
& \qw & \qw & \qw
& \qw & \qw
& \qw & \qw & \qw
& \qw & \qw
& \qw \\
\lstick{\ket{b_{5}}} & \ctrl{1} & \targ & \qw 
& \ctrl{-3} & \qw
& \ctrl{1} & \qw & \qw
& \qw & \qw & \qw
& \ctrl{-3} & \qw
& \targ & \qw
& \qw & \qw & \qw
& \qw & \qw
& \qw & \qw & \qw
& \qw & \qw & \qw
& \qw & \qw
& \qw & \qw & \qw
& \qw & \qw
& \qw  
& \rstick{\ket{(a+b)_{5}}} \\
\lstick{\ket{c_{6}}} & \targ & \qw & \qw
& \qw & \qw
& \targ & \qw & \qw
& \targ & \qw & \qw
& \qw & \qw
& \qw & \qw
& \qw & \qw & \qw
& \qw & \qw
& \qw & \qw & \qw
& \qw & \qw & \qw
& \qw & \qw
& \qw & \qw & \qw
& \qw & \qw
& \qw 
& \rstick{\ket{(a+b)_{6}}} \\
& & & \mbox{}
& & \mbox{}
& & & \mbox{}
& & & \mbox{}
& & \mbox{}
& & \mbox{}
& & & \mbox{}
& & \mbox{}
& & & \mbox{}
& & & \mbox{}
& & \mbox{}
& & & \mbox{}
& & \mbox{}
& \mbox{}
\gategroup{2}{4}{23}{4}{.1em}{--} %
\gategroup{2}{6}{23}{6}{.1em}{--} %
\gategroup{2}{9}{23}{9}{.1em}{--} %
\gategroup{2}{12}{23}{12}{.1em}{--} %
\gategroup{2}{14}{23}{14}{.1em}{--} %
\gategroup{2}{19}{23}{19}{.1em}{--} %
\gategroup{2}{21}{23}{21}{.1em}{--} %
\gategroup{2}{24}{23}{24}{.1em}{--} %
\gategroup{2}{27}{23}{27}{.1em}{--} %
\gategroup{2}{29}{23}{29}{.1em}{--} %
\gategroup{3}{32}{23}{32}{.1em}{--} %
}
\]
\caption{\label{fig:Original_Adder} 
An example of Draper et al.'s carry-lookahead adder.
This circuit adds two $6$-bit numbers $a$ and $b$, namely $\ket{a}\ket{b} \to \ket{a}\ket{a+b}$.
In this figure, we sort qubits from the lowest qubits to the highest qubits.
The labels at the top are the rounds including Toffoli gates.
Init means Initialization.
IP, IC, IG, and IInit mean Inverse P-rounds, Inverse C-rounds, Inverse G-rounds, and Inverse Initialization respectively.
}
\end{figure}
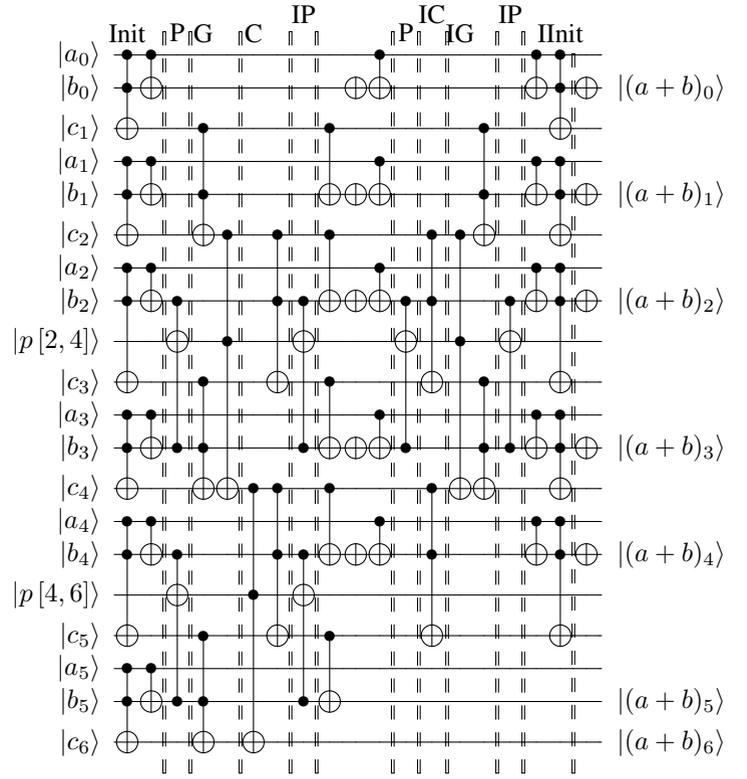

\subsection{Detailed Construction of Our Control Modular Adder}

In this section, we explain detail of our control modular adder.
We show the example figures of our control modular adder too.

\subsubsection{A C-Comparator}

Now, we explain the construction of a C-comparator in more detail.
In a C-comparator, we judge whether or not $b \geq d$, where $b$ is a quantum value and $d$ is a classical value.
As noted in Section III. A., we conduct this by calculating the carry out of the entire circuit $b + \left(2^{n} - d\right)$.
Our construction is given as follows:

\textbf{Initialization} 

If we conduct Initialization naively, we apply a Toffoli gate and a CNOT gate for each bit.
However, the compilation of a quantum algorithm often requires compilation (selection of the sequence of gates) to be adapted to the specific classical values that are inputs to the overall algorithm.
Because $2^n - d$ is a classical value, we can convert some Toffoli gates to CNOT gates and eliminate other gates.
Then, we calculate each $\left(2^{n} - d \right)_{i}~\left(0 \leq i \leq n-1 \right)$.
If $\left(2^{n} - d \right)_{i} = 1$,
\begin{enumerate}
\item We apply CNOT gates with the control qubit $\ket{b_{i}}$ and the target qubit $\ket{c_{i+1}}$.
\item We apply $X$ gates with on $\ket{b_{i}}$.
\end{enumerate}
These operations correspond to Toffoli gates or CNOT gates in the Initialization phase in Draper et al.'s construction, respectively.

\textbf{P-rounds and G-rounds} 

We conduct P-rounds and G-rounds similar to Draper et al.'s construction.

\textbf{Writing result on the \compq qubit} ($O(1)$ gates and $O(1)$ depth)

If we want to flip \compq when $b \geq d$, we apply Toffoli gates
with the control qubits of
\ctrlq and $\ket{g\left[0, n\right]}$,
and with the target qubit of
\compq.
If we want to flip \compq when $b < d$, we apply Toffoli gates similarly to $b \geq d$, but we apply NOT gates on $\ket{g\left[0, n\right]}$ before and after the Toffoli gate.

\textbf{Resetting qubits}

We conduct inverse G-rounds and inverse P-rounds similar to Draper et al.'s construction.
Moreover, we conduct the inverse of our Initialization.
Then, we reset all qubits except \compq as the initial values.

\subsubsection{A CC-adder}

First, we explain the construction of embedding in more detail.
We want to embed as follows:
\begin{itemize}
\item If \ctrlq is $1$ and \compq is $1$, we embed $2^{n}+a-N$.
\item If \ctrlq is $1$ and \compq is $0$, we embed $a$.
\item Otherwise, we embed no value.
\end{itemize}
Therefore, we embed on the second register on Figure~\ref{fig:Our_Madd} as follows:
\begin{itemize}
\item If \ctrlq is $1$ and $\left(2^{n}+a-N\right)_{i} = a_{i} = 1$, $i$-th qubit is $\ket{1}$.
\item If \ctrlq is $1$, \compq is $1$, $\left(2^{n}+a-N\right)_{i} = 1$, and $a_{i} = 0$, $i$-th qubit is $\ket{1}$.
\item If \ctrlq is $1$, \compq is $0$, $\left(2^{n}+a-N\right)_{i} = 0$, and $a_{i} = 1$, $i$-th qubit is $\ket{1}$.
\item Otherwise, we do nothing.
\end{itemize}
In the above condition, the values of $\left(2^{n}+a-N\right)_{i}$ and $a_{i}$ are classical information, and \ctrlq and \compq are quantum information.
Thus, embedding in the first condition can be realized by CNOT gates with the control qubit of \ctrlq.
Moreover, embedding in the second and third condition can be realized by Toffoli gates with the control qubits of \ctrlq and \compq.
However, the set of $i$ in each classical condition has no overlap.
Therefore, once we embed one of $i$, we can embed the remaining value as CNOT gates.
In each set, we have average $n/4$ elements requiring $n/4$ CNOT gates, $O(1)$ additional gates.
Thus, these embedding can be implemented by $3n/4$ CNOT gates.
Moreover, because we can run these simultaneously, embedding requires $\log n$ CNOT depth.
The reset of embedding can be implemented similarly.

Next, we explain the optimization in an adder.
In our calculation, there is no carry for $g\left[0, n\right]$ whether we subtract $N-a$ or add $a$.
Thus, we can disregard calculation of carry qubit $g\left[0, n\right]$.
To realize this, we omit calculation of $p\left[i, n\right]$ and $g\left[i, n\right]$ $(i < n)$.
Moreover, by using classicality of $a$ and $N$, we know that we embed no value in average $n/4$ qubits on the second register of Figure~\ref{fig:Our_Madd}.
In these qubits, we can omit Initialization, inverse Initialization, and CNOT gates with the control qubit of $\ket{a_{i}}$ and the target qubit of $\ket{b_{i}}$ in erasing carry.
By considering these optimizations, we reduce $n/2$ Toffoli gates and $3n/4$ CNOT gates.

The gate count and depth is shown in Table~\ref{tab:detail_gate_count_of_ModAdder}.

\begin{table*}
\caption{\label{tab:detail_gate_count_of_ModAdder} 
Gate count and depth of our proposed control modular adder.
We omit the rounds whose gate count is $O(1)$ and whose depth is $O(1)$.}
\centering
\begin{tabular}{c|c|cc|cc}
&
&\multicolumn{2}{c|}{Count} & \multicolumn{2}{c}{Depth} \\ \hline
Operation & Rounds 
& Toffoli & CNOT & Toffoli & CNOT \\\hline
& Initialization 
& $0$ & $0.5n$ & $0$ & $O(1)$ \\ 
& P 
& $n$ & $0$ & \multirow{2}{*}{$\}\log n$} & \multirow{2}{*}{$0$} \\ 
C-comparator 
& G 
& $n$ & $0$ &  &  \\ 
(twice) 
& Inverse G 
& $n$ & $0$ & \multirow{2}{*}{$\}\log n$} & \multirow{2}{*}{$0$} \\ 
& Inverse P
& $n$ & $0$ &  & \\ 
& Inverse Initialization
& $0$ & $0.5n$ & $0$ & $O(1)$ \\
\cline{2-6} 
& Total 
& $4n$ & $n$ & $2 \log n$ & $O(1)$ \\\hline
& Embedding 
& $O(1)$ & $0.75n$ & $O(1)$ & $\log n$ \\
& Initialization
& $0.75n$ & $0.75n$ & $O(1)$ & $O(1)$ \\ 
& P 
& $n$ & $0$ & \multirow{2}{*}{$\}\log n$} & \multirow{2}{*}{$0$} \\ 
& G 
& $n$ & $0$ & &  \\ 
& C 
& $n$ & $0$ & \multirow{2}{*}{$\}\log n$} & \multirow{2}{*}{$0$} \\ 
& Inverse P 
& $n$ & $0$ & & \\
& Calculating $\ket{a + b}$ 
& $0$ & $n$ & $0$ & $O(1)$ \\
CC-adder
& PE 
& $0$ & $0.75n$ & $0$ & $O(1)$ \\
& P 
& $n$ & $0$ & \multirow{2}{*}{$\}\log n$} & \multirow{2}{*}{$0$} \\ 
& Inverse C 
& $n$ & $0$ & & \\
& Inverse G 
& $n$ & $0$ & \multirow{2}{*}{$\}\log n$} & \multirow{2}{*}{$0$} \\ 
& Inverse P 
& $n$ & $0$ & & \\ 
& Inverse Initialization
& $0.75n$ & $0.75n$ & $O(1)$ & $O(1)$ \\ 
& Resetting
& $O(1)$ & $0.75n$ & $O(1)$ & $\log n$ \\
\cline{2-6} 
& Total 
& $9.5n$ & $4.75n$ & $4 \log n$ & $2 \log n$ \\\hline
Total 
& 
& $17.5n$ & $6.75n$ & $8\log n$ & $2 \log n$
\end{tabular}
\end{table*}

\subsubsection{Example of Our Control Modular Adder}

\begin{figure}
\[
\Qcircuit @C=.3em @R=1.5em {
& \mbox{\text{Init}} & & \mbox{}
& \mbox{\text{P}} & & \mbox{}
& \mbox{\text{G}} & &
& \mbox{} & & \mbox{} 
&  & \mbox{\text{IG}} & & \mbox{}
&  & \mbox{\text{IP}} & \mbox{}
& & \mbox{\text{IInit}}
&  \\
\lstick{\text{CTRL}} & \qw & \qw  & \qw 
& \qw & \qw & \qw 
& \qw & \qw & \qw
& \qw & \ctrl{21} & \qw
& \qw & \qw & \qw & \qw 
& \qw & \qw & \qw 
& \qw & \qw 
& \qw \\
\lstick{\ket{d_{0}}} & \qw & \qw & \qw 
& \qw & \qw & \qw 
& \qw & \qw & \qw
& \qw & \qw & \qw
& \qw & \qw & \qw
& \qw & \qw & \qw & \qw
& \qw & \qw 
& \qw \\
\lstick{\ket{b_{0}}} & \qw & \qw & \qw 
& \qw & \qw & \qw 
& \qw & \qw & \qw
& \qw & \qw & \qw
& \qw & \qw & \qw
& \qw & \qw & \qw & \qw
& \qw & \qw
& \qw \\
\lstick{\ket{c_{1}}} & \qw & \qw & \qw 
& \qw & \qw & \qw 
& \ctrl{2} & \qw & \qw
& \qw & \qw & \qw
&\qw & \qw & \ctrl{2}
& \qw & \qw & \qw & \qw
& \qw & \qw
& \qw \\
\lstick{\ket{d_{1}}} & \qw & \qw & \qw 
& \qw & \qw & \qw 
& \qw & \qw & \qw
& \qw & \qw & \qw
& \qw & \qw & \qw & \qw 
& \qw & \qw & \qw 
& \qw & \qw
& \qw \\
\lstick{\ket{b_{1}}} & \ctrl{1} & \targ  & \qw 
& \qw & \qw  & \qw 
& \ctrl{1} & \qw & \qw
& \qw & \qw & \qw
& \qw & \qw & \ctrl{1} & \qw 
& \qw & \qw & \qw 
& \targ & \ctrl{1}
& \qw \\
\lstick{\ket{c_{2}}} & \targ & \qw & \qw 
& \qw & \qw & \qw 
& \targ & \ctrl{3} & \qw
& \qw & \qw & \qw
& \qw & \ctrl{3} & \targ & \qw 
& \qw & \qw & \qw 
& \qw & \targ
& \qw \\
\lstick{\ket{d_{2}}} & \qw & \qw & \qw 
& \qw & \qw & \qw 
& \qw & \qw & \qw
& \qw & \qw & \qw
& \qw & \qw & \qw & \qw 
& \qw & \qw & \qw 
& \qw & \qw
& \qw \\
\lstick{\ket{b_{2}}} & \qw & \qw & \qw 
& \ctrl{1} & \qw & \qw 
& \qw & \qw & \qw
& \qw & \qw & \qw
& \qw & \qw & \qw & \qw 
& \qw & \ctrl{1} & \qw 
& \qw & \qw
& \qw \\
\lstick{\ket{p\left[2, 4\right]}} & \qw & \qw & \qw 
& \targ & \qw & \qw 
& \qw & \ctrl{4} & \qw
& \qw & \qw & \qw
& \qw & \ctrl{4} & \qw & \qw 
& \qw & \targ & \qw 
& \qw & \qw
& \qw \\
\lstick{\ket{c_{3}}} & \qw & \qw & \qw  
& \qw & \qw & \qw 
& \ctrl{2} & \qw & \qw
& \qw & \qw & \qw
& \qw & \qw & \ctrl{2} & \qw 
& \qw & \qw & \qw 
& \qw & \qw
& \qw \\
\lstick{\ket{d_{3}}} & \qw & \qw & \qw 
& \qw & \qw & \qw 
& \qw & \qw & \qw
& \qw & \qw & \qw
& \qw & \qw & \qw & \qw 
& \qw & \qw & \qw 
& \qw & \qw
& \qw \\
\lstick{\ket{b_{3}}} & \ctrl{1} & \targ & \qw  
& \ctrl{-3} & \qw & \qw 
& \ctrl{1} & \qw & \qw
& \qw & \qw & \qw
& \qw & \qw & \ctrl{1} & \qw 
& \qw & \ctrl{-3} & \qw 
& \targ & \ctrl{1}
& \qw \\
\lstick{\ket{c_{4}}} & \targ & \qw & \qw  
& \qw & \qw & \qw 
& \targ & \targ & \ctrl{3}
& \qw & \qw & \qw
& \ctrl{3} & \targ & \targ & \qw 
& \qw & \qw & \qw 
& \qw & \targ
& \qw \\
\lstick{\ket{d_{4}}} & \qw & \qw & \qw 
& \qw & \qw & \qw 
& \qw & \qw & \qw
& \qw & \qw & \qw
& \qw & \qw & \qw & \qw 
& \qw & \qw & \qw 
& \qw & \qw
& \qw \\
\lstick{\ket{b_{4}}} & \qw & \qw & \qw 
& \ctrl{1} & \qw & \qw 
& \qw & \qw & \qw
& \qw & \qw & \qw
& \qw & \qw & \qw & \qw 
& \qw & \ctrl{1} & \qw 
& \qw & \qw
& \qw \\
\lstick{\ket{p\left[4, 6\right]}} & \qw & \qw & \qw 
& \targ & \qw & \qw 
& \qw & \qw & \ctrl{4}
& \qw & \qw & \qw
& \ctrl{4} & \qw & \qw & \qw 
& \qw & \targ & \qw 
& \qw & \qw
& \qw \\
\lstick{\ket{c_{5}}} & \qw & \qw & \qw 
& \qw & \qw & \qw 
& \ctrl{2} & \qw & \qw
& \qw & \qw & \qw
& \qw & \qw & \ctrl{2} & \qw 
& \qw & \qw & \qw 
& \qw & \qw
& \qw \\
\lstick{\ket{d_{5}}} & \qw & \qw & \qw 
& \qw & \qw & \qw 
& \qw & \qw & \qw
& \qw & \qw & \qw
& \qw & \qw & \qw & \qw 
& \qw & \qw & \qw 
& \qw & \qw
& \qw \\
\lstick{\ket{b_{5}}} & \ctrl{1} & \targ & \qw 
& \ctrl{-3} & \qw & \qw 
& \ctrl{1} & \qw & \qw
& \qw & \qw & \qw
& \qw & \qw & \ctrl{1} & \qw 
& \qw & \ctrl{-3} & \qw 
& \targ & \ctrl{1}
& \qw \\
\lstick{\ket{c_{6}}} & \targ & \qw & \qw 
& \qw & \qw & \qw 
& \targ & \qw & \targ
& \qw & \ctrl{1} & \qw
& \targ & \qw & \targ & \qw 
& \qw & \qw & \qw 
& \qw & \targ
& \qw \\
\lstick{\text{COMP}} & \qw & \qw & \qw 
& \qw & \qw & \qw 
& \qw & \qw & \qw
& \qw & \targ & \qw
& \qw & \qw & \qw & \qw 
& \qw & \qw & \qw 
& \qw & \qw
& \qw \\
& & & \mbox{}
& & & \mbox{}
& & &
& \mbox{} & & \mbox{}
& & & & \mbox{} 
& & & \mbox{}
\gategroup{1}{4}{24}{4}{.1em}{--} %
\gategroup{1}{7}{24}{7}{.1em}{--} %
\gategroup{1}{11}{24}{13}{.1em}{.} %
\gategroup{1}{17}{24}{17}{.1em}{--} %
\gategroup{1}{20}{24}{20}{.1em}{--} %
}
\]
\caption{\label{fig:CFcomp} 
An example circuit of the first C-comparator for flipping the COMP qubit if $b \geq 22$.
To achieve this, We add $2^{6} - 22 = 42 = 101010_{2}$ and use the COMP qubit as the carry out of the adder.
The Init phase consists of pairs of gates, a CNOT and an $X$, on the second, fourth, and sixth groups of qubits including $\ket{d_{i}}$, $\ket{b_{i}}$, and $\ket{c_{i+1}}$  from the lowest bit.
This circuit is symmetric about the Toffoli gate surrounded by a dotted box.
Init, IP, IG, and IInit mean Initialization, Inverse P-rounds, Inverse C-rounds, Inverse G-rounds, and Inverse Initialization respectively.
}
\end{figure}
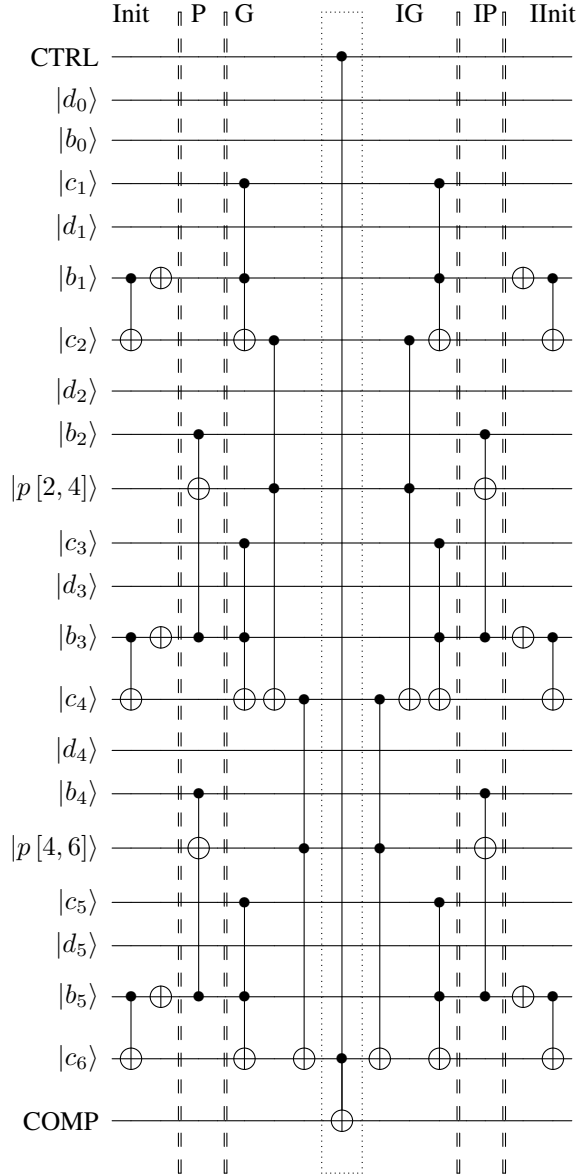

We show an example of a $6$-bit control modular adder when $N = 59$ and $a = 37$.
Circuits are given in Figures~\ref{fig:CFcomp}--\ref{fig:CLcomp}.

In these example figures registers are shown with low-order qubits at the top, in contrast to Figure~\ref{fig:Our_Madd}.
In this subsection, the register $\ket{b}$ contains a quantum value. 

The algorithm follows in this order:
\begin{enumerate}
\item Conduct a C-comparator with the control qubit \ctrlq.
Compare $\ket{b}$ and $N - a = 22$.
If $b \geq 22$, flip \compq.
This is implemented by adding $2^{6} - (N - a) = 42$ and using the carry out.
\item Conduct a CC-adder.
If both \ctrlq and \compq are $1$, subtract $N - a = 22$.
This is implemented by adding $2^{6} - (N - a) = 42$ without calculating carry $c_{6}$.
If \ctrlq is $1$ and \compq is $0$, add $a = 37$, otherwise, add no value.
\item Conduct a C-comparator with the control qubit \ctrlq.
Compare $\ket{b}$ and $a = 37$.
If $b < 37$, flip \compq.
This is implemented by calculating carry of adding $2^{6} - a = 27$.
\end{enumerate}
These steps correspond to Figure~\ref{fig:CFcomp},~\ref{fig:Padd}, and~\ref{fig:CLcomp} respectively.

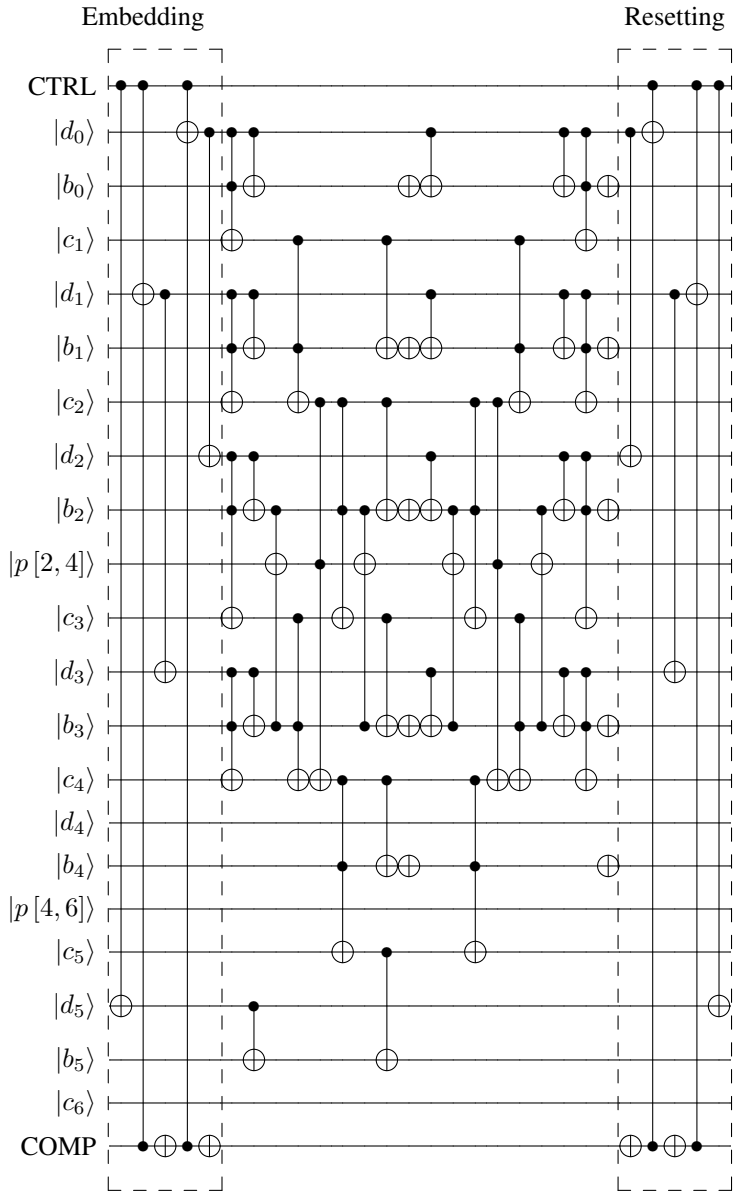
\begin{figure}
\[
\Qcircuit @C=.01em @R=1.2em {
& & & \mbox{Embedding} & & &  &
& & 
& 
& & 
& & 
& 
& 
& & 
& 
& & 
& & 
& 
& & 
& 
&
& & & \mbox{Resetting} & &
&
&  \\
\mbox{} &&& & & & & \mbox{}
& & 
& 
& & 
& & 
& 
& 
& & 
& 
& & 
& & 
& 
& & 
& 
& \mbox{} &&&&&& \mbox{}
&  \\
\lstick{\text{CTRL}} & \qw
& \ctrl{18} & \ctrl{4} & \qw & \ctrl{1} & \qw & \qw
& \qw & \qw
& \qw
& \qw & \qw
& \qw & \qw
& \qw
& \qw
& \qw & \qw
& \qw
& \qw & \qw
& \qw & \qw
& \qw
& \qw & \qw
& \qw
& \qw
& \qw & \ctrl{1} & \qw & \ctrl{4} & \ctrl{18}
& \qw
& \qw \\
\lstick{\ket{d_{0}}} & \qw
& \qw & \qw & \qw & \targ & \ctrl{6} & \qw
& \ctrl{1} & \ctrl{1} 
& \qw
& \qw & \qw
& \qw & \qw
& \qw
& \qw
& \qw & \ctrl{1}
& \qw
& \qw & \qw
& \qw & \qw
& \qw
& \ctrl{1} & \ctrl{1} 
& \qw
& \qw
& \ctrl{6} & \targ & \qw & \qw & \qw 
& \qw
& \qw \\
\lstick{\ket{b_{0}}} & \qw
& \qw & \qw & \qw & \qw & \qw & \qw
& \ctrl{1} & \targ 
& \qw
& \qw & \qw
& \qw & \qw
& \qw
& \qw
& \targ & \targ
& \qw
& \qw & \qw
& \qw & \qw
& \qw
& \targ & \ctrl{1}
& \targ 
& \qw
& \qw & \qw & \qw & \qw & \qw 
& \qw
& \qw \\
\lstick{\ket{c_{1}}} & \qw
& \qw & \qw & \qw & \qw & \qw & \qw
& \targ & \qw 
& \qw
& \ctrl{2} & \qw
& \qw & \qw
& \qw
& \ctrl{2}
& \qw & \qw
& \qw
& \qw & \qw
& \qw & \ctrl{2}
& \qw
& \qw & \targ
& \qw
& \qw
& \qw & \qw & \qw & \qw & \qw 
& \qw
& \qw \\
\lstick{\ket{d_{1}}} & \qw
& \qw & \targ & \ctrl{7} & \qw & \qw & \qw
& \ctrl{1} & \ctrl{1} 
& \qw
& \qw & \qw
& \qw & \qw
& \qw
& \qw
& \qw & \ctrl{1}
& \qw
& \qw & \qw
& \qw & \qw
& \qw
& \ctrl{1} & \ctrl{1}
& \qw
& \qw
& \qw & \qw & \ctrl{7} & \targ & \qw
& \qw
& \qw \\
\lstick{\ket{b_{1}}} & \qw
& \qw & \qw & \qw & \qw & \qw & \qw
& \ctrl{1} & \targ 
& \qw
& \ctrl{1} & \qw
& \qw & \qw
& \qw
& \targ
& \targ & \targ
& \qw
& \qw & \qw
& \qw & \ctrl{1}
& \qw
& \targ & \ctrl{1}
& \targ 
& \qw
& \qw & \qw & \qw & \qw & \qw
& \qw
& \qw  \\
\lstick{\ket{c_{2}}} & \qw
& \qw & \qw & \qw & \qw & \qw & \qw
& \targ & \qw 
& \qw
& \targ & \ctrl{3}
& \qw & \ctrl{2}
& \qw
& \ctrl{2}
& \qw & \qw
& \qw
& \ctrl{2} & \qw
& \ctrl{3} & \targ
& \qw
& \qw & \targ
& \qw
& \qw
& \qw & \qw & \qw & \qw & \qw
& \qw
& \qw \\
\lstick{\ket{d_{2}}} & \qw
& \qw & \qw & \qw & \qw & \targ & \qw
& \ctrl{1} & \ctrl{1} 
& \qw
& \qw & \qw
& \qw & \qw
& \qw
& \qw
& \qw & \ctrl{1}
& \qw
& \qw & \qw
& \qw & \qw
& \qw
& \ctrl{1} & \ctrl{1}
& \qw
& \qw
& \targ & \qw & \qw & \qw & \qw
& \qw
& \qw \\
\lstick{\ket{b_{2}}} & \qw
& \qw & \qw & \qw & \qw & \qw & \qw
& \ctrl{2} & \targ 
& \ctrl{1}
& \qw & \qw
& \qw & \ctrl{2}
& \ctrl{1}
& \targ
& \targ & \targ
& \ctrl{1}
& \ctrl{2} & \qw
& \qw & \qw
& \ctrl{1}
& \targ & \ctrl{2}
& \targ
& \qw
& \qw & \qw & \qw & \qw & \qw
& \qw
& \qw \\
\lstick{\ket{p\left[2, 4\right]}} & \qw
& \qw & \qw & \qw & \qw & \qw & \qw
& \qw & \qw
& \targ
& \qw & \ctrl{4}
& \qw & \qw
& \targ
& \qw
& \qw & \qw
& \targ
& \qw & \qw
& \ctrl{4} & \qw
& \targ
& \qw & \qw
& \qw
& \qw
& \qw & \qw & \qw & \qw & \qw
& \qw
& \qw \\
\lstick{\ket{c_{3}}} & \qw 
& \qw & \qw & \qw & \qw & \qw & \qw
& \targ & \qw 
& \qw
& \ctrl{2} & \qw
& \qw & \targ
& \qw
& \ctrl{2}
& \qw & \qw
& \qw
& \targ & \qw
& \qw & \ctrl{2}
& \qw
& \qw & \targ
& \qw
& \qw
& \qw & \qw & \qw & \qw & \qw
& \qw
& \qw \\
\lstick{\ket{d_{3}}} & \qw
& \qw & \qw & \targ & \qw & \qw & \qw
& \ctrl{1} & \ctrl{1} 
& \qw
& \qw & \qw
& \qw & \qw
& \qw
& \qw
& \qw & \ctrl{1}
& \qw
& \qw & \qw
& \qw & \qw
& \qw
& \ctrl{1} & \ctrl{1}
& \qw
& \qw
& \qw & \qw & \targ & \qw & \qw
& \qw
& \qw \\
\lstick{\ket{b_{3}}} & \qw
& \qw & \qw & \qw & \qw & \qw & \qw
& \ctrl{1} & \targ 
& \ctrl{-3}
& \ctrl{1} & \qw
& \qw & \qw
& \ctrl{-3}
& \targ
& \targ & \targ
& \ctrl{-3}
& \qw & \qw
& \qw & \ctrl{1}
& \ctrl{-3}
& \targ & \ctrl{1}
& \targ 
& \qw
& \qw & \qw & \qw & \qw & \qw
& \qw
& \qw \\
\lstick{\ket{c_{4}}} & \qw
& \qw & \qw & \qw & \qw & \qw & \qw
& \targ & \qw 
& \qw 
& \targ & \targ
& \qw & \ctrl{2}
& \qw
& \ctrl{2}
& \qw & \qw
& \qw
& \ctrl{2} & \qw
& \targ & \targ
& \qw
& \qw & \targ
& \qw
& \qw
& \qw & \qw & \qw & \qw & \qw
& \qw
& \qw \\
\lstick{\ket{d_{4}}} & \qw 
& \qw & \qw & \qw & \qw & \qw & \qw
& \qw & \qw
& \qw
& \qw & \qw
& \qw & \qw
& \qw
& \qw
& \qw & \qw
& \qw
& \qw & \qw
& \qw & \qw
& \qw
& \qw & \qw
& \qw
& \qw
& \qw & \qw & \qw & \qw & \qw
& \qw
& \qw \\
\lstick{\ket{b_{4}}} & \qw
& \qw & \qw & \qw & \qw & \qw & \qw
& \qw & \qw
& \qw
& \qw & \qw
& \qw & \ctrl{2}
& \qw
& \targ
& \targ & \qw
& \qw
& \ctrl{2} & \qw
& \qw & \qw
& \qw
& \qw & \qw
& \targ
& \qw
& \qw & \qw & \qw & \qw & \qw
& \qw
& \qw  \\
\lstick{\ket{p\left[4, 6\right]}} & \qw
& \qw & \qw & \qw & \qw & \qw & \qw
& \qw & \qw 
& \qw
& \qw & \qw
& \qw & \qw
& \qw
& \qw
& \qw & \qw
& \qw
& \qw & \qw
& \qw & \qw
& \qw
& \qw & \qw
& \qw
& \qw
& \qw & \qw & \qw & \qw & \qw
& \qw
& \qw \\
\lstick{\ket{c_{5}}} & \qw
& \qw & \qw & \qw & \qw & \qw & \qw
& \qw & \qw 
& \qw
& \qw & \qw
& \qw & \targ
& \qw
& \ctrl{2}
& \qw & \qw
& \qw
& \targ & \qw
& \qw & \qw
& \qw
& \qw & \qw
& \qw
& \qw
& \qw & \qw & \qw & \qw & \qw
& \qw
& \qw \\
\lstick{\ket{d_{5}}} & \qw
& \targ & \qw & \qw & \qw & \qw & \qw
& \qw & \ctrl{1} 
& \qw
& \qw & \qw
& \qw & \qw
& \qw
& \qw
& \qw & \qw
& \qw
& \qw & \qw
& \qw & \qw
& \qw
& \qw & \qw
& \qw
& \qw
& \qw & \qw & \qw & \qw & \targ
& \qw
& \qw \\
\lstick{\ket{b_{5}}} & \qw
& \qw & \qw & \qw & \qw & \qw & \qw
& \qw & \targ 
& \qw
& \qw & \qw
& \qw & \qw
& \qw
& \targ
& \qw & \qw
& \qw
& \qw & \qw
& \qw & \qw
& \qw
& \qw & \qw
& \qw
& \qw
& \qw & \qw & \qw & \qw & \qw
& \qw
& \qw \\
\lstick{\ket{c_{6}}} & \qw
& \qw & \qw & \qw & \qw & \qw & \qw
& \qw & \qw 
& \qw
& \qw & \qw
& \qw & \qw
& \qw
& \qw
& \qw & \qw
& \qw
& \qw & \qw
& \qw & \qw
& \qw
& \qw & \qw
& \qw
& \qw
& \qw & \qw & \qw & \qw & \qw
& \qw
& \qw \\
\lstick{\text{COMP}} & \qw
& \qw & \ctrl{-17} & \targ & \ctrl{-20} & \targ & \qw
& \qw & \qw 
& \qw
& \qw & \qw
& \qw & \qw
& \qw
& \qw
& \qw & \qw
& \qw
& \qw & \qw
& \qw & \qw
& \qw
& \qw & \qw
& \qw
& \qw 
& \targ & \ctrl{-20} & \targ & \ctrl{-17} & \qw 
& \qw
& \qw & \\
\mbox{} & & & & & & & \mbox{}
& &
& 
& &
& &
& 
& 
& & 
& 
& &
& &
&
& &
&
& \mbox{} & \mbox{} & \mbox{} & \mbox{} & \mbox{} & \mbox{} & \mbox{} & \mbox{} 
\gategroup{2}{1}{25}{8}{.1em}{--}
\gategroup{2}{29}{25}{35}{.1em}{--}
}
\]
\caption{\label{fig:Padd} 
An example of the CC-adder.
If both \ctrlq and \compq are $1$, we subtract $N - a = 22$.
This is implemented by adding $2^{6} - (N - a) = 42 = 101010_{2}$ without calculating carry $c_{6}$.
If \ctrlq is $1$ and \compq is $0$, we add $a = 37 = 100101_{2}$.
Based on these, we conduct embedding and resetting.
The remaining part is an adder, and we omit the calculation of $p[i, 6]$ and $g[i, 6]$ $(i < 6)$.
}
\end{figure}

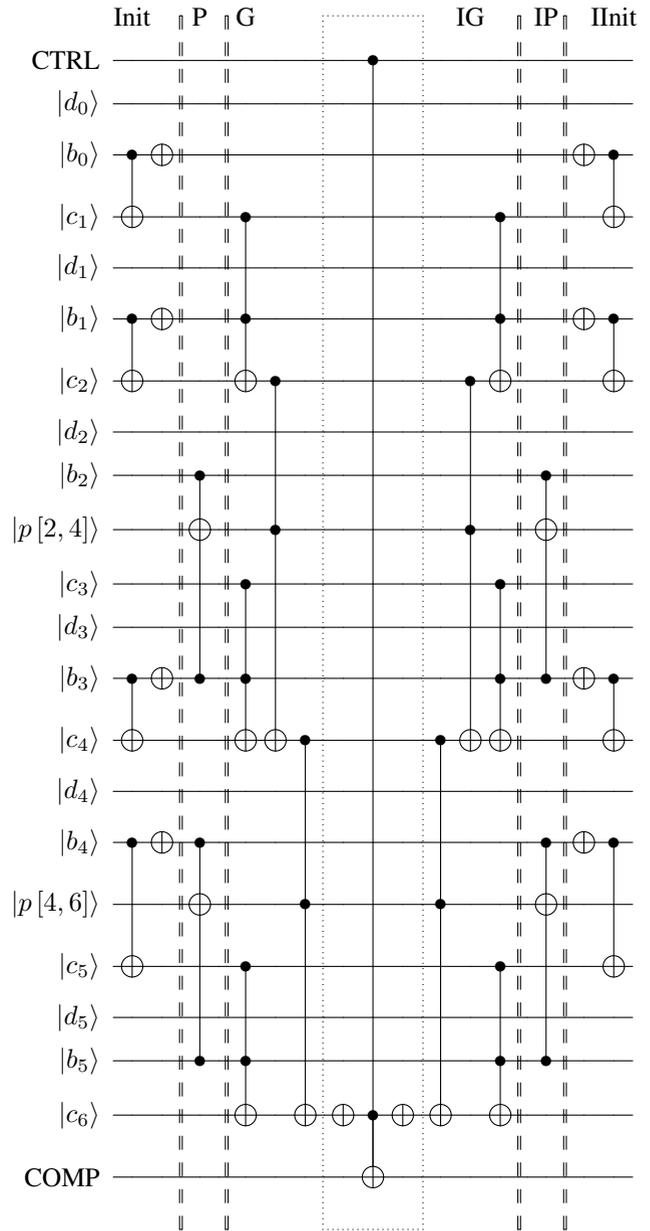
\begin{figure}
\[
\Qcircuit @C=.3em @R=1.5em {
& \mbox{\text{Init}} & & \mbox{}
& \mbox{\text{P}} & & \mbox{}
& \mbox{\text{G}} & &
& \mbox{} & & & & \mbox{} 
&  & \mbox{\text{IG}} & & \mbox{}
&  & \mbox{\text{IP}} & \mbox{}
& & \mbox{\text{IInit}}
&  \\
\lstick{\text{CTRL}} & \qw & \qw & \qw
& \qw & \qw & \qw
& \qw & \qw & \qw
& \qw &\qw & \ctrl{21} & \qw & \qw
& \qw & \qw & \qw & \qw
& \qw & \qw & \qw
& \qw & \qw 
& \qw \\
\lstick{\ket{d_{0}}} & \qw & \qw & \qw
& \qw & \qw & \qw
& \qw & \qw & \qw
& \qw & \qw & \qw & \qw & \qw
& \qw & \qw & \qw & \qw
& \qw & \qw & \qw
& \qw & \qw
& \qw \\
\lstick{\ket{b_{0}}} & \ctrl{1} & \targ & \qw
& \qw & \qw & \qw
& \qw & \qw & \qw
& \qw & \qw & \qw & \qw & \qw
& \qw & \qw & \qw & \qw
& \qw & \qw & \qw
& \targ & \ctrl{1}
& \qw \\
\lstick{\ket{c_{1}}} & \targ & \qw & \qw 
& \qw & \qw & \qw
& \ctrl{2} & \qw & \qw
& \qw & \qw & \qw & \qw & \qw
&\qw & \qw & \ctrl{2} & \qw
& \qw & \qw & \qw
& \qw & \targ
& \qw \\
\lstick{\ket{d_{1}}} & \qw & \qw & \qw
& \qw & \qw & \qw
& \qw & \qw & \qw
& \qw & \qw & \qw & \qw & \qw
& \qw & \qw & \qw & \qw
& \qw & \qw & \qw
& \qw & \qw
& \qw \\
\lstick{\ket{b_{1}}} & \ctrl{1} & \targ & \qw
& \qw & \qw & \qw
& \ctrl{1} & \qw & \qw
& \qw & \qw & \qw & \qw & \qw
& \qw & \qw & \ctrl{1} & \qw
& \qw & \qw & \qw
& \targ & \ctrl{1}
& \qw \\
\lstick{\ket{c_{2}}} & \targ & \qw  & \qw
& \qw & \qw & \qw
& \targ & \ctrl{3} & \qw
& \qw & \qw & \qw & \qw & \qw
& \qw & \ctrl{3} & \targ & \qw
& \qw & \qw & \qw
& \qw & \targ
& \qw \\
\lstick{\ket{d_{2}}} & \qw & \qw & \qw
& \qw & \qw & \qw
& \qw & \qw & \qw
& \qw & \qw & \qw & \qw & \qw
& \qw & \qw & \qw & \qw
& \qw & \qw & \qw
& \qw & \qw
& \qw \\
\lstick{\ket{b_{2}}} & \qw & \qw & \qw
& \ctrl{1} & \qw & \qw
& \qw & \qw & \qw
& \qw & \qw & \qw & \qw & \qw
& \qw & \qw & \qw & \qw
& \qw & \ctrl{1} & \qw
& \qw & \qw
& \qw \\
\lstick{\ket{p\left[2, 4\right]}} & \qw & \qw & \qw
& \targ & \qw & \qw
& \qw & \ctrl{4} & \qw
& \qw & \qw & \qw & \qw & \qw
& \qw & \ctrl{4} & \qw & \qw
& \qw & \targ & \qw
& \qw & \qw
& \qw \\
\lstick{\ket{c_{3}}} & \qw & \qw & \qw
& \qw & \qw & \qw
& \ctrl{2} & \qw & \qw
& \qw & \qw & \qw & \qw & \qw
& \qw & \qw & \ctrl{2} & \qw
& \qw & \qw & \qw
& \qw & \qw
& \qw \\
\lstick{\ket{d_{3}}} & \qw & \qw & \qw
& \qw & \qw & \qw
& \qw & \qw & \qw
& \qw & \qw & \qw & \qw & \qw
& \qw & \qw & \qw & \qw
& \qw & \qw & \qw
& \qw & \qw
& \qw \\
\lstick{\ket{b_{3}}} & \ctrl{1} & \targ & \qw 
& \ctrl{-3} & \qw & \qw
& \ctrl{1} & \qw & \qw
& \qw & \qw & \qw & \qw & \qw
& \qw & \qw & \ctrl{1} & \qw
& \qw & \ctrl{-3} & \qw
& \targ & \ctrl{1}
& \qw \\
\lstick{\ket{c_{4}}} & \targ & \qw & \qw
& \qw & \qw & \qw
& \targ & \targ & \ctrl{3}
& \qw & \qw & \qw & \qw & \qw
& \ctrl{3} & \targ & \targ & \qw
& \qw & \qw & \qw
& \qw & \targ
& \qw \\
\lstick{\ket{d_{4}}} & \qw & \qw & \qw
& \qw & \qw & \qw
& \qw & \qw & \qw
& \qw & \qw & \qw & \qw & \qw
& \qw & \qw & \qw & \qw
& \qw & \qw & \qw
& \qw & \qw
& \qw \\
\lstick{\ket{b_{4}}} & \ctrl{2} & \targ & \qw
& \ctrl{1} & \qw & \qw
& \qw & \qw & \qw
& \qw & \qw & \qw & \qw & \qw
& \qw & \qw & \qw & \qw
& \qw & \ctrl{1} & \qw
& \targ & \ctrl{2}
& \qw \\
\lstick{\ket{p\left[4, 6\right]}} & \qw & \qw & \qw
& \targ & \qw & \qw
& \qw & \qw & \ctrl{4}
& \qw & \qw & \qw & \qw & \qw
& \ctrl{4} & \qw & \qw & \qw
& \qw & \targ & \qw
& \qw & \qw
& \qw \\
\lstick{\ket{c_{5}}} & \targ & \qw & \qw 
& \qw & \qw & \qw
& \ctrl{2} & \qw & \qw
& \qw & \qw & \qw & \qw & \qw
& \qw & \qw & \ctrl{2} & \qw
& \qw & \qw & \qw
& \qw & \targ
& \qw \\
\lstick{\ket{d_{5}}} & \qw & \qw & \qw
& \qw & \qw & \qw
& \qw & \qw & \qw
& \qw & \qw & \qw & \qw & \qw
& \qw & \qw & \qw & \qw
& \qw & \qw & \qw
& \qw & \qw
& \qw \\
\lstick{\ket{b_{5}}} & \qw & \qw & \qw 
& \ctrl{-3} & \qw & \qw
& \ctrl{1} & \qw & \qw
& \qw & \qw & \qw & \qw & \qw
& \qw & \qw & \ctrl{1} & \qw
& \qw & \ctrl{-3} & \qw
& \qw & \qw
& \qw \\
\lstick{\ket{c_{6}}} & \qw& \qw & \qw
& \qw & \qw & \qw
& \targ & \qw & \targ
& \qw & \targ & \ctrl{1} & \targ & \qw
& \targ & \qw & \targ & \qw
& \qw & \qw & \qw
& \qw & \qw
& \qw \\
\lstick{\text{COMP}} & \qw & \qw & \qw 
& \qw & \qw & \qw
& \qw & \qw & \qw
& \qw & \qw & \targ & \qw & \qw
& \qw & \qw & \qw & \qw
& \qw & \qw & \qw
& \qw & \qw
& \qw \\
& & & \mbox{}
& & & \mbox{}
& & &
&\mbox{} & & & & \mbox{}
& & & & \mbox{}
& & & \mbox{}
\gategroup{1}{4}{24}{4}{.1em}{--}
\gategroup{1}{7}{24}{7}{.1em}{--}
\gategroup{1}{11}{24}{15}{.1em}{.}
\gategroup{1}{19}{24}{19}{.1em}{--}
\gategroup{1}{22}{24}{22}{.1em}{--}
}
\]
\caption{\label{fig:CLcomp} 
An example of the last C-comparator.
We flip the COMP qubit if $b < 37$.
This is achieved by adding $2^{6} - 37 = 27 = 011011_{2}$ and using the carry out.
First, we apply pairs of gates, a CNOT and an $X$ gate, on the first, second, fourth, and fifth groups of qubits.
In contrast  to Figure~\ref{fig:CFcomp}, we apply $X$ gates before and after the center Toffoli gate.
This circuit is symmetric about the Toffoli gate surrounded by a dotted box.
Init, IP, IG, and IInit means Initialization, Inverse P-rounds, Inverse C-rounds, Inverse G-rounds, and Inverse Initialization respectively.
}
\end{figure}

\subsection{Detailed Calculation of $T$-depth}

In this section, we analyze the $T$-depth of our $T$-optimal control modular adder.
We assume that we run \grt with the same timing, and each \grt has $T$-depth $2$ from Figure~\ref{fig:toffoli_Gid}.
We focus on the parts that can be run concurrently.
Except for Initialization, we run
\begin{itemize}
\item P-rounds and G-rounds simultaneously,
\item C-rounds and inverse P-rounds simultaneously,
\item P-rounds and inverse C-rounds simultaneously, and
\item inverse G-rounds and inverse P-rounds simultaneously.
\end{itemize}
In the first and third steps, we run many $T$~gates simultaneously at the start and fewer $T$~gates as the calculation progresses.
In the second and fourth steps, we run only a few $T$~gates simultaneously initially and more as the calculation progresses.
Thus, there is a difference in the number of $T$~gates we can run simultaneously.

As noted in Section IV. A., we define $n_{T}$ as the upper-bound of the number of $T$~gates running simultaneously, and we calculate $T$-depth based on $n_{T}$ as in Figure~\ref{fig:KQT}.
In each round, there are parts where we can run more than $n_{T}$~$T$ gates.
However, by setting $n_{T}$, we run these $T$ gates separately.
Compared to this, in the parts having less than $n_{T}$~$T$~gates, we can run these $T$ gates simultaneously.

First, we consider the parts having fewer than $n_{T}$~$T$~gates, which happens when we run P-rounds and G-rounds simultaneously,
C-rounds and inverse P-rounds simultaneously,
P-rounds and inverse C-rounds simultaneously, and
inverse G-rounds and inverse P-rounds simultaneously.
In these rounds, if we have no restriction on running $T$ gates, patterns are given as follows:
\begin{itemize}
\item In the first and the third cases, the number of $T$ gates we can run simultaneously decreases by one half as the calculation progresses.
Thus, in the latter part of the calculation, we run fewer than $n_{T}$~$T$ gates simultaneously.
This part has $T$-depth $2\log n_{T}$ and $n_{T}$~$T$~gates in total.
\item In the second and the fourth cases, the number of $T$ gates we can run simultaneously doubles as the calculation progresses.
Thus, in the former part of the calculation, we run less than $n_{T}$ $T$ gates simultaneously.
This part has $T$-depth $2\log n_{T}$ and $n_{T}$~$T$~gates in total.
\end{itemize}
We have $6$ parts each with a small number of $T$ gates, as follows:
\begin{itemize}
\item P-rounds and G-rounds in the first C-comparator,
\item P-rounds and G-rounds in the CC-adder,
\item C-rounds and inverse P-rounds in the CC-adder,
\item P-rounds and inverse C-rounds in the CC-adder,
\item inverse G-rounds and inverse P-rounds in the CC-adder, and
\item P-rounds and G-rounds in the final C-comparator
\end{itemize}
Thus, we consume $12 \log n_{T}$ $T$-depth and $6n_{T}$~$T$~gates in these.

Next, we consider the remaining parts.
In these parts, we run $T$~gates $n_{T}$ each.
The number of total $T$~gates is $43n$ from Table~\ref{tab:FTQC_gate_count}, and we run
$43n - 6n_{T}$~$T$ gates.
Thus, $T$-depth of this part is given by
\begin{align}
\dfrac{2\left(43n - 6n_{T}\right)}{n_{T}} = \dfrac{86n}{n_{T}} - 12.
\end{align}

In conclusion, $T$-depth is given by
\begin{align}
\dfrac{86n}{n_{T}} + 12 \log n_{T} - 12.
\end{align}

\subsection{Detailed Gate Count on FTQ or NISQ}

In this section, we detail the $T$~gate count on FTQ or NISQ.
The FTQ count is shown in Table~\ref{tab:detail_FTQC_gate_count}.
The detailed CNOT gate count on NISQ is shown in Table~\ref{tab:detail_NISQ_gate_count}.
The detailed CNOT depth count and $\text{KQ}_{\text{CX}}$ on NISQ are shown in Table~\ref{tab:detail_NISQ_depth_count}.

\begin{table*}
\caption{\label{tab:detail_FTQC_gate_count} 
The breakdown of Toffoli count and $T$-count of our control modular adder.
Tof means the number of Toffoli gates in each round.
Gate means the type of using relative-phase Toffoli gates in each round.
Cost means the number of $T$ gates in each relative-phase Toffoli gate.
Count means $T$-count in each round.
We omit the rounds whose $T$-count is $O(1)$.
Inv means Inverse, C-comp means a C-comparator, CC-add means a CC-adder, and Init means Initialization.
}
\centering
\begin{tabular}{c|c|c|ccc|ccc|ccc|ccc}
& &
& \multicolumn{12}{c}{Toffoli Decomposition} \\\hline
& & 
& & &
& \multicolumn{3}{c|}{Thapliyal et al.~\cite{TMK2020}} 
& \multicolumn{3}{c|}{Thapliyal et al.~\cite{TMK2020}} 
& & & \\
& & 
& \multicolumn{3}{c|}{Draper et al.~\cite{DKRS2006}} 
& \multicolumn{3}{c|}{(qubit-optimize)} 
& \multicolumn{3}{c|}{($T$-optimize)} 
& \multicolumn{3}{c}{\textbf{Ours}}  \\\hline
Operation & Rounds & Tof
& gate & cost & count
& gate & cost & count
& gate & cost & count
& gate & cost & count \\\hline
& P & $n$
& \st & $7$ & $7n$
& \grt & $4$ & $4n$
& \grt & $4$ & $4n$
& \grt & $4$ & $4n$ \\ 
C-comp & G & $n$
& \st & $7$ & $7n$
& \st & $7$ & $7n$
& \pgrt & $4$ & $4n$
& \grt & $4$ & $4n$ \\ 
(twice) & InvG & $n$
& \st & $7$ & $7n$
& \st & $7$ & $7n$
& \pgrt & $4$ & $4n$
& \igrt & $0$ & $0$ \\
& InvP & $n$
& \st & $7$ & $7n$
& \igrt & $0$ & $0$
& \igrt & $0$ & $0$
& \igrt & $0$ & $0$ \\ 
\cline{2-15} & Total & $4n$
& -- & -- & $28n$
& -- & -- & $18n$
& -- & -- & $12n$
& -- & -- & $8n$ \\\hline
& Init & $0.75n$ 
& \st & $7$ & $5.25n$
& \grt & $4$ & $3n$
& \grt & $4$ & $3n$ 
& \grt & $4$ & $3n$ \\ 
& P & $n$ 
& \st & $7$ & $7n$
& \grt & $4$ & $4n$
& \grt & $4$ & $4n$ 
& \grt & $4$ & $4n$ \\ 
& G & $n$ 
& \st & $7$ & $7n$
& \st & $7$ & $7n$
& \pgrt & $4$ & $4n$ 
& \pgrt & $4$ & $4n$ \\ 
& C & $n$ 
& \st & $7$ & $7n$
& \st & $7$ & $7n$
& \pgrt & $4$ & $4n$ 
& \pgrt & $4$ & $4n$ \\ 
& InvP & $n$ 
& \st & $7$ & $7n$
& \igrt & $0$ & $0$
& \igrt & $0$ & $0$ 
& \igrt & $0$ & $0$ \\ 
CC-add & P & $n$ 
& \st & $7$ & $7n$
& \grt & $4$ & $4n$
& \grt & $4$ & $4n$ 
& \grt & $4$ & $4n$ \\ 
& InvC & $n$ 
& \st & $7$ & $7n$
& \st & $7$ & $7n$
& \pgrt & $4$ & $4n$ 
& \pgrt & $4$ & $4n$ \\ 
& InvG & $n$ 
& \st & $7$ & $7n$
& \st & $7$ & $7n$
& \pgrt & $4$ & $4n$ 
& \pgrt & $4$ & $4n$ \\ 
& InvP & $n$ 
& \st & $7$ & $7n$
& \igrt & $0$ & $0$
& \igrt & $0$ & $0$ 
& \igrt & $0$ & $0$ \\ 
& InvInit & $0.75n$ 
& \st & $7$ & $5.25n$
& \igrt & $0$ & $0$
& \igrt & $0$ & $0$ 
& \igrt & $0$ & $0$ \\ 
\cline{2-15} & Total & $9.5n$
& -- & -- & $66.5n$
& -- & -- & $39n$
& -- & -- & $27n$
& -- & -- & $27n$ \\\hline
Total & & $17.5n$  
& -- & -- & $122.5n$
& -- & -- & $75n$
& -- & -- & $51n$
& -- & -- & $43n$ 
\end{tabular}
\end{table*}

\begin{table*}
\caption{\label{tab:detail_NISQ_gate_count}
The breakdown of Toffoli count and CNOT count of our control modular adder.
Gate means the type of using relative-phase Toffoli gates in each round.
Cost means the number of CNOT gates in each relative-phase Toffoli gate.
Count means CNOT count in each round.
We do not show the rounds whose CNOT count is $O(1)$.
Inv means Inverse, C-comp means a C-comparator, CC-add means a CC-adder, Init means Initialization, Embed means Embedding, Calc means Calculating of $\ket{a+b}$, and Reset means Resetting.
}
\centering
\begin{tabular}{c|c|cc|ccc|ccc|ccc|ccc}
& & &
& \multicolumn{12}{c}{Toffoli Decomposition} \\\hline
& & &
& & & 
& \multicolumn{3}{c|}{Thapliyal et al.~\cite{TMK2020}} 
& \multicolumn{3}{c|}{Thapliyal et al.~\cite{TMK2020}} 
& & & \\
& & &
& \multicolumn{3}{c|}{Draper et al.~\cite{DKRS2006}} 
& \multicolumn{3}{c|}{(qubit-optimize)} 
& \multicolumn{3}{c|}{($T$-optimize)} 
& \multicolumn{3}{c}{\textbf{Ours}}  \\\hline
Operation & Rounds & &
& gate & cost & count
& gate & cost & count
& gate & cost & count
& gate & cost & count \\\hline
& Init & CNOT & $0.5n$
& -- & -- & $0.5n$
& -- & -- & $0.5n$
& -- & -- & $0.5n$
& -- & -- & $0.5n$ \\ 
& P & Toffoli & $n$
& \st & $6$ & $6n$
& \grt & $6$ & $6n$
& \grt & $6$ & $6n$
& \tcx & $3$ & $3n$ \\
& G & Toffoli & $n$
& \st & $6$ & $6n$
& \st & $6$ & $6n$
& \pgrt & $8$ & $8n$
& \tcx & $3$ & $3n$ \\
C-Comp & InvG & Toffoli & $n$
& \st & $6$ & $6n$
& \st & $6$ & $6n$
& \pgrt & $8$ & $8n$
& \itcx & $3$ & $3n$ \\
(twice) & InvP & Toffoli & $n$
& \st & $6$ & $6n$
& \igrt & $1$ & $n$
& \igrt & $1$ & $n$
& \itcx & $3$ & $3n$ \\
& InvInit & CNOT & $0.5n$
& -- & -- & $0.5n$
& -- & -- & $0.5n$
& -- & -- & $0.5n$
& -- & -- & $0.5n$ \\
\cline{2-16} 
& Total & &
& -- & -- & $25n$
& -- & -- & $20n$
& -- & -- & $24n$
& -- & -- & $13n$ \\\hline
& Embed & CNOT & $0.75n$
& -- & -- & $0.75n$
& -- & -- & $0.75n$
& -- & -- & $0.75n$
& -- & -- & $0.75n$ \\
& \multirow{2}{*}{Init} & Toffoli & $0.75n$
& \st & $6$ & $4.5n$
& \grt & $6$ & $4.5n$
& \grt & $6$ & $4.5n$
& \fcx & $4$ & $3n$ \\ 
& & CNOT & $0.75n$
& -- & -- & $0.75n$
& -- & -- & $0.75n$
& -- & -- & $0.75n$
& -- & -- & $0.75n$ \\
& P & Toffoli & $n$
& \st & $6$ & $6n$
& \grt & $6$ & $6n$
& \grt & $6$ & $6n$
& \tcx & $3$ & $3n$ \\ 
& G & Toffoli & $n$
& \st & $6$ & $6n$
& \st & $6$ & $6n$
& \pgrt & $8$ & $8n$
& \fcx & $4$ & $4n$ \\ 
& C & Toffoli & $n$
& \st & $6$ & $6n$
& \st & $6$ & $6n$
& \pgrt & $8$ & $8n$
& \fcx & $4$ & $4n$ \\ 
& InvP & Toffoli & $n$
& \st & $6$ & $6n$
& \igrt & $1$ & $n$
& \igrt & $1$ & $n$
& \itcx & $3$ & $3n$ \\ 
& Calc & CNOT & $n$
& -- & -- & $n$
& -- & -- & $n$
& -- & -- & $n$
& -- & -- & $n$ \\
CC-add & PE & CNOT & $0.75n$
& -- & -- & $0.75n$
& -- & -- & $0.75n$
& -- & -- & $0.75n$
& -- & -- & $0.75n$ \\
& P & Toffoli & $n$
& \st & $6$ & $6n$
& \grt & $6$ & $6n$
& \grt & $6$ & $6n$
& \tcx & $3$ & $3n$ \\ 
& InvC & Toffoli & $n$
& \st & $6$ & $6n$
& \st & $6$ & $6n$
& \pgrt & $8$ & $8n$
& \ifcx & $4$ & $4n$ \\ 
& InvG & Toffoli & $n$
& \st & $6$ & $6n$
& \st & $6$ & $6n$
& \pgrt & $8$ & $8n$
& \ifcx & $4$ & $4n$ \\ 
& InvP & Toffoli & $n$
& \st & $6$ & $6n$
& \igrt & $1$ & $n$
& \igrt & $1$ & $n$
& \itcx & $3$ & $3n$ \\ 
& \multirow{2}{*}{InvInit} & Toffoli & $0.75n$
& \st & $6$ & $4.5n$
& \igrt & $1$ & $0.75n$
& \igrt & $1$ & $0.75n$
& \fcx & $4$ & $3n$ \\ 
& & CNOT & $0.75n$
& -- & -- & $0.75n$
& -- & -- & $0.75n$
& -- & -- & $0.75n$
& -- & -- & $0.75n$ \\
& Reset & CNOT & $0.75n$
& -- & -- & $0.75n$
& -- & -- & $0.75n$
& -- & -- & $0.75n$
& -- & -- & $0.75n$ \\
\cline{2-16} 
& Total & &
& -- & -- & $61.75n$
& -- & -- & $48n$
& -- & -- & $56n$
& -- & -- & $38.75n$ \\\hline
Total & & &  
& -- & -- & $111.75n$
& -- & -- & $88n$
& -- & -- & $104n$
& -- & -- & $64.75n$ 
\end{tabular}
\end{table*}

\begin{table*}
\caption{\label{tab:detail_NISQ_depth_count}
The breakdown of Toffoli count and CNOT-depth of our control modular adder.
Gate means the type of using relative-phase Toffoli gates in each round.
Cost means the number of CNOT gates in each relative-phase Toffoli gate.
Depth means CNOT-depth in each round.
We omit the rounds whose CNOT-depth is $O(1)$.
Inv means Inverse, C-comp means a C-comparator, CC-add means a CC-adder, Init means Initialization, Embed means Embedding, and Reset means Resetting.
}
\centering
\scalebox{0.88}{
\begin{tabular}{c|c|cc|ccc|ccc|ccc|ccc}
& & &
& \multicolumn{12}{c}{Toffoli Decomposition} \\\hline
& & &
& & & 
& \multicolumn{3}{c|}{Thapliyal et al.~\cite{TMK2020}} 
& \multicolumn{3}{c|}{Thapliyal et al.~\cite{TMK2020}} 
& & & \\
& & &
& \multicolumn{3}{c|}{Draper et al.~\cite{DKRS2006}} 
& \multicolumn{3}{c|}{(qubit-optimize)} 
& \multicolumn{3}{c|}{($T$-optimize)} 
& \multicolumn{3}{c}{\textbf{Ours}}  \\\hline
Operation & Rounds & &
& gate & cost & depth
& gate & cost & depth
& gate & cost & depth
& gate & cost & depth \\\hline
& P & Toffoli & \multirow{2}{*}{$\}\log n$}
& \st & $6$ & \multirow{2}{*}{$\}6\log n$}
& \grt & $6$ & \multirow{2}{*}{$\}6\log n$}
& \grt & $6$ & \multirow{2}{*}{$\}8\log n$}
& \tcx & $3$ & \multirow{2}{*}{$\}3\log n$} \\
& G & Toffoli &
& \st & $6$ & 
& \st & $6$ & 
& \pgrt & $8$ & 
& \tcx & $3$ & \\
C-Comp & InvG & Toffoli & \multirow{2}{*}{$\}\log n$}
& \st & $6$ & \multirow{2}{*}{$\}6\log n$}
& \st & $6$ & \multirow{2}{*}{$\}6\log n$}
& \pgrt & $8$ & \multirow{2}{*}{$\}8\log n$}
& \itcx & $3$ & \multirow{2}{*}{$\}3\log n$} \\
(twice) & InvP & Toffoli &
& \st & $6$ & 
& \igrt & $1$ & 
& \igrt & $1$ & 
& \itcx & $3$ & \\
\cline{2-16} 
& Total & &
& -- & -- & $12\log n$
& -- & -- & $12\log n$
& -- & -- & $16\log n$
& -- & -- & $6\log n$ \\\hline
& Embed & CNOT & $\log n$
& -- & -- & $\log n$
& -- & -- & $\log n$
& -- & -- & $\log n$
& -- & -- & $\log n$ \\
& P & Toffoli & \multirow{2}{*}{$\}\log n$}
& \st & $6$ & \multirow{2}{*}{$\}6\log n$}
& \grt & $6$ & \multirow{2}{*}{$\}6\log n$}
& \grt & $6$ & \multirow{2}{*}{$\}8\log n$}
& \tcx & $3$ & \multirow{2}{*}{$\}4\log n$} \\ 
& G & Toffoli &
& \st & $6$ & 
& \st & $6$ & 
& \pgrt & $8$ & 
& \fcx & $4$ &  \\ 
& C & Toffoli & \multirow{2}{*}{$\}\log n$}
& \st & $6$ & \multirow{2}{*}{$\}6\log n$}
& \st & $6$ & \multirow{2}{*}{$\}6\log n$}
& \pgrt & $8$ & \multirow{2}{*}{$\}8\log n$}
& \fcx & $4$ & \multirow{2}{*}{$\}4\log n$} \\ 
& InvP & Toffoli & 
& \st & $6$ & 
& \igrt & $1$ &
& \igrt & $1$ & 
& \itcx & $3$ &  \\ 
CC-add & P & Toffoli & \multirow{2}{*}{$\}\log n$}
& \st & $6$ & \multirow{2}{*}{$\}6\log n$}
& \grt & $6$ & \multirow{2}{*}{$\}6\log n$}
& \grt & $6$ & \multirow{2}{*}{$\}8\log n$}
& \tcx & $3$ & \multirow{2}{*}{$\}4\log n$} \\ 
& InvC & Toffoli & 
& \st & $6$ & 
& \st & $6$ & 
& \pgrt & $8$ & 
& \ifcx & $4$ & \\ 
& InvG & Toffoli & \multirow{2}{*}{$\}\log n$}
& \st & $6$ & \multirow{2}{*}{$\}6\log n$}
& \st & $6$ & \multirow{2}{*}{$\}6\log n$}
& \pgrt & $8$ & \multirow{2}{*}{$\}8\log n$}
& \ifcx & $4$ & \multirow{2}{*}{$\}4\log n$} \\ 
& InvP & Toffoli & 
& \st & $6$ & 
& \igrt & $1$ & 
& \igrt & $1$ & 
& \itcx & $3$ &  \\ 
& Reset & CNOT & $\log n$
& -- & -- & $\log n$
& -- & -- & $\log n$
& -- & -- & $\log n$
& -- & -- & $\log n$ \\
\cline{2-16} 
& Total & &
& -- & -- & $26\log n$
& -- & -- & $26\log n$
& -- & -- & $34\log n$
& -- & -- & $18\log n$ \\\hline
Total & & &  
& -- & -- & $50\log n$
& -- & -- & $50\log n$
& -- & -- & $66\log n$
& -- & -- & $30\log n$ \\
\#qubits & & &  
& -- & -- & $4n$
& -- & -- & $4n$
& -- & -- & $4.5n$
& -- & -- & $4n$ \\
$\text{KQ}_{\text{CX}}$ & & &
& -- & -- & $200n\log n$
& -- & -- & $200n\log n$
& -- & -- & $297n\log n$
& -- & -- & $120n\log n$
\end{tabular}
}
\end{table*}

\subsection{A Distillation Circuit for a $T$ gate}

A distillation circuit for a $T$ gate is given as Figure~\ref{fig:Agate}.

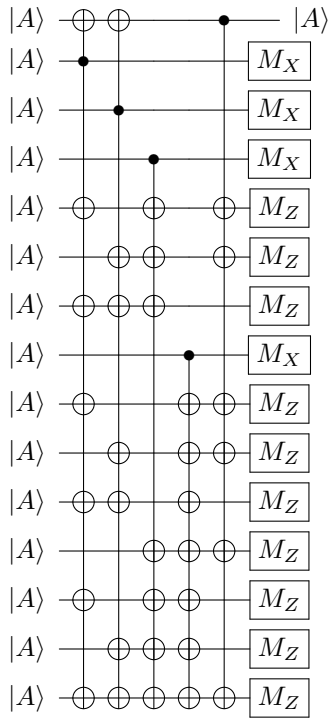
\begin{figure}
\[
\Qcircuit @C=.5em @R=.4em {
  \lstick{\ket{A}} & \targ & \targ & \qw & \qw & \ctrl{14} & \rstick{\ket{A}} \qw \\
  \lstick{\ket{A}} & \ctrl{-1} \qwx[13] & \qw & \qw & \qw & \qw & \gate{M_{X}} \\
  \lstick{\ket{A}} & \qw & \ctrl{-2} \qwx[12] & \qw & \qw & \qw & \gate{M_{X}} \\
  \lstick{\ket{A}} & \qw & \qw & \ctrl{11} & \qw & \qw & \gate{M_{X}} \\
  \lstick{\ket{A}} & \targ & \qw & \targ & \qw & \targ & \gate{M_{Z}} \\
  \lstick{\ket{A}} & \qw &\targ & \targ & \qw & \targ & \gate{M_{Z}} \\
  \lstick{\ket{A}} & \targ &\targ & \targ & \qw & \qw & \gate{M_{Z}} \\
  \lstick{\ket{A}} & \qw & \qw & \qw & \ctrl{7} & \qw & \gate{M_{X}} \\
  \lstick{\ket{A}} & \targ & \qw & \qw & \targ & \targ & \gate{M_{Z}} \\
  \lstick{\ket{A}} & \qw & \targ & \qw & \targ & \targ & \gate{M_{Z}} \\
  \lstick{\ket{A}} & \targ & \targ & \qw & \targ & \qw & \gate{M_{Z}} \\
  \lstick{\ket{A}} & \qw & \qw & \targ & \targ & \targ & \gate{M_{Z}} \\
  \lstick{\ket{A}} & \targ & \qw & \targ & \targ & \qw & \gate{M_{Z}} \\
  \lstick{\ket{A}} & \qw & \targ & \targ & \targ & \qw & \gate{M_{Z}} \\
  \lstick{\ket{A}} & \targ & \targ & \targ & \targ & \targ & \gate{M_{Z}}
}
\]
\caption{\label{fig:Agate} 
A distillation circuit of $\ket{A}$~\cite{FSG2009}. 
By this distillation circuit, we reduce the error rate of $\ket{A}$ from $p$ to $35p^{3}$.}
\end{figure}

\section*{Acknowledgment}
The first author is supported by a JSPS Fellowship for Young Scientists.
This work was supported by JSPS KAKENHI Grant Numbers JP20J11754, JP21H03440, MEXT Quantum Leap Flagship Program Grant Number JPMXS0118067285, and JST CREST Grant Number JPMJCR14D6, Japan.
This paper is written based on Chapter~6 of the dissertation by Oonishi~\cite{Oon2020}.

\bibliographystyle{IEEEtran}
\bibliography{apssamp}%

\begin{IEEEbiography}
[{\includegraphics[width=1in,height=1.25in,clip,keepaspectratio]{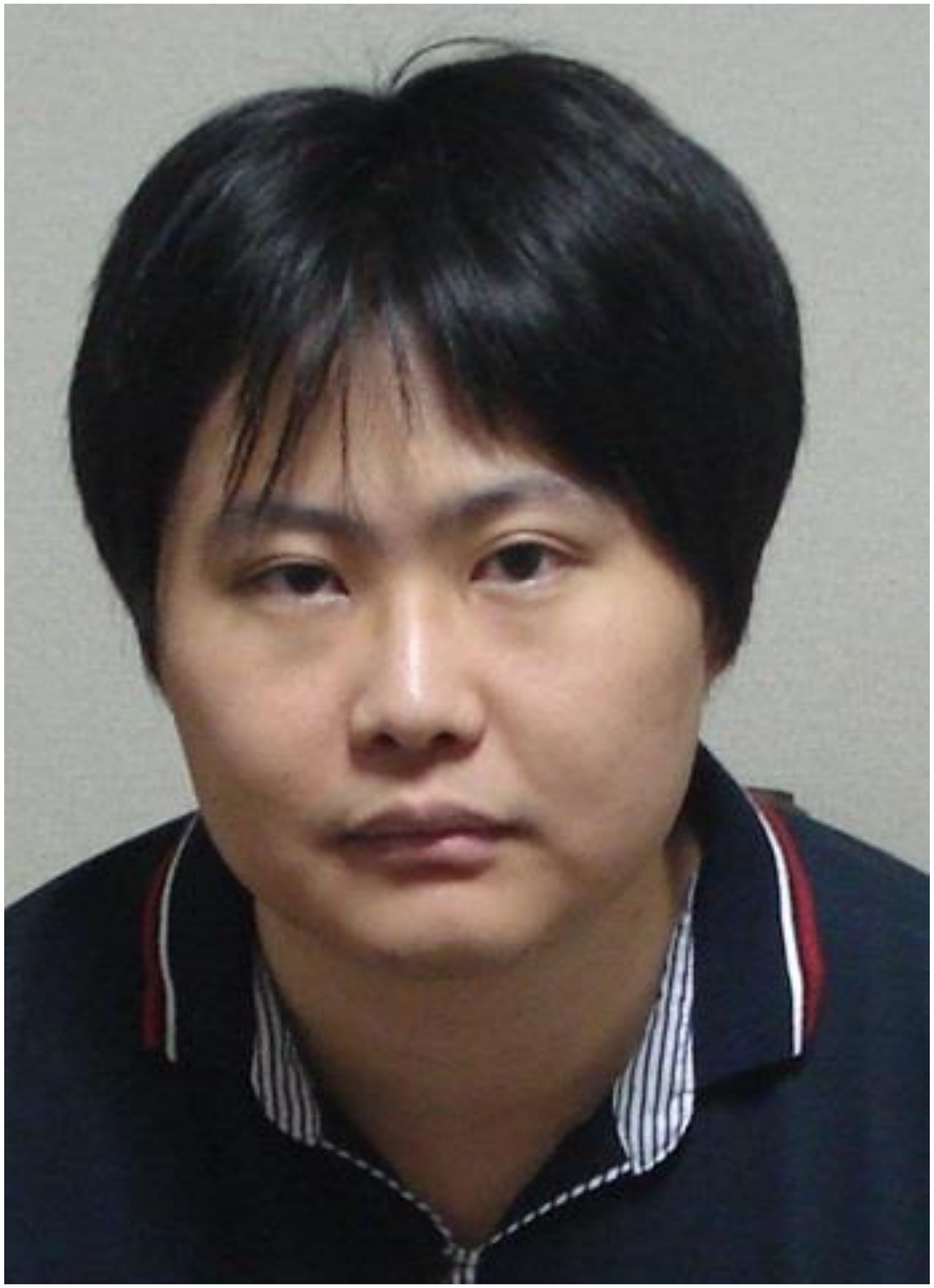}}]
{Kento~Oonishi} received his B.S. in engineering, M.S. in information science and technology, and Ph.D. in mathematical informatics from the University of Tokyo in 2016, 2018, and 2021.
His major is security evaluation of public-key cryptography against side-channel and quantum attacks. During his Ph.D., he studied quantum computation at Keio University. He is currently a researcher at Mitsubishi Electric. His research interest includes cryptography, quantum computation, and artificial intelligence.
\end{IEEEbiography}

\begin{IEEEbiography}
[{\includegraphics[width=1in,height=1.25in,clip,keepaspectratio]{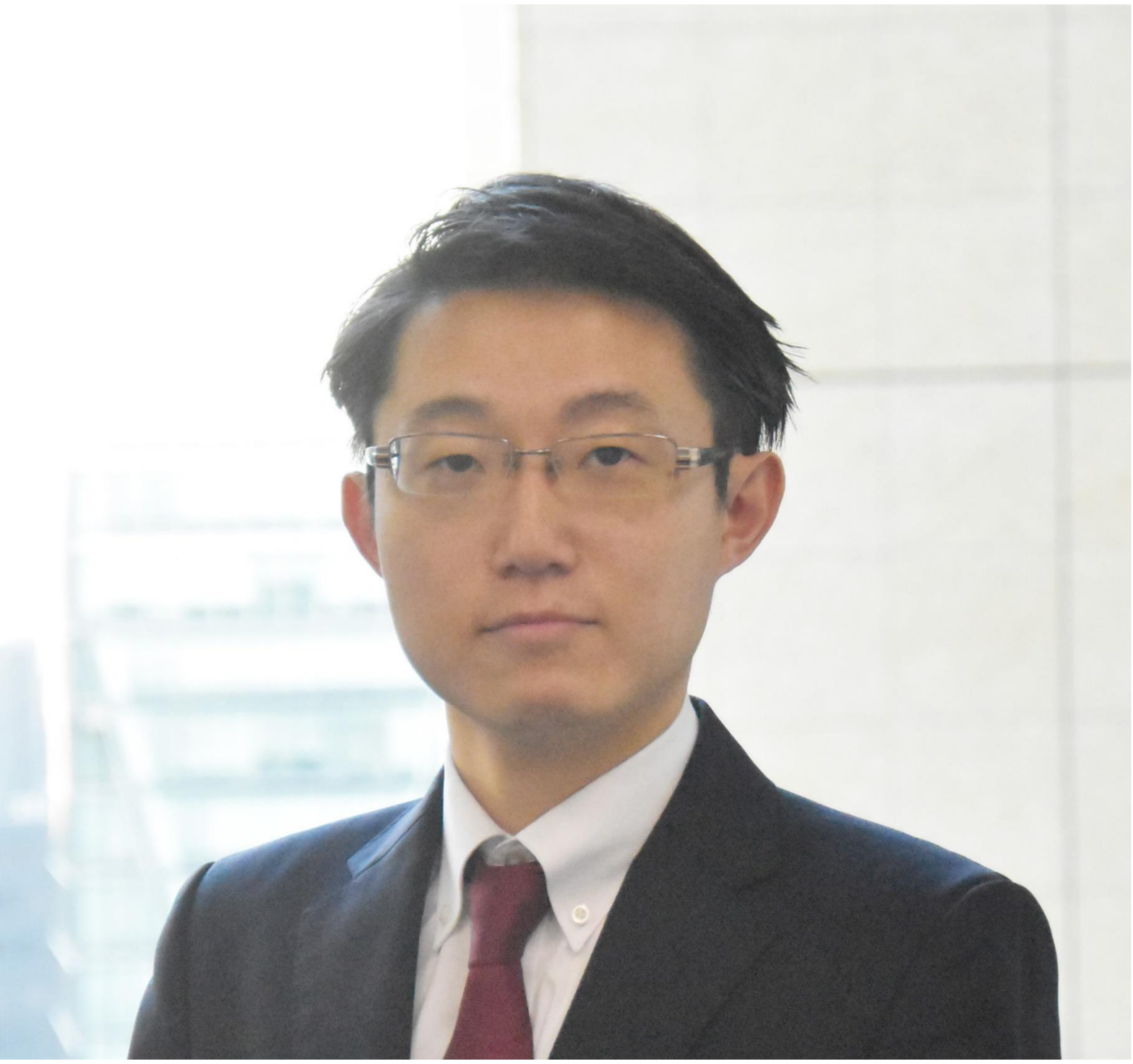}}]{Tomoki Tanaka} is a Vice President of Mitsubishi UFJ Financial Group, Inc. (MUFG). He received his B.S. in Science in 2009 and his M.S. in Mathematical Sciences in 2011 from Nagoya University. His major is Topology, especially Knot theory. In 2011, he entered MUFG and from 2018, MUFG participated in IBM Quantum Network Hub @ Keio University and he joined this project as a project researcher of Keio Quantum Computing Center. He researches quantum computing for using financial applications, such as derivatives simulation, risk management, optimization and machine learning. Contact him at tomoki\_tanaka@mufg.jp.
\end{IEEEbiography}

\begin{IEEEbiography}
    [{\includegraphics[width=1in,height=1.25in,clip,keepaspectratio]{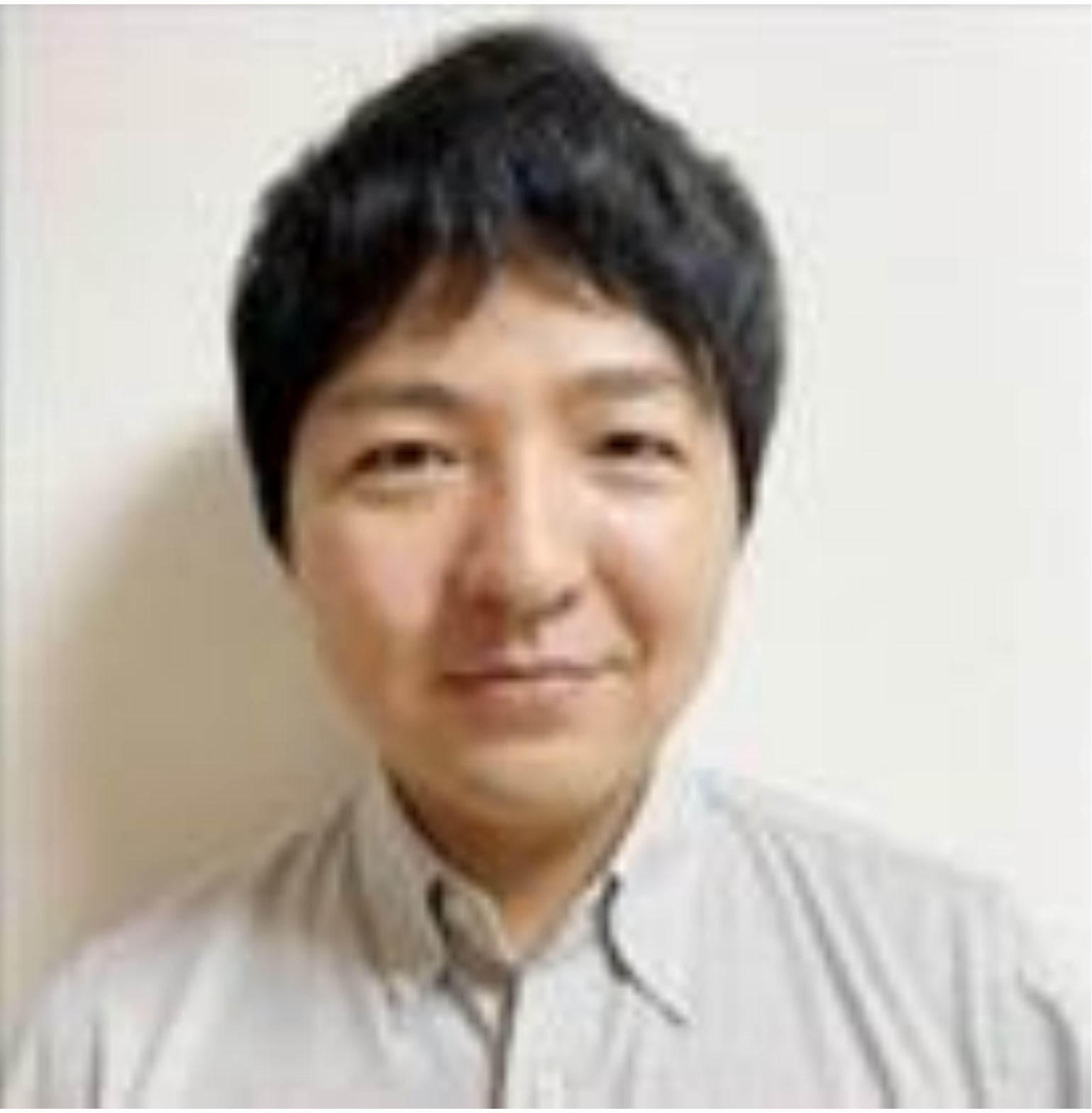}}]
    {Shumpei~Uno} is a Chief Consultant of Mizuho Research \& Technologies, Ltd(MHRT).
	He holds a M.Sc and a Ph.D in Particle Physics from Nagoya University.
During the Ph.D, he formulated quantum electrodynamics on finite volume lattice in order to accurately predict the light quark masses.
 He entered MHRT in 2011 and become a project researcher of Keio Quantum Computing Center in 2018.
 He researches quantum computing for using financial applications, such as derivatives simulation, risk management, optimization and machine learning. Contact him at shumpei.uno@mizuho-ir.co.jp.
\end{IEEEbiography}

\begin{IEEEbiography}[{\includegraphics[width=1in,height=1.25in,clip,keepaspectratio]{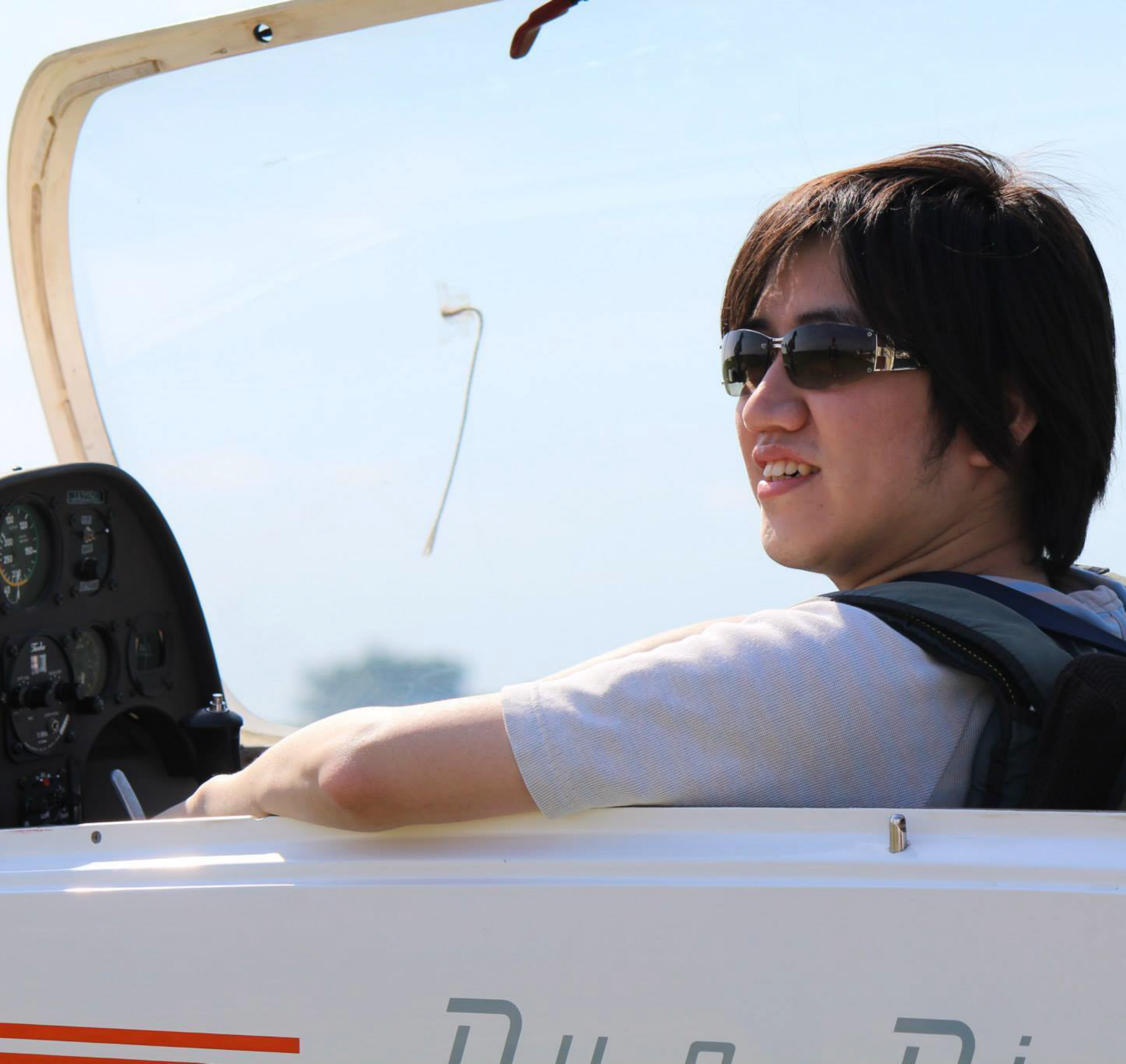}}]{Takahiko~Satoh} is a project assistant professor of Keio University Quantum Computing Center. He studied at Keio University and the University of Tokyo in 2015. He received a PhD in computer science from UT. His research field is quantum computing and quantum networking, particularly quantum network coding, NISQ algorithm design, and Quantum Internet security. He is a member of the Physical Society of Japan (JPS).
Contact him at satoh@sfc.wide.ad.jp.
\end{IEEEbiography}

\begin{IEEEbiography}[{\includegraphics[width=1in,height=1.25in,clip,keepaspectratio]{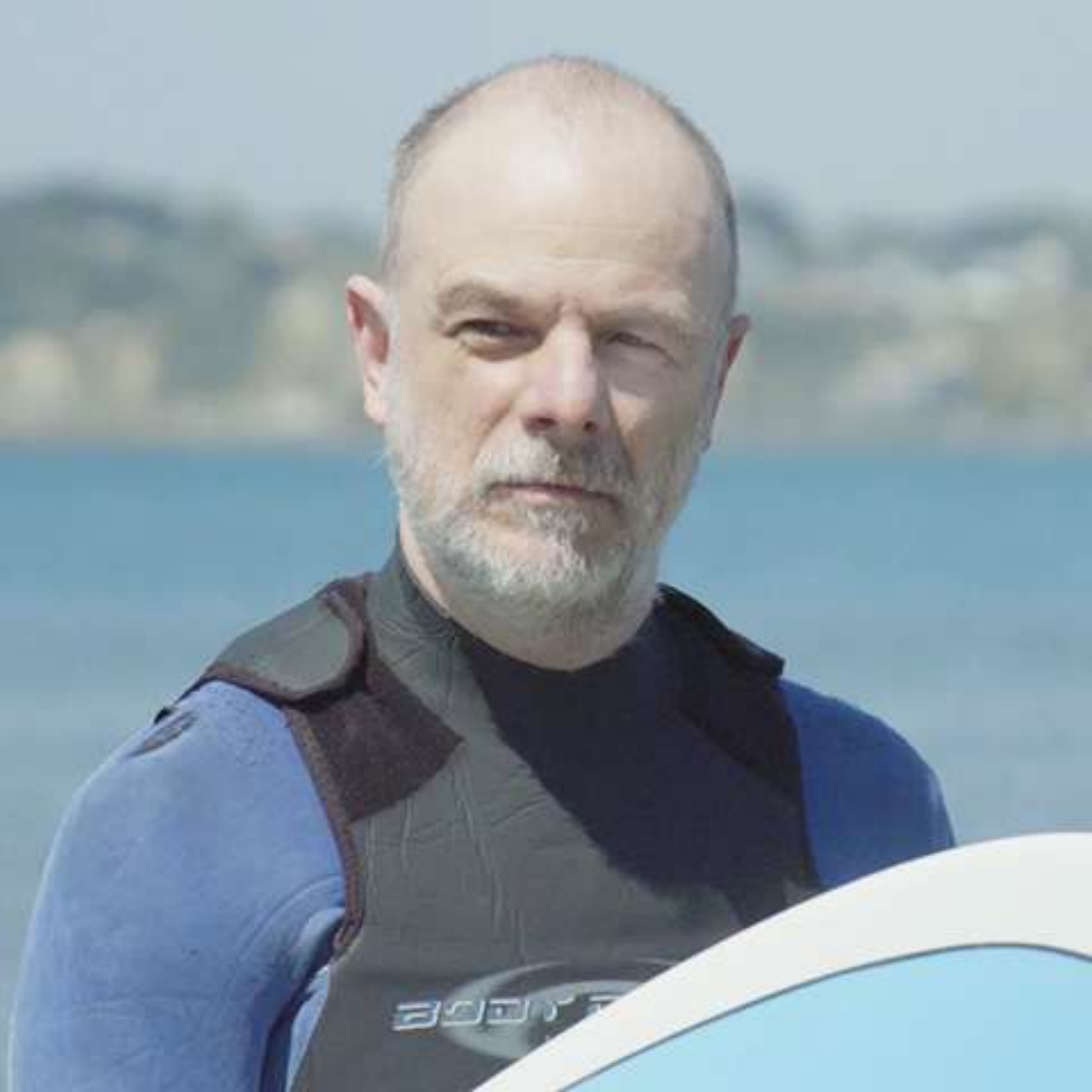}}]
{Rodney~Van~Meter} (Senior Member) is a  professor of Environment and Information Studies at Keio University's Shonan Fujisawa Campus. He is vice center chair of the Keio Quantum Computing Center, a board member of the WIDE Project, and a member of the Quantum Internet Task Force. Besides quantum networking and quantum computing, his research interests include storage systems, networking, and post-Moore's law computer architecture. Van Meter received a PhD in computer science from Keio University. He is member of ACM, the American Physical Society, and the American Association for the Advancement of Science (AAAS). Contact him at rdv@sfc.wide.ad.jp.

\end{IEEEbiography}

\begin{IEEEbiography}[{\includegraphics[width=1in,height=1.25in,clip,keepaspectratio]{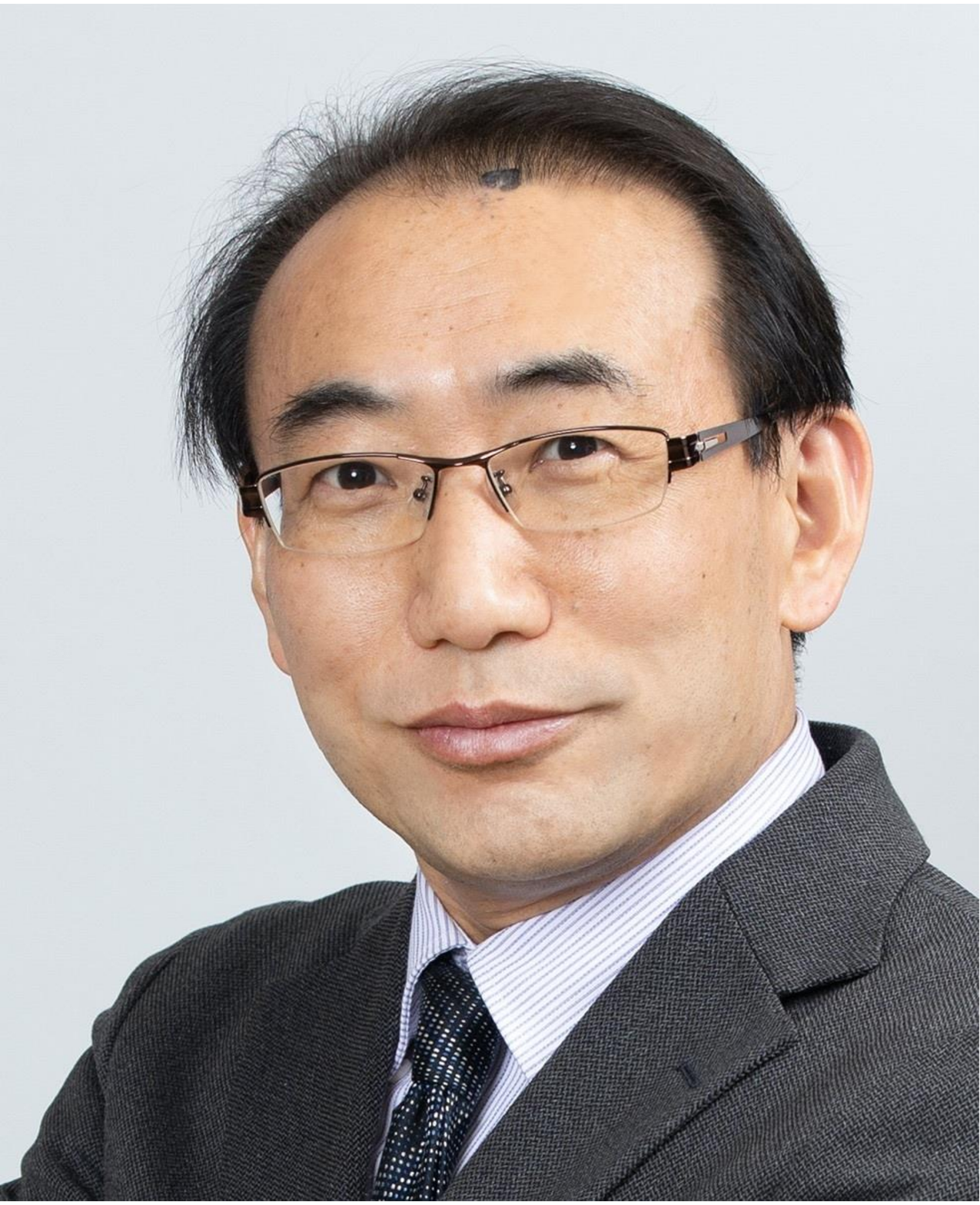}}]{Noboru~Kunihiro} received his B. E., M. E. and Ph.D. in mathematical engineering and information physics from the University of Tokyo in 1994, 1996 and 2001, respectively.
He has been a professor of University of Tsukuba since 2019.
He was a researcher of NTT Communication Science Laboratories from 1996 to 2002.
He was an associate professor of the University of Electro-Communications from 2002 to 2008.
He was an associate professor of the University of Tokyo from 2008 to 2019.
His research interest includes cryptography, information security and quantum computation.

\end{IEEEbiography}

\EOD

\end{document}